\documentclass[structabstract]{aa}  %

\usepackage{xcolor}
\usepackage{soul}
\usepackage{graphicx}

\usepackage{mathrsfs}

\usepackage{hyperref}
\hypersetup{
    colorlinks = true,
    citecolor ={blue}
}

\usepackage{tablefootnote}
\usepackage{longtable}
\usepackage{lscape}
\usepackage{natbib}

\usepackage{bm}
\usepackage{txfonts}
\usepackage{supertabular,booktabs}

\usepackage{subfigure}

\usepackage{txfonts}
\usepackage{natbib}
\bibpunct{(}{)}{;}{a}{}{,} % to follow the A&A style
\begin{document}  

\title{An analysis of spectroscopic, seismological, astrometric, and photometric masses of pulsating white dwarf stars}

\author{Leila M. Calcaferro\inst{1,2},
Alejandro H. C\'orsico\inst{1,2}, 
Murat Uzundag\inst{3},
Leandro G. Althaus\inst{1,2},
S. O. Kepler\inst{4}, \and
Klaus Werner\inst{5}}  
\offprints{acorsico@fcaglp.unlp.edu.ar} 

\institute{$^1$ Grupo  de Evoluci\'on  Estelar y  Pulsaciones,  Facultad de 
           Ciencias Astron\'omicas  y Geof\'{\i}sicas, Universidad
           Nacional de La Plata, Paseo del Bosque s/n, (1900) La
           Plata, Argentina\\   
           $^{2}$ Instituto de Astrof\'{\i}sica
           La Plata, CONICET-UNLP, Paseo del Bosque s/n, (1900) La
           Plata,
           Argentina\\ 
           $^{3}$ Institute of Astronomy, KU Leuven, Celestijnenlaan 200D, 3001, Leuven, Belgium\\
           $^4$ Instituto de F\'{\i}sica da Universidade Federal do Rio Grande do Sul, 91501-970 Porto Alegre, Brazil\\
           $^5$ Institut f\"ur Astronomie und Astrophysik, Kepler Center for Astro and Particle Physics, Eberhard 
           Karls Universit\"at, Sand 1, 72076 T\"ubingen, Germany\\    
\email{acorsico@fcaglp.unlp.edu.ar}}
\date{Received, }  

\abstract {A central challenge in the field of stellar astrophysics
  lies in accurately determining the mass of stars, particularly when
  dealing with isolated ones. However, for pulsating white dwarf
  stars, the task becomes more tractable due to the availability of
  multiple approaches such as spectroscopy, asteroseismology,
  astrometry, and photometry, each providing valuable insights
  into their mass properties.}  {Numerous asteroseismological studies
  of white dwarfs have been published, focusing on determining
    stellar mass using pulsational spectra and comparing it with
    spectroscopic mass, which uses surface temperature and
    gravity. The objective of this work is to compare these mass
    values in detail and, in turn, to compare them with the mass
    values derived using astrometric parallaxes/distances and
    photometry data from {\it Gaia}, employing two methods, the
    astrometric and the photometric ones.}  {Our analysis encompasses
  a selection of pulsating white dwarfs with different surface
  chemical abundances that define the main classes of variable white
  dwarfs. We calculated their spectroscopic masses, compiled
  seismological masses, and determined astrometric masses. We
    also derived photometric masses, when possible. Subsequently, we
  compared all the sets of stellar masses obtained through these
  different methods. To ensure consistency and robustness in our
  comparisons, we used identical white dwarf models and evolutionary
  tracks across all  four methods.}  {The analysis suggests a
  general consensus among the four methods regarding the masses
  of pulsating white dwarfs with hydrogen-rich atmospheres,
  known as DAV or ZZ Ceti stars, especially
  for objects with masses below approximately $0.75 M_{\sun}$,
  although notable disparities emerge for certain massive stars. For
  pulsating white dwarf stars with helium-rich atmospheres,
  called DBV or V777 Her stars, we find
  that astrometric masses generally exceed seismological,
  spectroscopic,  and photometric masses. Finally,
  while there
  is agreement among the sets of stellar masses for pulsating
  white dwarfs with carbon-, oxygen-, and helium-rich atmospheres
   (designated as GW Vir stars), 
  outliers exist where mass determinations by various methods show
  significant discrepancies.}  {Although a general agreement exists
  among different methodologies for estimating the mass of pulsating
  white dwarfs, significant discrepancies are prevalent in many
  instances. This shows the need to redo the determination of
  spectroscopic parameters, the parallax, and/or improve
  asteroseismological models for many stars.}
 
  \authorrunning{Calcaferro et al.}
  \titlerunning{Stellar masses of pulsating  white dwarfs}
  \maketitle
%----------------------------------------------------------------  
   
\section{Introduction}  
\label{intro}  

The mass of stars stands as a fundamental quantity in stellar
astrophysics, as it shapes the entire life cycle of stars, from their
birth to their death \citep[see,
  e.g.,][]{2004sipp.book.....H,2013sse..book.....K}. Stellar masses
cover a vast range, extending from approximately $0.08$ to about $150$
times the mass of the Sun ($M_{\sun}$) and even beyond. The accurate
determination of stellar mass is pivotal for a myriad of studies of
formation, evolution, ages, and distances of stellar populations, as
well as investigations into the chemical composition of stars,
supernovae, asteroseismology, exoplanets, and more. Nevertheless,
precisely measuring stellar mass poses challenges, particularly for
isolated stars lacking companions. For an exhaustive exploration of
the various methodologies employed in measuring stellar mass, we
recommend consulting the comprehensive review article by
\cite{2021A&ARv..29....4S} and the references provided therein.

White dwarf (WD) stars represent the predominant fate among stars in
the Universe
\citep[e.g.][]{2010A&ARv..18..471A,2022PhR...988....1S}. In fact, most
stars whose progenitor masses are below $8-10.5 M_{\sun}$, depending
on metallicity, will end their evolution as WDs
\citep[e.g.][]{2014MNRAS.441..582D}.  The observed mass range of WDs
spans from approximately $0.17 M_{\sun}$
\citep[SDSS~J091709.55+463821.8;][]{2007ApJ...660.1451K} to around
$1.35 M_{\sun}$
\citep[ZTF~J190132.9+145808.7;][]{2021Natur.595...39C}. The precise
measurement of WD masses is pivotal in numerous astrophysical
studies. It plays a crucial role in the assessment of the
initial-to-final mass relationship of WD \citep{1977A&A....59..411W,
  2008MNRAS.387.1693C, 2018ApJ...860L..17E,
  2019ApJ...871L..18C}. Additionally, WD masses are crucial to compute
the WD luminosity function, which serves as a valuable tool to infer
the age, structure, and evolution of the Galactic disk, as well as the
nearest open and globular clusters
\citep{2001PASP..113..409F,2009ApJ...697..965B,2010Natur.465..194G,2015MNRAS.448.1779B,2013MNRAS.433..243C,2016MNRAS.456.3729C,2016NewAR..72....1G,2017ApJ...837..162K}. 

In select situations, the mass of a WD can be measured directly using
model-independent methods. This is notably observed in astrometric WD
binaries that possess precise orbital parameters that allow a dynamic
determination of their mass \citep[e.g.][]{2015ApJ...813..106B,
  2017ApJ...840...70B, 2017ApJ...848...16B}, also in detached
eclipsing binaries \citep[e.g.][]{2017MNRAS.470.4473P}, and by means
of the gravitational redshift of spectral lines
\citep[e.g.][]{2019A&A...627L...8P}. In the vast majority of cases,
however, WDs are discovered in isolation, necessitating the estimation
of their stellar masses through model-dependent methods. The key
methods we will discuss in this study include the determination of the
spectroscopic mass, the seismological mass, the astrometric mass,
and the photometric mass. 

The spectroscopic mass of WDs is derived using the effective
temperature and surface gravity, acquired by fitting the atmosphere
models $\chi^2$ to stellar spectra with line profiles.  This is called
the spectroscopic technique, which has historically been the most
successful technique to obtain the atmospheric parameters $T_{\rm
  eff}$ and $\log g$ of WDs. These parameters are commonly referred to
as spectroscopic $T_{\rm eff}$ and $\log g$
\citep{1992ApJ...394..228B, 2005ApJS..156...47L,
  2009ApJ...696.1755T}. The process of assessing the stellar mass from
$T_{\rm eff}$ and $\log g$ involves the utilisation of evolutionary
tracks of WDs in the gravity vs. effective temperature plane, often
referred to as "Kiel diagrams".  An illustrative example of the
derivation of spectroscopic masses for a large sample of WDs based on
$\log g$ and $T_{\rm eff}$ can be found in the research conducted by
\cite{2013ApJS..204....5K}.

The seismological masses of WDs are determined by asteroseismology, a
technique that involves comparing the pulsation spectra observed in
the $g$ (gravity) mode in variable WDs with the theoretical spectra
calculated on the appropriate grids of the WD models
\citep{2008ARA&A..46..157W,2008PASP..120.1043F,2010A&ARv..18..471A,2019A&ARv..27....7C}. Continuous
observations from space, exemplified by missions such as {\it CoRoT},
{\it Kepler}, and {\it TESS}, have significantly advanced the field of
WD asteroseismology
\citep{2020FrASS...7...47C,2022BAAA...63...48C,2022MNRAS.511.1574R,2023MNRAS.518.1448R}. Asteroseismology
has been shown to be effective in obtaining stellar masses of isolated
pulsating WDs \citep{2012MNRAS.420.1462R,
  2018Natur.554...73G}. Seismological models can be constructed by
fitting individual periods for each pulsating star, allowing the
derivation of the seismological mass. In instances where a constant
period spacing is discernible in the observed pulsation spectrum, the
seismological mass can also be determined by comparing this period
spacing with the uniform period spacings calculated for theoretical
models \citep[see, for
  instance,][]{1987fbs..conf..297K,2021A&A...645A.117C}. This
particular method relies on the spectroscopic effective
temperature of the star and its associated uncertainties.

The astrometric masses of WDs can be determined by calculating the
theoretical distances of WD models associated with evolutionary tracks
of varying masses. This process involves utilizing the apparent
magnitude of a WD and the absolute magnitude of the models. The
calculated theoretical distance for different stellar masses is then
compared with the astrometric distance of the WD, derived from its
parallax, thus allowing for a mass estimate. The availability of
  accurate measurements from {\it Gaia}  \citep{2020yCat.1350....0G}
  gives this technique particular relevance. It is important to note
that the determination of stellar masses for WDs through this method
is also dependent on the spectroscopic effective temperature of
the star.  

Finally, the photometric masses of WDs can be assessed by fitting
  photometry and astrometric parallaxes or distances, employing
  synthetic fluxes from model atmospheres, to constrain the WD
  $T_{\rm eff}$ and radius
  \citep{1997ApJS..108..339B,2001ApJS..133..413B,2019ApJ...876...67B,2019MNRAS.482.4570G}. The
  subsequent use of the mass-radius relationships yields the stellar
  mass. The accurate trigonometric {\it Gaia} pallarax/distance measurements, coupled with
  large photometric surveys such as the Sloan Digital Sky Survey
  \citep[SDSS;][]{2000AJ....120.1579Y}, and the Panoramic
  Survey Telescope and Rapid Response System
  \citep[Pan-STARRS;][]{2016arXiv161205560C}, have made of this a very
  precise technique to measure the mass of isolated WDs \citep[see
    e.g.,][]{2019ApJ...876...67B,2019ApJ...871..169G,2019MNRAS.482.4570G,2019MNRAS.482.5222T}.

Many asteroseismological analyses providing the stellar masses of pulsating WDs have been carried out so far, and these estimates have been compared with spectroscopic or photometric determinations of stellar mass. 
It has been concluded that these sets of masses generally agree with each other. However, a detailed comparative analysis has not been carried out until now. The main objective of the present analysis is to determine to what extent the different methods used to derive the stellar mass of isolated pulsating WDs are consistent with each other. In this paper, we undertake a comparative analysis of WD stellar masses employing the methods previously described. In particular, we make extensive use of astrometric distance estimates provided by {\it Gaia}, which allow us to make a new estimate of the stellar mass of isolated WDs.  Our approach involves using identical evolutionary tracks and model WDs across all procedures. Specifically, for evaluating spectroscopic, astrometric, and photometric masses, we employ the same evolutionary tracks associated with the sets of WD stellar models utilized in asteroseismological analyses to derive seismological masses. This strategy ensures consistency and robustness when comparing the four stellar mass estimates. 

The paper is structured as follows. In Section \ref{sec:samples}, we provide a brief overview of the samples of stars analyzed in this study. The specific objects considered are listed in the Appendix \ref{sec:appendix}. Section \ref{sec:mass_derivation} outlines the four methods used to derive stellar mass, while Section \ref{sec:analysis} is dedicated to presenting a comparative analysis of the mass determinations obtained for our sample of stars. We discuss potential reasons for discrepancies among the different mass determinations in Section \ref{discussion}. Finally, we offer a summary and draw conclusions from our findings in Section \ref{sec:conclusions}.

\section{Samples of stars}
\label{sec:samples}

\begin{figure}
\centering
 \includegraphics[clip,width=1.0\linewidth]{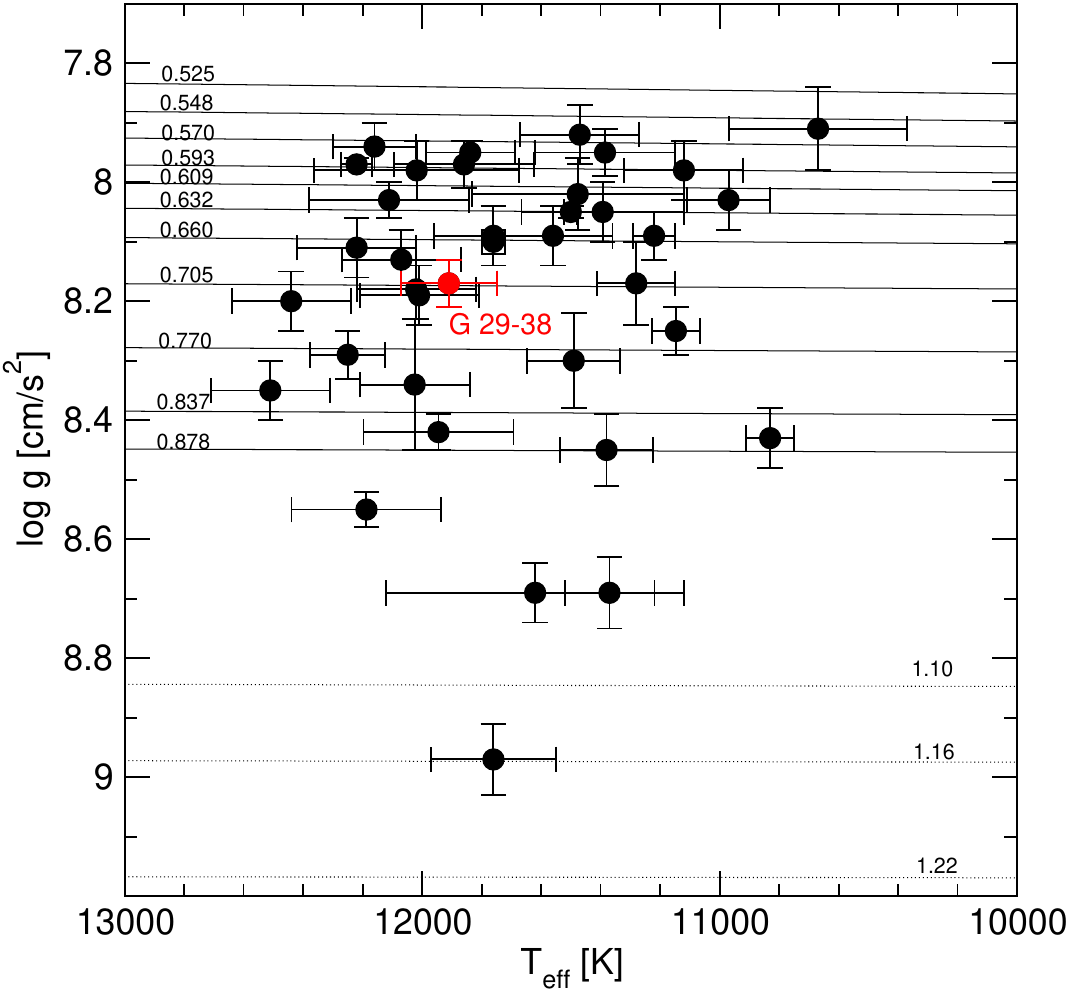}
	\caption{Location of the sample of DAV stars considered in this work on the 
 $T_{\rm eff}-\log g$ plane, depicted with black circles. Solid curves 
 show the CO-core DA WD evolutionary tracks from \cite{2010ApJ...717..183R},
  and dotted curves display the ultra-massive ONe-core DA WD
  evolutionary tracks from \cite{2019A&A...625A..87C}, for different 
  stellar masses. The location of the DAV star G~29$-$38 is emphasized with a red symbol.}
	\label{fig:01}
\end{figure}

\begin{figure}
\centering
 \includegraphics[clip,width=1.0\linewidth]{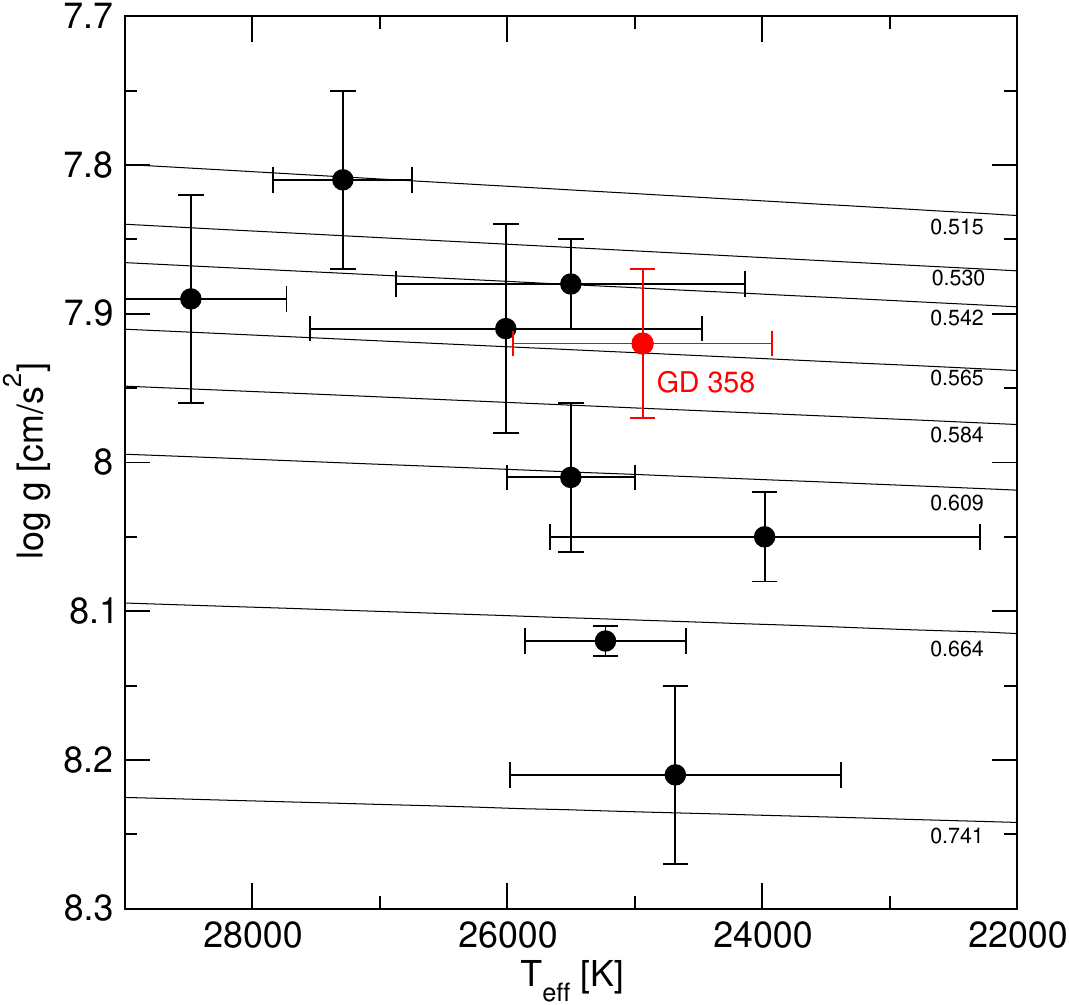}
	\caption{Location of the sample of DBV stars considered in this work 
 on the $T_{\rm eff}-\log g$ diagram, marked with  black circles. 
 Thin solid curves show the CO-core DB WD
  evolutionary tracks from \cite{2009ApJ...704.1605A} for
  different stellar masses. The location of the DBV star GD~358 is emphasized with a red symbol.}
	\label{fig:02}
\end{figure}

\begin{figure}
\centering
 \includegraphics[clip,width=1.0\linewidth]{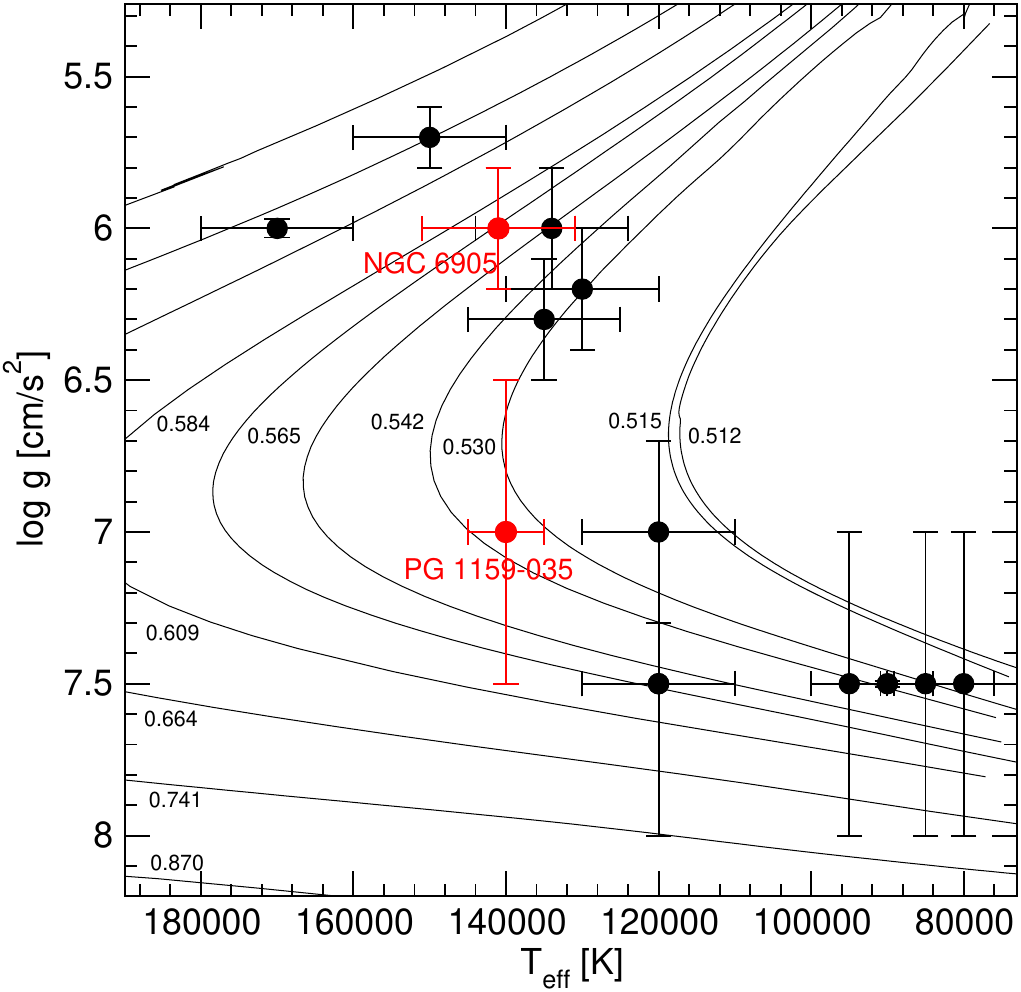}
 \caption{Location of the sample of GW Vir variable stars
   considered in this work
  in the  $T_{\rm eff}- \log g$
  plane depicted with black circles. Thin   solid   curves   show   the
  CO-core PG~1159 evolutionary tracks from  \cite{2006A&A...454..845M}
  for  different  stellar  masses. The locations of the GW Vir stars PG~1159$-$035 (DOV type) and NGC~6904 (PNNV type) are marked with red symbols.}
	\label{fig:03}
\end{figure}

We have selected samples of three types of pulsating WD stars: DAV\footnote{We do not consider in this study the pulsating low-mass and extremely low-mass helium-core WDs \citep[also known as ELMVs; see e.g.,][]{2019A&ARv..27....7C}, that are also hydrogen-rich atmosphere WDs, the results of which will be presented in a separate work (Calcaferro et al. 2024, in preparation).} (spectral type DA, with hydrogen-rich atmospheres), DBV (spectral type DB, with helium-rich atmospheres), and GW~Vir (spectral types PG~1159  and [WC], with oxygen-, carbon- and helium-rich atmospheres) stars, respectively. Within the category of GW Vir stars, we include DOV-type stars (GW Vir stars that lack a nebula) and PNNV-type stars (GW Vir stars that are still surrounded by a nebula). The stars discussed in this study have been seismologically analysed using evolutionary models generated with the {\tt LPCODE} evolutionary code \citep{2022A&A...663A.167A} and pulsation periods of $g$ modes calculated with the {\tt LP-PUL} pulsation code \citep{2006A&A...454..863C}, both developed by the La Plata Group\footnote{\url{http://evolgroup.fcaglp.unlp.edu.ar/}}. In Tables \ref{table:sample_DAVS}, \ref{table:sample_DBVS}, and \ref{table:sample_GWVIR}, we present the details of the stars included in our study. The tables provide information such as star names, equatorial coordinates, apparent magnitude $V$, and DR3 {\it Gaia} apparent magnitudes $G$, $G_{\rm BP}$, and $G_{\rm RP}$. In addition, they list spectral type, spectroscopic effective temperature, and surface gravity,  DR3 {\it Gaia} parallax \citep{2023A&A...674A...1G}, and geometric distance from \cite{2021AJ....161..147B}. For DBV and GW Vir stars (Tables \ref{table:sample_DBVS}, and \ref{table:sample_GWVIR}), we have included an extra column that indicates interstellar extinction ($A_V$). It should be noted that the GW~Vir star PG~2131+066 lacks parallax data from {\it Gaia}, so we could not obtain the distance from \cite{2021AJ....161..147B}. Instead, we have included the distance derived by \cite{2000ApJ...545..429R} using the spectroscopic parallax of the nearby M star to which PG~2131+066 appears to compose a binary.

The extinction values listed in Tables~\ref{table:sample_DBVS} and \ref{table:sample_GWVIR} were obtained using the following approach. We used the Python package {\tt dustmaps}\footnote{\url{https://github.com/gregreen/dustmaps}} to derive reddening values $E(B-V)$ for each target location in the sky, based on the 3D reddening map {\tt Bayestar17} \citep{2018MNRAS.478..651G}. The central $E(B-V)$ value was directly extracted from these maps, serving as our primary estimate of extinction at each target location. Furthermore, we determined the upper and lower percentiles of the extinction values (typically at 84.1\% and 15.9\%, respectively) to establish uncertainty limits. Subsequently, we computed the error in reddening by measuring the difference between the central value and the upper and lower percentiles.
We then calculated the extinction values for the $V$ band ($A_V$) using the formula $A_V = R_V\  E(B-V)$, using $R_V = 3.2$ \citep{2004ASPC..309...33F}. For DBV stars located below a declination of $-30$ degrees, we used the 2D "SFD" dust maps within the {\tt dustmaps} package, based on the catalogue compiled by \cite{1998ApJ...500..525S}, as the {\tt Bayestar17} catalogue does not cover this range. Our $A_V$ values for GW Vir stars closely match those reported by \cite{2023ApJS..269...32S}, which is reasonable given that both sets of extinctions were derived from the {\tt Bayestar17} map.

Furthermore, we used {\tt Bayestar17} to determine the extinction values for DAVs (not included in Table \ref{table:sample_DAVS}) and compared them with the values obtained from the Montreal WD database \citep{2017ASPC..509....3D}. The results showed a strong agreement between the two sets of values. Due to the negligible impact of extinction on DAVs owing to their proximity to the Sun, we did not account for it in our calculations when determining astrometric masses.

\begin{figure}
\centering
 \includegraphics[clip,width=1.0\linewidth]{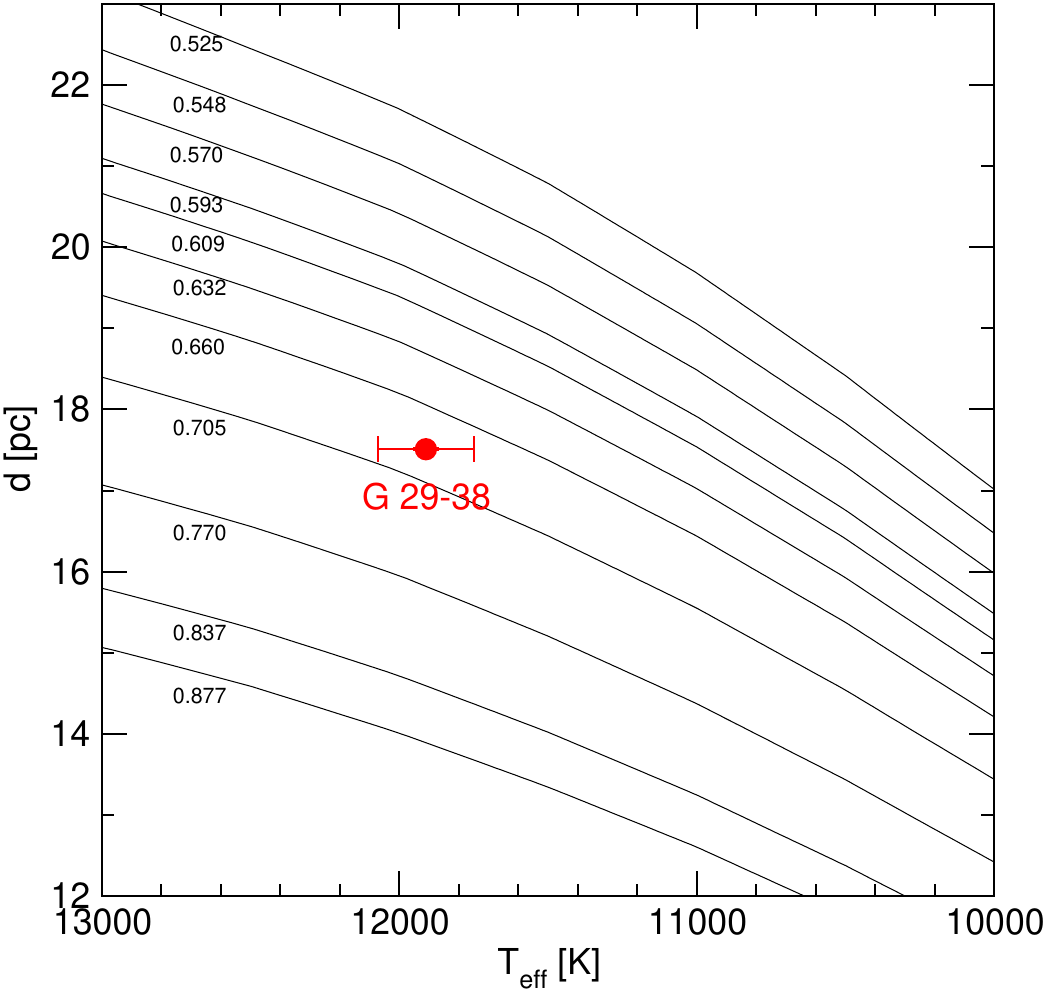}
	\caption{Distance vs effective temperature curves corresponding to evolutionary sequences of DA WD models extracted from \cite{2010ApJ...717..183R} with different stellar masses  and for the apparent magnitude $V$ of the DAV star G~29$-$38, whose location is indicated with a red circle with error bars. This star is characterized by $T_{\rm eff}= 11\,910 \pm 162$ K and $d_{\rm BJ}= 17.51 \pm 0.01$ pc. The uncertainty in the distance is so small that it is contained within the symbol. From linear interpolation,  the astrometric mass of G~29$-$38 is $M_{\rm Astr}= 0.684^{+0.011}_{-0.014} M_{\sun}$.}
	\label{fig:g29-38}
\end{figure}

\begin{figure}
\centering
 \includegraphics[clip,width=1.0\linewidth]{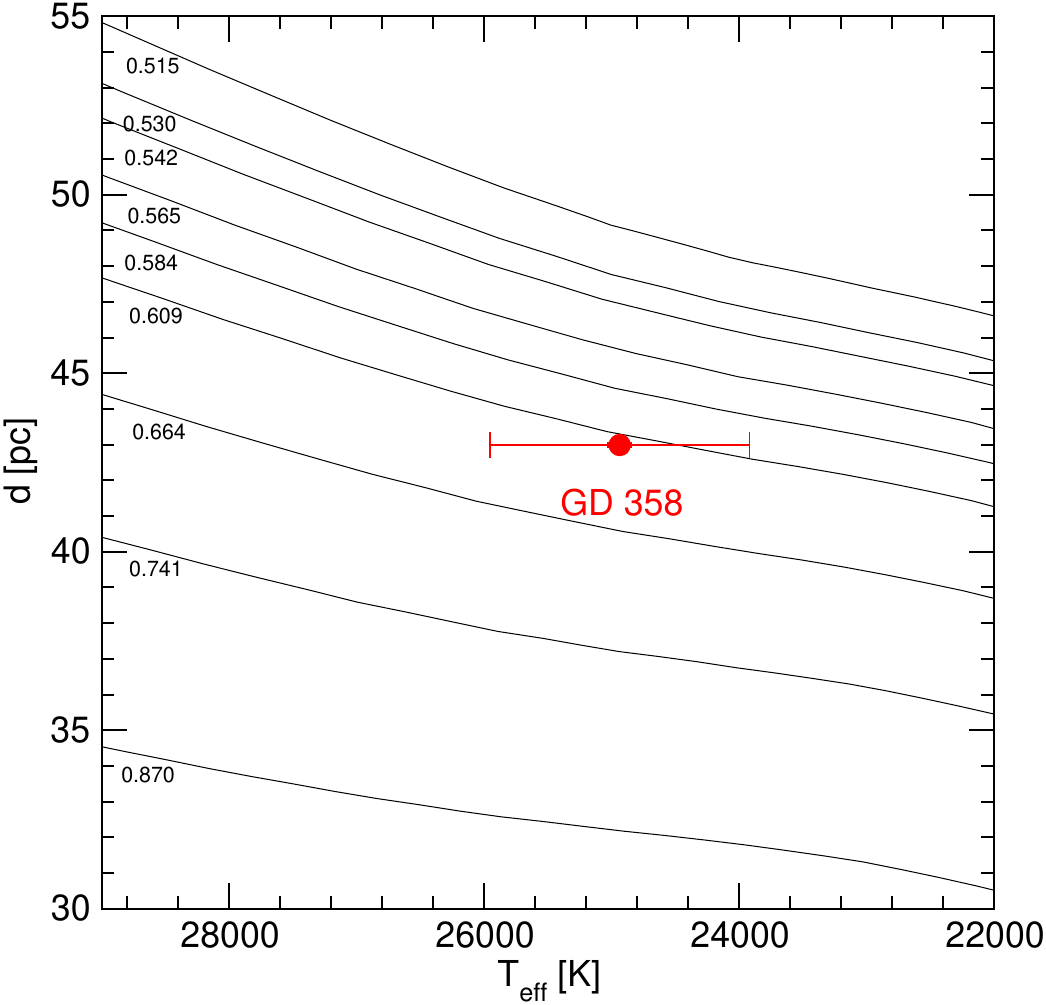}
	\caption{Distance vs effective temperature curves corresponding to evolutionary sequences of DB WD models extracted from \cite{2009ApJ...704.1605A} with different stellar masses and the apparent magnitude $V$ of the 
 DBV star GD~358 (red circle with error bars), characterized by $T_{\rm eff}= 24\,937 \pm 1018$ K and $d_{\rm BJ}= 42.99 \pm 0.05$ pc. The uncertainty in the distance is so small that it is contained within the symbol. From linear interpolation,  the astrometric mass of GD~358 is $M_{\rm Astr}= 0.616^{+0.016}_{-0.017} M_{\sun}$.}
	\label{fig:gd358}
\end{figure}

\begin{figure}
\centering
 \includegraphics[clip,width=1.0\linewidth]{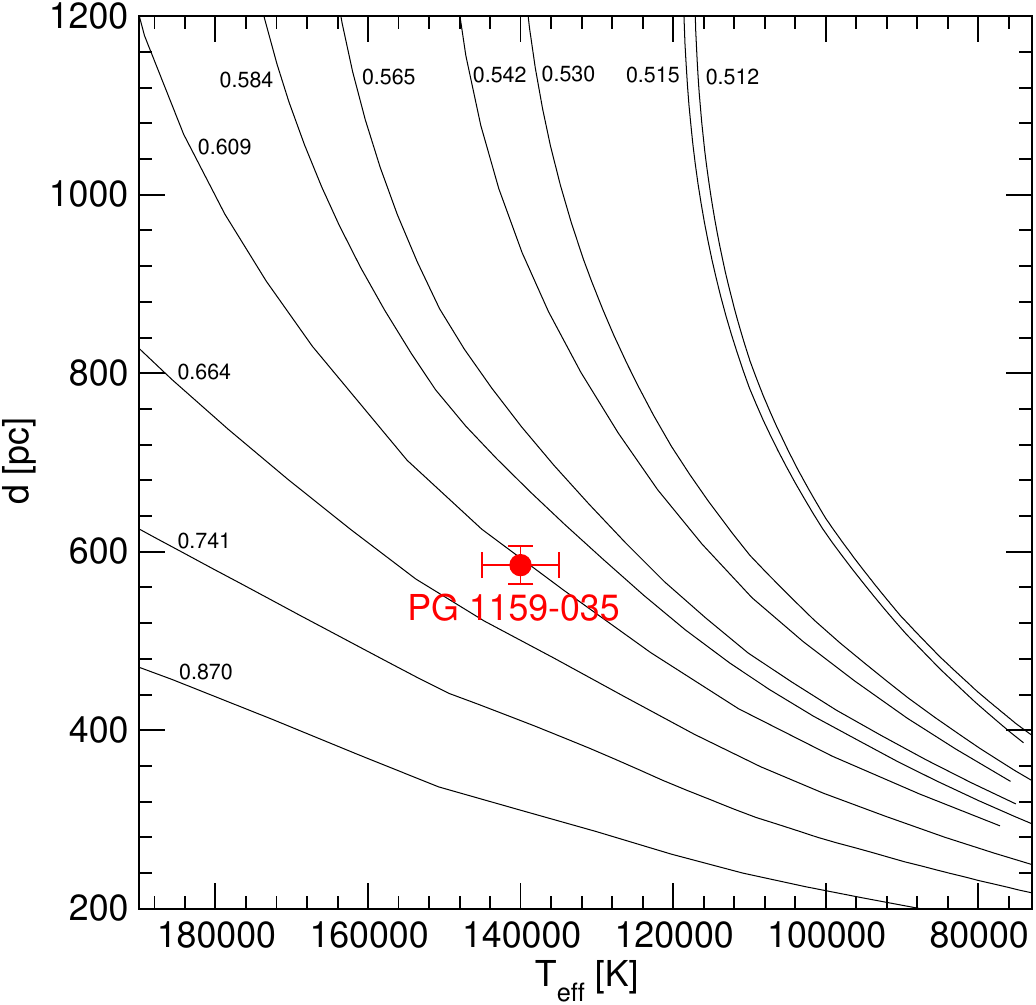}
	\caption{Distance vs effective temperature curves corresponding to evolutionary sequences of PG~1159 models extracted from \cite{2006A&A...454..845M} with different stellar masses and the apparent magnitude $V$ corresponding to the GW Vir star (DOV-type) PG~1159$-$035. The location of this star is marked with a red circle with error bars. The star is characterized by $T_{\rm eff}= 140\,000 \pm 5000$ K and $d_{\rm BJ}= 585^{+20}_{-21}$ pc. From linear interpolation,  the astrometric mass of PG~1159$-$035  is $M_{\rm Astr}= 0.614^{+0.028}_{-0.020} M_{\sun}$.}
	\label{fig:pg1159}
\end{figure}

\begin{figure}
\centering
 \includegraphics[clip,width=1.0\linewidth]{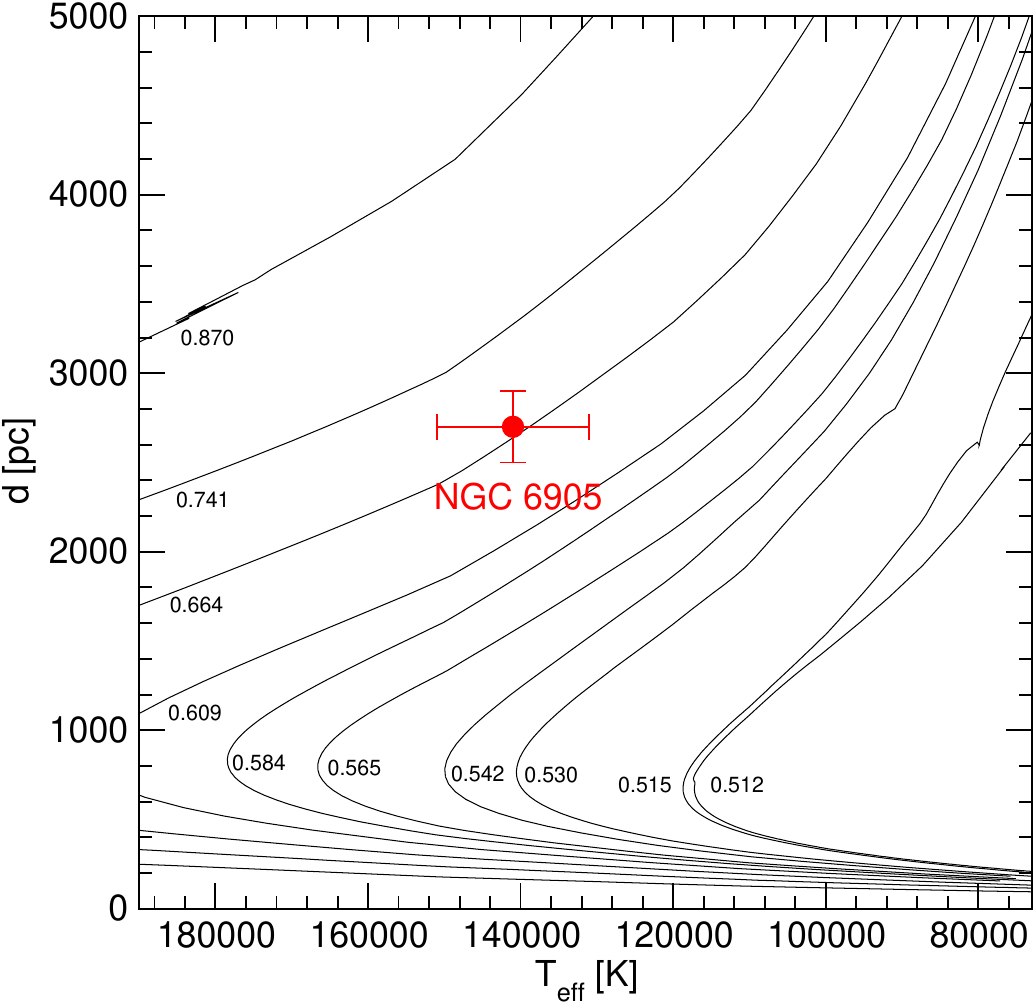}
	\caption{Distance vs effective temperature curves corresponding to evolutionary sequences of PG1159 models extracted from \cite{2006A&A...454..845M} with different stellar masses and the apparent magnitude $V$ corresponding to the GW Vir star (PNNV-type) NGC~6905 (red circle with error bars), characterized by $T_{\rm eff}= 141\,000 \pm 10000$ K and $d_{\rm BJ}= 2700\pm200$ pc. From linear interpolation,  the astrometric mass of NGC~6905  is $M_{\rm Astr}= 0.668^{+0.063}_{-0.044} M_{\sun}$.}
	\label{fig:ngc6905}
\end{figure}

It is widely recognised that DA WDs are formed with a range of hydrogen (H) envelope thicknesses rather than a single value, spanning the range $-14 \lesssim \log(M_{\rm H}/M_{\star}) \lesssim -3$. The stellar radius and surface gravity of the DA WDs are significantly influenced by the thickness of the H envelope. As a result, the evolutionary tracks in the $T_{\rm eff}$ vs. $\log g$ diagram for a given stellar mass vary, with models featuring thinner H envelopes displaying higher gravities \citep{2019MNRAS.484.2711R}. Although spectroscopic mass tabulations of DA WDs are commonly found, it is important to note that these values are derived from evolutionary tracks corresponding to DA WD models with "canonical" envelopes, representing the thickest possible H envelopes. We have observed a recurring error in several asteroseismological studies of DAVs, where seismological masses are often compared with spectroscopic masses. This error arises from comparing seismological masses derived from WD models with varying thicknesses of H envelopes, ranging from thick (canonical) to thin, with spectroscopic masses obtained from evolutionary tracks associated with WD models featuring only canonical H envelopes. This comparison is evidently incorrect.

Considering the various potential thicknesses of the H envelope, the
determination of the stellar mass of DA WDs becomes degenerate without
any external constraints. To simplify our analysis and avoid
complications, we focus exclusively on studying DAVs in which
asteroseismological models ---the DA WD models that most accurately
replicate the observed periods--- are characterised by canonical H
envelopes. This approach significantly reduces the number of DAVs
available for our analysis. For DBVs and GW Vir stars, we disregard
the possibility of thinner He and He/C/O envelopes, respectively, than
the canonical ones.

\section{Mass derivation}
\label{sec:mass_derivation}

In this section, we describe the methods that we use to determine the
mass of selected DAV, DBV, and GW Vir stars (Tables
\ref{table:sample_DAVS}, \ref{table:sample_DBVS}, and
\ref{table:sample_GWVIR}). It is important to clarify that the
spectroscopic, seismological, and astrometric masses are not entirely
independent, since the methods to obtain them rely on the effective temperature
($T_{\rm eff}$) derived from spectroscopy. Since
  the photometric method determines $T_{\rm eff}$ as part of its own
  fitting process, the photometric mass is independent of the
  spectroscopic $T_{\rm eff}$, at variance with the other methods.
Furthermore, it is
crucial to highlight that all  four sets of masses are
model-dependent. Spectroscopic masses are derived from observed
spectra, but are based on atmospheric models to determine $T_{\rm
  eff}$ and $\log g$,  and WD  evolutionary tracks of
  different masses to derive the stellar mass. Seismological masses are
obtained from measuring
the pulsation periods of $g$ modes using photometric techniques, but
they depend on the evolutionary pulsation models of WDs and the
specific asteroseismological technique employed (e.g., period fits
using different quality functions). Lastly, both astrometric and
  photometric masses are derived from {\it Gaia}'s parallaxes or their
  derived distances, combined with apparent magnitudes also measured
  by {\it Gaia}. However, these methods also rely on the modelling:
  the astrometric method uses evolutionary tracks that provide
  luminosity in terms of $T_{\rm eff}$ and bolometric corrections
  derived from atmospheric models which are necessary for converting
  to absolute magnitudes, while the photometric method depends on
  theoretical mass-radius relations and synthetic fluxes also derived
  from atmospheric models. Table \ref{tabla-methods} provides a
summary of the points discussed here.

\begin{table*}
\caption{Characteristics of the different methods considered in this work to assess the stellar mass of pulsating WDs.}
\centering
\begin{tabular}{l|ll}
\hline
\hline
\noalign{\smallskip}
Method of derivation & Observable & Theoretical \\
\noalign{\smallskip}
\hline 
\noalign{\smallskip}
Spectroscopy ($M_{\rm Spec}$) &  Electromagnetic spectra   &   Atmosphere models ($T_{\rm eff}, \log g$) \\
                                            &                             &   Evolutionary tracks ($T_{\rm eff} - \log g$ diagrams) \\       
\noalign{\smallskip}
\hline
\noalign{\smallskip}
Asteroseismology ($M_{\rm Seis}$)       &  $g$-mode periods ($\Pi$)      &  Evolution/pulsation models \\
                                            &  $g$-mode constant period spacing ($\Delta \Pi$) &  $T_{\rm eff}$ and $\log g$ to select seismological models \\  
\noalign{\smallskip}
\hline
\noalign{\smallskip}
Astrometry ($M_{\rm Astr}$) &  {\it Gaia}'s parallax/distance ($d_{\rm BJ}$)  &  Evolutionary tracks ($\log(L_{\star}/L_{\sun})$)\\
                                          &  {\it Gaia}'s apparent magnitudes  ($G, G_{\rm BP}, G_{\rm RP}$)&  Atmosphere models (bolometric correction, $BC$) \\ 
                                          &                                   &  Spectroscopic $T_{\rm eff}$ \\     
\noalign{\smallskip}
\hline
\noalign{\smallskip}
Photometry ($M_{\rm Phot}$) & {\it Gaia}'s parallax/distance ($d_{\rm BJ}$)  & Atmosphere models (synthetic fluxes, $H_\nu$)  \\
                                         & {\it Gaia}'s apparent  magnitudes ($G, G_{\rm BP}, G_{\rm RP}$)  &  Mass-radius relationships ($M_{\star}-R_{\star}$)  \\ 
\noalign{\smallskip}
\hline
\end{tabular}
\label{tabla-methods}
\end{table*}

\subsection{Spectroscopic masses}
\label{sec:spectroscopic_masses}

We determine the WD spectroscopic masses by the conventional method, that is, by interpolating the surface parameters $\log g$ and $T_{\rm eff}$ derived with the spectroscopic technique and presented in Tables \ref{table:sample_DAVS}, \ref{table:sample_DBVS} and \ref{table:sample_GWVIR}, on the evolutionary tracks provided by \cite{2010ApJ...717..183R} (CO-core DA WDs) and \cite{2019A&A...625A..87C} (ONe-core ultra-massive DA WDs) for DAVs, \cite{2009ApJ...704.1605A} for DBVs and \cite{2006A&A...454..845M} for GW Vir stars. The effective temperature and gravity of most of DAV and DBV stars of our samples have been extracted from the tabulations of \cite{2019A&ARv..27....7C} (see Tables \ref{table:sample_DAVS} and \ref{table:sample_DBVS}). These parameters have been corrected for 3D effects  \citep{2013A&A...559A.104T,2018MNRAS.481.1522C}. In the case of GW Vir stars, we have extracted the $T_{\rm eff}$ and $\log g$ values from various authors, as indicated in Table \ref{table:sample_GWVIR}. Figures \ref{fig:01}, \ref{fig:02}, and \ref{fig:03} display these evolutionary tracks in the $T_{\rm eff}- \log g$ diagram for each kind of pulsating WD, including the objects analysed in this work. In these figures, we emphasise some representative examples with red symbols and names. For DAVs, we consider only evolutionary tracks corresponding to DA WD models featuring canonical H envelope thicknesses. The derived spectroscopic masses corresponding to DAV, DBV, and GW Vir stars are provided in the second column of Tables \ref{table:masses-dav-sample}, \ref{table:masses-dbv-sample}, and \ref{table:masses-gwvir-sample}, respectively.

\subsection{Seismological masses}
\label{sec:seismological_masses}

We collect the seismological masses of pulsating WDs from asteroseismological models derived in the studies by \cite{2012MNRAS.420.1462R}, \cite{2013ApJ...779...58R}, \cite{2017ApJ...851...60R}, \cite{2019MNRAS.490.1803R}, \cite{2019A&A...632A.119C}, \cite{2022MNRAS.511.1574R}, \cite{2023MNRAS.518.1448R}, and \cite{2023MNRAS.526.2846U} for DAVs;
\cite{2012A&A...541A..42C}, \cite{2014A&A...570A.116B},  \cite{2019A&A...632A..42B},  \cite{2022A&A...659A..30C}, and \cite{2022A&A...668A.161C} for DBVs; and  \cite{2007A&A...475..619C},  \cite{2009A&A...499..257C}, \cite{2014MNRAS.442.2278K},  \cite{2016A&A...589A..40C},  \cite{2021A&A...645A.117C},  \cite{2021A&A...655A..27U}, \cite{2022ApJ...936..187O}, and \cite{2024arXiv240216642C}  for  GW Vir stars. The seismological masses
corresponding to DAV, DBV, and GW Vir stars derived from seismological models are provided in the third column of Tables \ref{table:masses-dav-sample}, \ref{table:masses-dbv-sample}, and \ref{table:masses-gwvir-sample}. In the case of DBV and GW Vir stars,  in many cases it has also been possible to estimate a seismological mass based on period spacing. These seismological masses are
provided in the fourth column of Tables \ref{table:masses-dbv-sample} and \ref{table:masses-gwvir-sample}. We emphasise that the sample of DAV stars is comprised solely of objects with asteroseismological models characterised by a canonical (thick) H envelope. This explains why our sample of DAVs (Table \ref{table:sample_DAVS}) is smaller than the total number of DAVs analysed seismologically to date.

\begin{table*}
\centering
\caption{Stellar masses of the DAV star sample. The first column displays the star names, the second column corresponds to the spectroscopic stellar mass ($M_{\rm Spec}$), the third column shows the seismological mass ($M_{\rm Seis}$), the fourth column shows the number of $g$-mode periods employed to derive the seismological mode ($N_{\Pi}$), the fifth column corresponds to the astrometric mass ($M_{\rm Astr}$), and the sixth column corresponds to the photometric mass ($M_{\rm Phot}$). It is important to note that the sample is comprised solely of DAV stars with asteroseismological models characterised by a canonical H-envelope thickness.  In the case of the stars analyzed by \cite{2022MNRAS.511.1574R} and \cite{2023MNRAS.518.1448R}, the uncertainties in the seismological masses are not available, as they have not been explicitly provided by the authors.}
\begin{tabular}{lcccccc}
\hline
\noalign{\smallskip}
Star            & $M_{\rm Spec}$ & $M_{\rm Seis}$ & $N_{\Pi}$ & $M_{\rm Astr}$ & $M_{\rm Phot}$\\
                 & [$M_{\sun}$] & [$M_{\sun}$] &  & [$M_{\sun}$]  & $[M_{\sun}]$\\
\hline
\noalign{\smallskip}
GD~244          & $0.656\pm0.030$ & $0.593\pm0.012^{(1)}$ & 5 & $0.651^{+0.025}_{-0.026}$ & $0.614^{+0.027}_{-0.025}$\\
\noalign{\smallskip}
G~226$-$29      & $0.815\pm0.032$ & $0.770\pm0.034^{(1)}$ & 1 & $0.815^{+0.015}_{-0.012}$  & $0.770^{+0.015}_{-0.014}$\\
\noalign{\smallskip}
HS~0507+0434B   & $0.715\pm0.031$ & $0.660\pm0.023^{(1)}$ & 4 & $0.715^{+0.021}_{-0.027}$ & $0.675^{+0.028}_{-0.033}$\\
\noalign{\smallskip}
EC~11507$-$1519 & $0.722\pm0.031$ & $0.705\pm0.033^{(1)}$ & 2 & $0.691^{+0.018}_{-0.020}$ & $0.620^{+0.025}_{-0.024}$\\
\noalign{\smallskip}
L~19$-$2        & $0.680\pm0.030$ & $0.705\pm0.033^{(1)}$ & 5 & $0.691^{+0.013}_{-0.017}$ & $0.668^{+0.015}_{-0.015}$\\
\noalign{\smallskip}
MCT~2148$-$2911 & $0.591\pm0.005$ & $0.632\pm0.014^{(1)}$ & 1 & $0.650^{+0.023}_{-0.021}$ & $0.589^{+0.027}_{-0.025}$\\
\noalign{\smallskip}
EC~14012$-$1446 & $0.709\pm0.031$ & $0.632\pm0.014^{(1)}$ & 9 & $0.697^{+0.036}_{-0.039}$ & $0.632^{+0.054}_{-0.048}$ \\
\noalign{\smallskip}
EC~23487$-$2424 & $0.655\pm0.030$ & $0.770\pm0.034^{(1)}$ & 3 & $0.667^{+0.031}_{-0.033}$ & $0.620^{+0.034}_{-0.031}$\\
\noalign{\smallskip}
GD~165          & $0.668\pm0.030$ & $0.632\pm0.014^{(1)}$ & 4 & $0.660^{+0.018}_{-0.016}$ & $0.639^{+0.018}_{-0.017}$\\
\noalign{\smallskip}
Ross~808           & $0.593\pm0.028$ & $0.705\pm0.033^{(1)}$ & 17 & $0.615^{+0.024}_{-0.025}$ & $0.591^{+0.020}_{-0.019}$ \\
\noalign{\smallskip}
HL~Tau$-$76     & $0.563\pm0.026$ & $0.548\pm0.012^{(1)}$ & 12 & $0.571^{+0.025}_{-0.026}$ & $0.532^{+0.027}_{-0.029}$ \\
\noalign{\smallskip}
GALEX~J0048+1521    & $0.702\pm0.043$ & $0.949\pm0.014^{(2)}$ & 6 & $0.821^{+0.067}_{-0.081}$ & $0.852^{+0.066}_{-0.060}$ \\
\noalign{\smallskip}
SDSS~J0843+0431     & $0.655\pm0.024$ & $0.837\pm0.021^{(2)}$ & 6 & $0.662^{+0.028}_{-0.028}$ & $0.594^{+0.026}_{-0.025}$\\
\noalign{\smallskip}
GALEX~J1257+0124     & $0.782\pm0.051$ & $0.705\pm0.023^{(2)}$ & 8 & $0.668^{+0.059}_{-0.064}$ & $0.744^{+0.053}_{-0.044}$\\
\noalign{\smallskip}
2QZ~J1323+0103     & $0.876\pm0.038$ & $0.917\pm0.020^{(2)}$ & 15 & $1.049^{+0.037}_{-0.033}$ & $0.820^{+0.040}_{-0.036}$\\
\noalign{\smallskip}
GALEX~J1612+0830     & $0.776\pm0.025$ & $0.705\pm0.023^{(2)}$ & 1 & $0.874^{+0.023}_{-0.037}$  & $0.830^{+0.024}_{-0.023}$\\
\noalign{\smallskip}
SDSS~J1641+3521     & $0.808\pm0.069$ & $0.721\pm0.025^{(2)}$ & 2 & $0.721^{+0.046}_{-0.059}$  & $0.596^{+0.049}_{-0.035}$\\
\noalign{\smallskip}
GALEX~J1650+3010     & $0.863\pm0.032$ & $1.024\pm0.013^{(2)}$ & 3 & $0.877^{+0.028}_{-0.028}$ & $0.824^{+0.027}_{-0.026}$ \\
\noalign{\smallskip}
GALEX~J2208+0654     & $0.750\pm0.025$ & $0.949\pm0.014^{(2)}$ & 2 & $0.832^{+0.028}_{-0.037}$ & $0.825^{+0.033}_{-0.031}$\\
\noalign{\smallskip}
KIC~11911480         & $0.575\pm0.021$ & $0.548\pm0.010^{(3)}$ & 5 & $0.692^{+0.025}_{-0.026}$ & $0.597^{+0.029}_{-0.023}$\\
\noalign{\smallskip}
GD~1212              & $0.620\pm0.029$ & $0.632\pm0.014^{(3)}$ & 7 & $0.641^{+0.018}_{-0.018}$ & $0.609^{+0.017}_{-0.016}$\\
\noalign{\smallskip}
GALEX~J1345$-$0055  & $0.662\pm0.012$ & $0.686\pm0.011^{(4)}$ & 2 & $0.636^{+0.017}_{-0.015}$  & $0.611^{+0.016}_{-0.015}$\\
\noalign{\smallskip}
SDSS~J2159+1322& $1.012\pm0.034$ & $0.917\pm0.040^{(4)}$ & 3 & $0.881^{+0.069}_{-0.070}$ & $0.928^{+0.060}_{-0.059}$\\
\noalign{\smallskip}
BPM~37093      & $1.012\pm0.029$ & $1.160\pm0.014^{(5)}$ & 8 & $1.045^{+0.029}_{-0.038}$  & $1.010^{+0.010}_{-0.010}$\\
\noalign{\smallskip}
GD~518         & $1.158\pm0.028$ & $1.220\pm0.030^{(5)}$ & 3 & $1.117^{+0.017}_{-0.013}$ & $1.106^{+0.020}_{-0.020}$\\
\noalign{\smallskip}
TIC~8445665    & $0.578\pm0.021$ & $0.675^{(6)}$  & 5 & $0.614^{+0.035}_{-0.034}$ & $0.580^{+0.038}_{-0.034}$\\
\noalign{\smallskip}
TIC~46847635   & $0.595\pm0.028$ & $0.686^{(6)}$   & 1 & $0.628^{+0.032}_{-0.032}$ & $0.597^{+0.047}_{-0.041}$\\
\noalign{\smallskip}
TIC~167486543  & $0.934\pm0.017$ & $0.820^{(6)}$   & 2 & $0.959^{+0.026}_{-0.025}$ & $0.930^{+0.023}_{-0.023}$\\
\noalign{\smallskip}
TIC~441500792  & $0.632\pm0.029$ & $0.705^{(6)}$   & 3 & $0.660^{+0.044}_{-0.046}$ & $0.630^{+0.048}_{-0.042}$\\
\noalign{\smallskip}
TIC~442962289  & $0.858\pm0.019$ & $0.837^{(6)}$   & 3 & $0.884^{+0.028}_{-0.036}$ & $0.851^{+0.030}_{-0.030}$\\
\noalign{\smallskip}
TIC~686044219  & $0.616\pm0.035$ & $0.639^{(6)}$   & 3 & $0.649^{+0.047}_{-0.052}$ & $0.616^{+0.059}_{-0.050}$\\
\noalign{\smallskip}
TIC~712406809  & $0.556\pm0.037$ & $0.646^{(6)}$   & 5 & $0.589^{+0.049}_{-0.048}$ & $0.563^{+0.066}_{-0.062}$\\
\noalign{\smallskip}
TIC~20979953   & $0.590\pm0.022$ & $0.593^{(6)}$   & 3 & $0.631^{+0.031}_{-0.029}$ & $0.591^{+0.038}_{-0.034}$\\
\noalign{\smallskip}
TIC~55650407   & $0.580\pm0.011$ & $0.570^{(6)}$   & 4 & $0.610^{+0.015}_{-0.018}$ & $0.580^{+0.017}_{-0.016}$\\
\noalign{\smallskip}
TIC~282783760  & $0.623\pm0.017$ & $0.593^{(6)}$   & 3 & $0.662^{+0.028}_{-0.030}$ & $0.621^{+0.031}_{-0.029}$\\
\noalign{\smallskip}
BPM~31594     & $0.632\pm0.006$  & $0.632^{(7)}$   & 6 & $0.643^{+0.015}_{-0.016}$ & $0.622^{+0.032}_{-0.029}$\\
\noalign{\smallskip}
G~29$-$38     & $0.703\pm0.025$  &  $0.632\pm0.030^{(8)}$ & 38 & $0.684^{+0.026}_{-0.022}$ & $0.629^{+0.027}_{-0.025}$\\
\noalign{\smallskip}
\hline
\noalign{\smallskip}
\end{tabular}
\tablefoot{(1) \cite{2012MNRAS.420.1462R}, (2) \cite{2013ApJ...779...58R}, 
(3) \cite{2017ApJ...851...60R}, (4) \cite{2019MNRAS.490.1803R}, (5) \cite{2019A&A...632A.119C}, 
(6) \cite{2022MNRAS.511.1574R},  
(7) \cite{2023MNRAS.518.1448R}, (8) \cite{2023MNRAS.526.2846U}.}
\label{table:masses-dav-sample}
\end{table*}

\begin{table*}
\centering
\caption{Similar to Table~\ref{table:masses-dav-sample}, but for the sample of DBV stars. The first column displays the star names, the second column corresponds to the spectroscopic stellar mass ($M_{\rm Spec}$), the third column shows the seismological mass ($M_{\rm Seis}$), The fourth column corresponds to the seismological mass inferred from the measured period spacing ($M_{\rm Seis (ps)}$),  the fifth column corresponds to the number of $g$-mode periods employed to derive the seismological model and the period spacing ($N_{\Pi}$), and the sixth and seventh columns are the astrometric ($M_{\rm Astr}$) and photometric ($M_{\rm Phot}$) masses, respectively.}
\begin{tabular}{lcccccc}
\hline
\noalign{\smallskip}
Star  & $M_{\rm Spec}$ & $M_{\rm Seis}$ &   $M_{\rm Seis (ps)}$ & $N_{{\Pi}}$ & $M_{\rm Astr}$ & $M_{\rm Phot}$\\
      & [$M_{\sun}$]   & [$M_{\sun}$]   &  [$M_{\sun}$]        & & [$M_{\sun}$] &  $[M_{\sun}]$\\
\hline
\noalign{\smallskip}
KIC~8626021    & $0.553\pm0.037$  &  $0.664\pm0.077^{(1)}$  & $0.696\pm0.031^{(1)}$ & 5 & $0.747^{+0.059}_{-0.057}$ & $0.680^{+0.049}_{-0.039}$\\
\noalign{\smallskip}
KUV~05134+2605 & $0.726\pm0.039$  &  $0.840\pm0.010^{(2)}$  & $0.850\pm0.050^{(2)}$ & 16 &  $0.798^{+0.033}_{-0.034}$ & $0.731^{+0.035}_{-0.033}$\\
\noalign{\smallskip}
TIC~257459955 &  $0.542\pm0.019$  &  $0.609\pm0.055^{(3)}$  & $0.621\pm0.057^{(3)}$ & 10 & $0.651^{+0.024}_{-0.027}$  & $0.596^{+0.022}_{-0.020}$\\
\noalign{\smallskip}
GD~358        &   $0.561\pm0.028$ &  $0.584\pm0.025^{(4)}$  & $0.588\pm0.024^{(4)}$  & 19 & $0.616^{+0.025}_{-0.026}$ & $0.579^{+0.034}_{-0.031}$\\ 
\noalign{\smallskip}
PG~1351+489     & $0.559\pm0.040$ &  $0.664\pm0.013^{(5)}$ &  $0.740$ to $0.870^{(5)}$  & 4 & $0.648^{+0.047}_{-0.047}$ & $0.598^{+0.038}_{-0.035}$\\
\noalign{\smallskip}
EC~20058$-$5234        & $0.611\pm0.029$ &  $0.664\pm0.013^{(5)}$  & $0.530$ to $0.550^{(5)}$ & 11 & $0.618^{+0.020}_{-0.020}$ & $0.556^{+0.021}_{-0.018}$ \\
\noalign{\smallskip}
EC~04207$-$4748       & $0.516\pm0.026$ &  $0.542\pm0.017^{(5)}$ & $0.565\pm0.020^{(5)}$ & 4 & $0.660^{+0.020}_{-0.019}$ & $0.580^{+0.020}_{-0.018}$\\
\noalign{\smallskip}
WD~J1527$-$4502 & $0.673\pm0.007$ &   $0.542\pm0.023^{(5)}$      & $\cdots$ &  4 & $0.694^{+0.018}_{-0.021}$ & $0.574^{+0.028}_{-0.026}$\\
\noalign{\smallskip} % TIC 150808542
L~7$-$44        & $0.631\pm0.020$ &   $0.565\pm0.023^{(5)}$      & $\cdots$    & 6 & $0.651^{+0.031}_{-0.030}$ & $0.594^{+0.026}_{-0.024}$\\  
\noalign{\smallskip}
\hline
\noalign{\smallskip}
\end{tabular}
\tablefoot{(1) \cite{2012A&A...541A..42C}, (2) \cite{2014A&A...570A.116B}, 
(3) \cite{2019A&A...632A..42B}, (4) \cite{2022A&A...659A..30C}, (5) \cite{2022A&A...668A.161C}.}
\label{table:masses-dbv-sample}
\end{table*}

\begin{table*}
\centering
\caption{Same as Table~\ref{table:masses-dbv-sample}, but for the sample of GW Vir stars.}
\begin{tabular}{lccccc}
\hline
\noalign{\smallskip}
Star  & $M_{\rm Spec}$ & $M_{\rm Seis}$  &  $M_{\rm Seis (ps)}$ & $N_{{\Pi}}$ & $M_{\rm Astr}$ \\
      & [$M_{\sun}$]   & [$M_{\sun}$]    & [ $M_{\sun}$]        & & [$M_{\sun}$] \\
\hline
\noalign{\smallskip}
PG~0122+200  & $0.526^{+0.172}_{-0.060}$  &  $0.556\pm0.014^{(1)}$ &  $0.567\pm0.013^{(1)}$ & 9 & $0.524^{+0.028}_{-0.014}$ \\
\noalign{\smallskip}
PG~2131+066 & $0.547^{+0.174}_{-0.074}$ & $0.589\pm0.024^{(2)}$ &  $0.578\pm0.022^{(2)}$ & 7 & $0.527^{+0.082}_{-0.047}$ \\
\noalign{\smallskip}
PG~1707+427 &  $0.536^{+0.172}_{-0.070}$& $0.542\pm0.014^{(2)}$ &  $0.566\pm0.024^{(2)}$ & 8 & $0.510^{+0.025}_{-0.035}$  \\
\noalign{\smallskip}
SDSS~J0754+0852  &$0.512^{+0.080}_{-0.018}$  & $0.556\pm0.014^{(3)}$  &  $\cdots$ & 3 &  $0.688^{+0.184}_{-0.168}$  \\ %(3) 
\noalign{\smallskip} %	GALEX J075415.1+085232
SDSS~J0349$-$0059 & $0.540^{+0.039}_{-0.064}$ &  $0.542\pm0.023^{(4)}$ & $0.535\pm0.004^{(4)}$ & 10 & $0.561^{+0.047}_{-0.033}$  \\%	KUV 03467-0108   
\noalign{\smallskip}
RX~J2117+3412 & $0.710^{+0.072}_{-0.059}$  & $0.565\pm0.024^{(5)}$ &  $0.569\pm0.015^{(5)}$ & 31 & $0.572^{+0.018}_{-0.018}$  \\ %TIC 117070953
\noalign{\smallskip} 
HS~2324+3944 & $0.531^{+0.049}_{-0.011}$  &  $0.664\pm0.077^{(5)}$ &  $0.727\pm0.017^{(5)}$ & 21 & $0.536^{+0.028}_{-0.015}$  \\
\noalign{\smallskip}
NGC~6905  &  $0.582^{+0.114}_{-0.049}$  & $\cdots$  &  $0.506$ to $0.818^{(5)}$   & 5 &   $0.668^{+0.067}_{-0.048}$       \\  
\noalign{\smallskip}
NGC~2371  & $0.534^{+0.044}_{-0.014}$ &  $0.664\pm0.077^{(5)}$  & $0.760\pm0.005^{(5)}$ & 10 &  $0.546^{+0.083}_{-0.018}$ \\
\noalign{\smallskip}
NGC~1501  & $0.562^{+0.093}_{-0.035}$  &  $0.609\pm0.055^{(5)}$ &  $0.586$ to $0.636^{(5)}$ & 24 & $0.779^{+0.056}_{-0.071}$   \\
\noalign{\smallskip}
TIC~333432673  & $0.580^{+0.174}_{-0.069}$ &  $0.589 \pm 0.020^{(6)}$ &  $0.600\pm0.011^{(6)}$ & 5  & $0.556^{+0.045}_{-0.027}$ \\
\noalign{\smallskip}
TIC~095332541  & $0.580^{+0.174}_{-0.069}$ &        $\cdots$    & $0.550$ to $0.570^{(6)}$  & 7 &  $0.601^{+0.047}_{-0.042}$  \\   
\noalign{\smallskip}
PG~1159$-$035      & $0.539^{+0.070}_{-0.010}$ & $0.565\pm0.024^{(7)}$ &  $0.536$ to $0.581^{(7)}$ & 41  & $0.614^{+0.031}_{-0.023}$  \\  
\noalign{\smallskip}
NGC~246          & $0.743^{+0.129}_{-0.106}$  & $0.570\pm0.024^{(8)}$ &  $0.568\pm0.012^{(8)}$ & 17 & $0.603^{+0.026}_{-0.029}$  \\    %(8)
\noalign{\smallskip}
\hline
\noalign{\smallskip}
\end{tabular}
\tablefoot{(1) \cite{2007A&A...475..619C}, (2) \cite{2009A&A...499..257C}, (3) \cite{2014MNRAS.442.2278K}, (4) \cite{2016A&A...589A..40C}, (5) \cite{2021A&A...645A.117C}, (6) \cite{2021A&A...655A..27U}, (7) \cite{2022ApJ...936..187O}, (8) \cite{2024arXiv240216642C}.}
\label{table:masses-gwvir-sample}
\end{table*}

\subsection{Astrometric masses}
\label{sec:astrometric_masses}

We calculate astrometric masses using the following procedure. We build distance curves versus $T_{\rm eff}$ by using Pogson's law, which states that
the distance of a star is given by $\log d = (V-M_V +5-A_V)/5$, where $V$ and $M_V$ represent the visual apparent and absolute magnitudes, respectively, and $A_V$ denotes the interstellar absorption in the $V$ band. Using the bolometric correction ($BC$) from the WD grids of the Montreal Group\footnote{\url{https://www.astro.umontreal.ca/~bergeron/CoolingModels/}} \citep{Bergeron_1995,2006AJ....132.1221H,2020ApJ...901...93B}, the absolute visual magnitude is calculated as $M_V= M_B- BC$, where $M_B=  M_{B \sun} - 2.5\ \log(L_{\star}/L_{\sun})$ and
$M_{B \sun}= 4.74$ \citep{2000asqu.book.....C}. The function $\log(L_{\star}/L_{\sun})$ in terms of $T_{\rm eff}$ for different values of $M_{\star}$ is obtained from the {\tt LPCODE} WD evolutionary tracks provided by \cite{2010ApJ...717..183R} (CO-core DA WDs) and \cite{2019A&A...625A..87C} (ONe-core ultra-massive DA WDs) for DAVs, \cite{2009ApJ...704.1605A} for DBVs, and \cite{2006A&A...454..845M} for GW Vir stars. Following \cite{2023ApJS..269...32S}, the apparent visual magnitude $V$ is evaluated based on the magnitudes $G$, $G_{\rm BP}$ and $G_{\rm RP}$ {\it Gaia} using the expression: 

\begin{eqnarray}
V= G + 0.02704 - 0.01424\ (G_{\rm BP}-G_{\rm RP}) \nonumber \\
+ 0.21560\ (G_{\rm BP}-G_{\rm RP})^2  \nonumber \\
- 0.01426\ (G_{\rm BP}-G_{\rm RP})^3, 
\end{eqnarray}

\noindent extracted from Table 5.9 of {\it Gaia} DR3 documentation\footnote{\url{https://gea.esac.esa.int/archive/documentation/GDR3/Data_processing/chap_cu5pho/cu5pho_sec_photSystem/}}. 
The uncertainties of $G$, $G_{\rm BP}$, and $G_{\rm RP}$ produce errors in $V$. To estimate the uncertainties $\Delta G$, $\Delta G_{\rm BP}$ and $\Delta G_{\rm RP}$, we use the following expressions:

\begin{eqnarray}
\Delta G= 1.089\ \frac{G}{(S/N)}, \nonumber \\
\Delta G_{\rm BP}= 1.089\ \frac{G_{\rm BP}}{(S/N)},  \nonumber \\
\Delta G_{\rm RP}= 1.089\ \frac{G_{\rm RP}}{(S/N}),
\end{eqnarray}

\noindent where $(S/N)$ is approximately equal to the parameter \texttt{phot\_g\_mean\_flux\_over\_error} of the {\it Gaia} database\footnote{\url{https://dc.zah.uni-heidelberg.de/gaia/q3/cone/info\#note-e}}. We propagate the errors of $G$, $G_{\rm BP}$ and $G_{\rm RP}$ to $V$ to obtain $\Delta V$ in the usual way (see the fourth column of Tables \ref{table:sample_DAVS}, \ref{table:sample_DBVS}, and \ref{table:sample_GWVIR}). We generate curves of $d$ as a function of $T_{\rm eff}$ for each stellar mass $M_{\star}$. We identify the target star based on its specified $T_{\rm eff}$ and the distance extracted from  \cite{2021AJ....161..147B} on the $d$ vs $T_{\rm eff}$ diagrams. By interpolating between these curves, we estimate the astrometric mass, $M_{\rm Astr}$. Each star will have its own set of $d$ vs. $T_{\rm eff}$ curves, as these curves depend on the star's magnitude $V$.
The resulting astrometric masses are provided in the fifth column of Table \ref{table:masses-dav-sample} for DAVs, and the sixth column of Tables \ref{table:masses-dbv-sample} and \ref{table:masses-gwvir-sample} for DBV and GW Vir stars.

In Fig. \ref{fig:g29-38}, we present a $T_{\rm eff}-d$ diagram illustrating the case of the DAV star G~29$-$38, while Fig. \ref{fig:gd358} shows the DBV star GD~358. Furthermore, Figs. \ref{fig:pg1159} and \ref{fig:ngc6905} highlight the GW Vir stars PG~1159$-$035 (DOV type) and NGC~6905 (PNNV type), respectively. For DAVs and DBVs, which are typically in close proximity to the Sun, the uncertainties in the distances provided by \cite{2021AJ....161..147B} are minimal, resulting in error bars that are contained within the symbols in Figs. \ref{fig:g29-38} and \ref{fig:gd358}. Consequently, the astrometric mass of these stars is primarily affected by uncertainties in $T_{\rm eff}$. In contrast, GW Vir stars are much more distant, leading to more significant uncertainties in their distances. This is evident in Figs. \ref{fig:pg1159} and \ref{fig:ngc6905}. For these stars, the uncertainties in the astrometric masses arise from both $T_{\rm eff}$ errors and $d_{\rm BJ}$. Despite these challenges, the uncertainties in astrometric masses of GW Vir stars are generally comparable to those of seismological masses, and both are notably smaller than the uncertainties associated with spectroscopic masses (refer to Table \ref{table:masses-gwvir-sample} and Sect. \ref{sec:GWVir} for further details).

\subsection{Photometric masses}
\label{sec:photometric_masses}

In the photometric method \citep[see,
    e.g.,][]{1997ApJS..108..339B,2019ApJ...876...67B}, the spectral
  energy distribution of a star is compared with model atmospheres
  using synthetic photometry. Magnitudes are converted to fluxes and
  compared with model fluxes averaged over the same filter
  bandpasses. The effective temperature ($T_{\rm eff}$) and solid
  angle ($\pi(R_{\star}/D)^2$) are considered free parameters, and the
  atmospheric composition is typically assumed to be pure H or He. If
  the star's distance ($D$) is known, often from trigonometric
  parallax measurements, its radius ($R_{\star}$) can be directly
  obtained, allowing for the calculation of the star's mass
  using WD mass-radius relationships
    ($M_{\star}-R_{\star}$).

The combination of accurate parallaxes from {\it Gaia} and
  optical photometry (e.g., SDSS {\it ugriz}, Pan-STARRS {\it grizy},
  or {\it Gaia}) has made the photometric method a robust
    tool for determining WD parameters \citep[see,
    e.g.,][]{2019ApJ...876...67B,2019ApJ...871..169G,2019MNRAS.482.4570G,2021MNRAS.508.3877G,2019MNRAS.482.5222T}. Notably,
  \cite{2019ApJ...871..169G,2019ApJ...882..106G} provided a thorough
  comparison of the spectroscopic and photometric parameters of a
  large sample of DA and DB WDs. 

In this work, we adopt a straightforward approach to determine the
photometric mass of the DAV and DBV stars in our sample.  We employ
the catalogue from \cite{2021MNRAS.508.3877G}, where $T_{\rm eff}$,
$\log g$, and $M_{\star}$ for a large selection of WDs were derived
using the photometric method based on {\it Gaia} astrometry and
photometry. To obtain the stellar radius ---a parameter not tabulated
in their catalogue--- we combined their parameters with the same
mass-radius relationships from \cite{2020ApJ...901...93B} that they
employed, yielding a model-independent value. Next, we apply our {\tt
  LPCODE} WD mass-radius relationships  to compute the photometric
mass ($M_{\rm phot}$), ensuring consistency with the other mass
estimates in this study. The results are listed in the last columns of
Tables~\ref{table:masses-dav-sample} and
\ref{table:masses-dbv-sample}. Regarding our sample of GW Vir stars,
to our very best knowledge, there are no photometric mass or radius
determinations. We hope that future studies will provide these
determinations to complete the picture. However, the issue in this case
is that these stars are exceptionally hot, resulting in an energy
distribution in the optical spectrum that adheres to the Rayleigh-Jeans
approximation. Consequently, this distribution becomes largely
insensitive to effective temperature, as illustrated, for example,
in Figure 3 of \cite{2020ApJ...901...93B}.
Therefore, it seems unlikely that photometric parameters will be
obtainable for these stars, at least not through the use of
optical photometry.

\section{Analysis}
\label{sec:analysis} 

In this section, we assess the consistency of spectroscopic, seismological, astrometric, and photometric (although only for DAVs and DBVs) masses through comparisons. We also examine seismological masses derived from period spacing for DBVs and GW Vir stars whenever feasible. To gauge the linear correlation between two sets of masses, we employ the Pearson coefficient $r$ \citep[see, for instance,][]{benesty2009pearson}. This coefficient ranges from $-1$ to $+1$, with $0$ indicating that there is no linear association, and a strong correlation as $r$ approaches $1$ in absolute value. However, a high correlation does not necessarily imply a good agreement between masses, as $r$ measures the strength of the relationship, not the agreement itself. For a more suitable assessment of the agreement, we turn to the Bland-Altman analysis \citep{1983TheStatistician...32..307A,2015BiochemMed...25..141G}. This involves plotting mass differences against their average values, helping to identify biases and outliers. The agreement limits are defined as $\pm 1.96 \sigma$, where $\sigma$ represents the standard deviation of the differences. In our analysis, we adopt a stricter criterion, using $\pm 1\sigma$ to define agreement limits, enabling the identification of outlier WD stars with significantly different masses derived from various methods.

\subsection{DAV stars}
\label{sec:DAVs}

\begin{figure}
\centering
  \includegraphics[clip,width=1.0\linewidth]{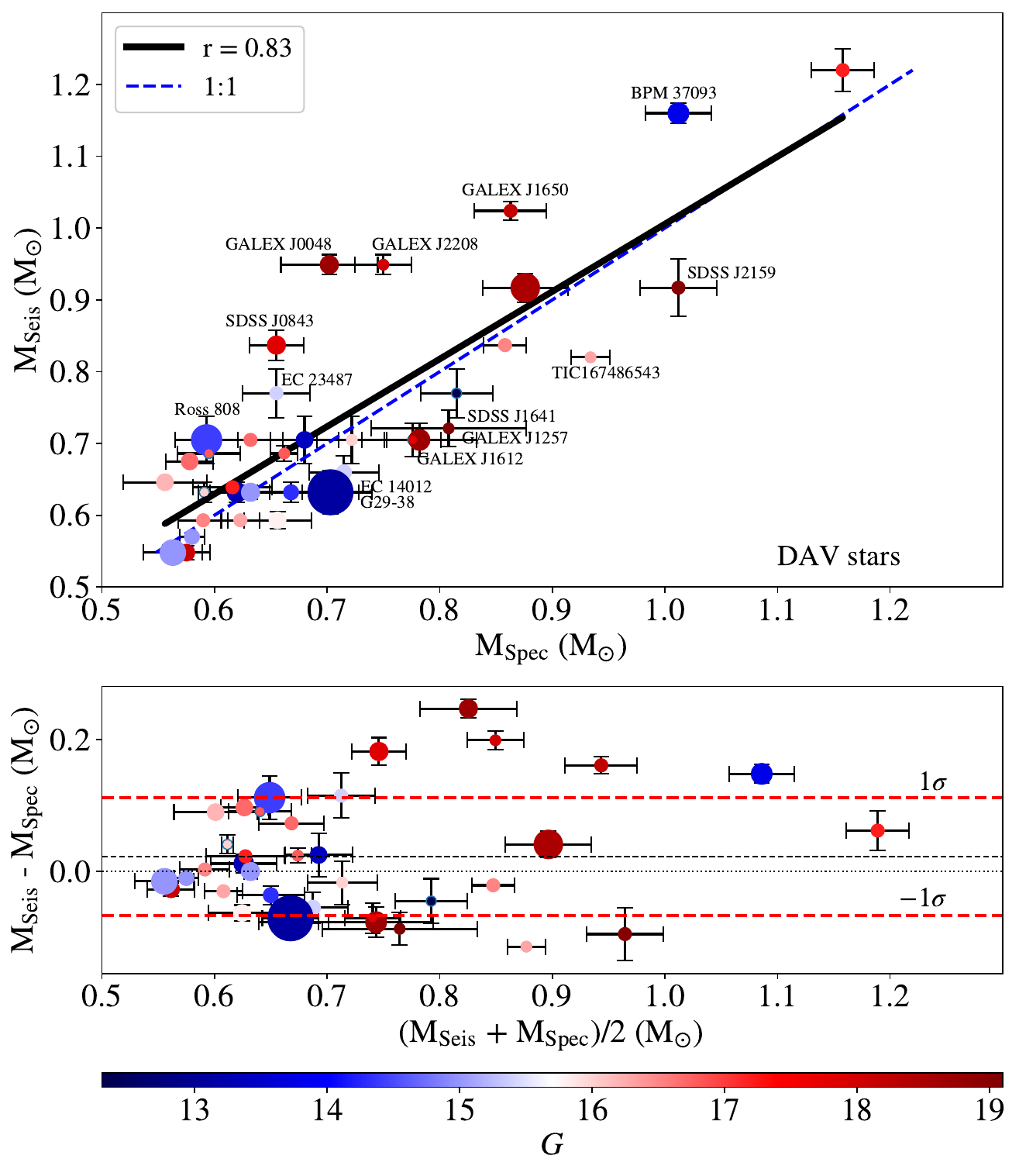}
	\caption{Upper panel: dispersion diagram showing the comparison between the stellar masses of DAVs obtained from spectroscopy and  those corresponding to seismological models (see Table \ref{table:masses-dav-sample}). The blue dashed line indicates the 1:1 correspondence between the two sets of stellar mass values. Stars whose mass estimates exhibit substantial discrepancies (i.e., deviate noticeably from the 1:1 line of correspondence) are labeled with their respective names. The Pearson coefficient, which measures the correlation between $M_{\rm Spec}$ and $M_{\rm Seis}$, is $r = +0.83$, revealing a strong correlation (the thick black line corresponds to the Pearson correlation fit). Lower panel: Bland-Altman diagram showing the mass difference in terms of the average mass for each object. The black short-dashed line corresponds to the mean difference, $\langle \Delta M_{\star} \rangle= +0.023 M_{\sun}$, whereas the two red dashed lines represent the limits of agreement, $\langle\Delta M_{\star}\rangle \pm 0.089 M_{\sun}$, considering the deviation of $\pm 1\sigma$. In both panels, the size of each symbol is proportional to the number of $g$-mode pulsation periods used to derive the seismological model (fourth column of Table \ref{table:masses-dav-sample}), and the colour palette indicates the apparent magnitude {\it Gaia} $G$ (fifth column of Table \ref{table:sample_DAVS}) of each star.}\label{fig:Mspec_vs_Mseis_DA}
\end{figure}

\begin{figure}
\centering
  \includegraphics[clip,width=1.0\linewidth]{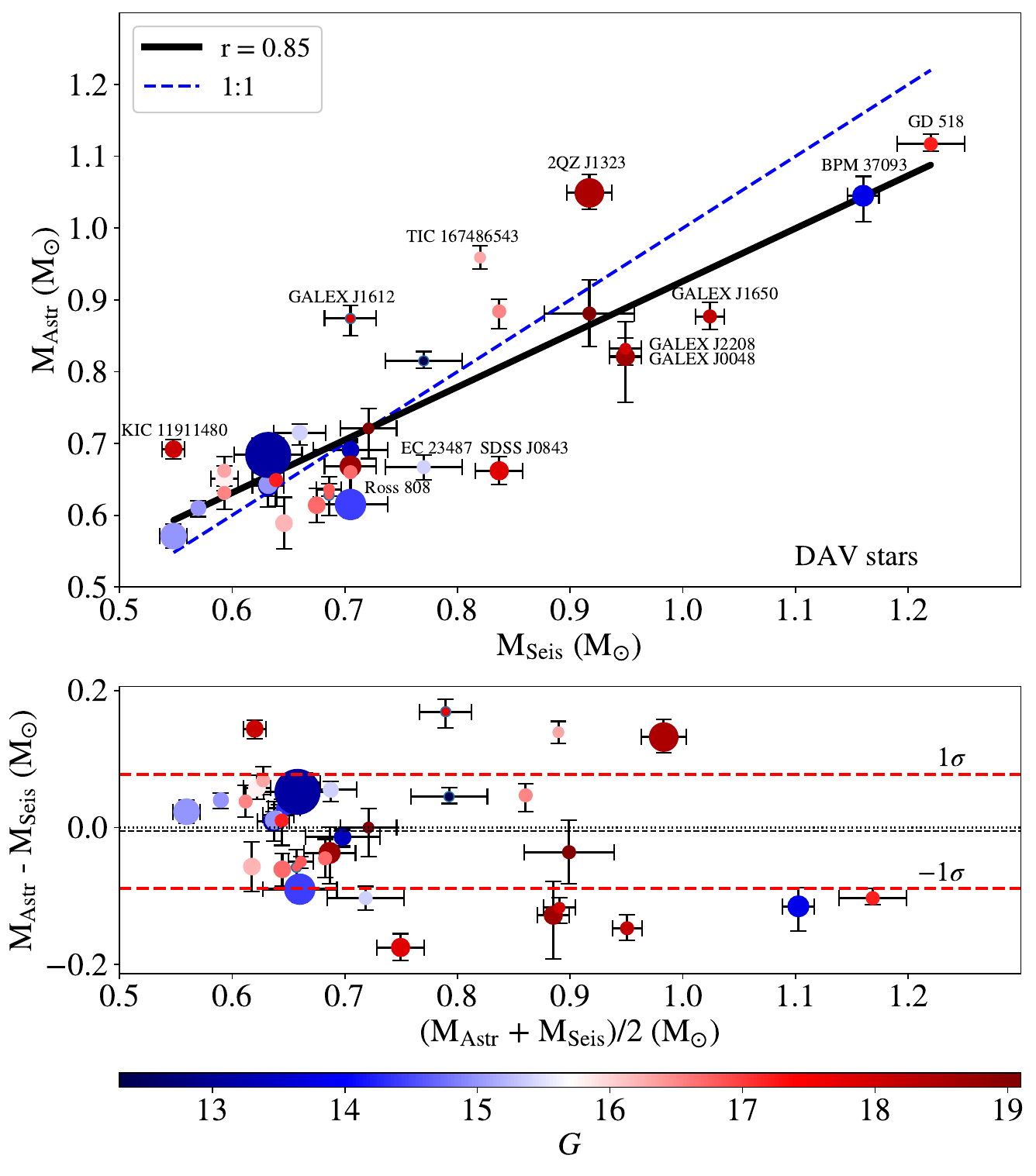}
	\caption{Similar to Fig. \ref{fig:Mspec_vs_Mseis_DA}, but for the comparison between seismological and astrometric masses.}
	\label{fig:Mseis_vs_Mastr_DA}
\end{figure}

\begin{figure}
\centering
 \includegraphics[clip,width=1.0\linewidth]{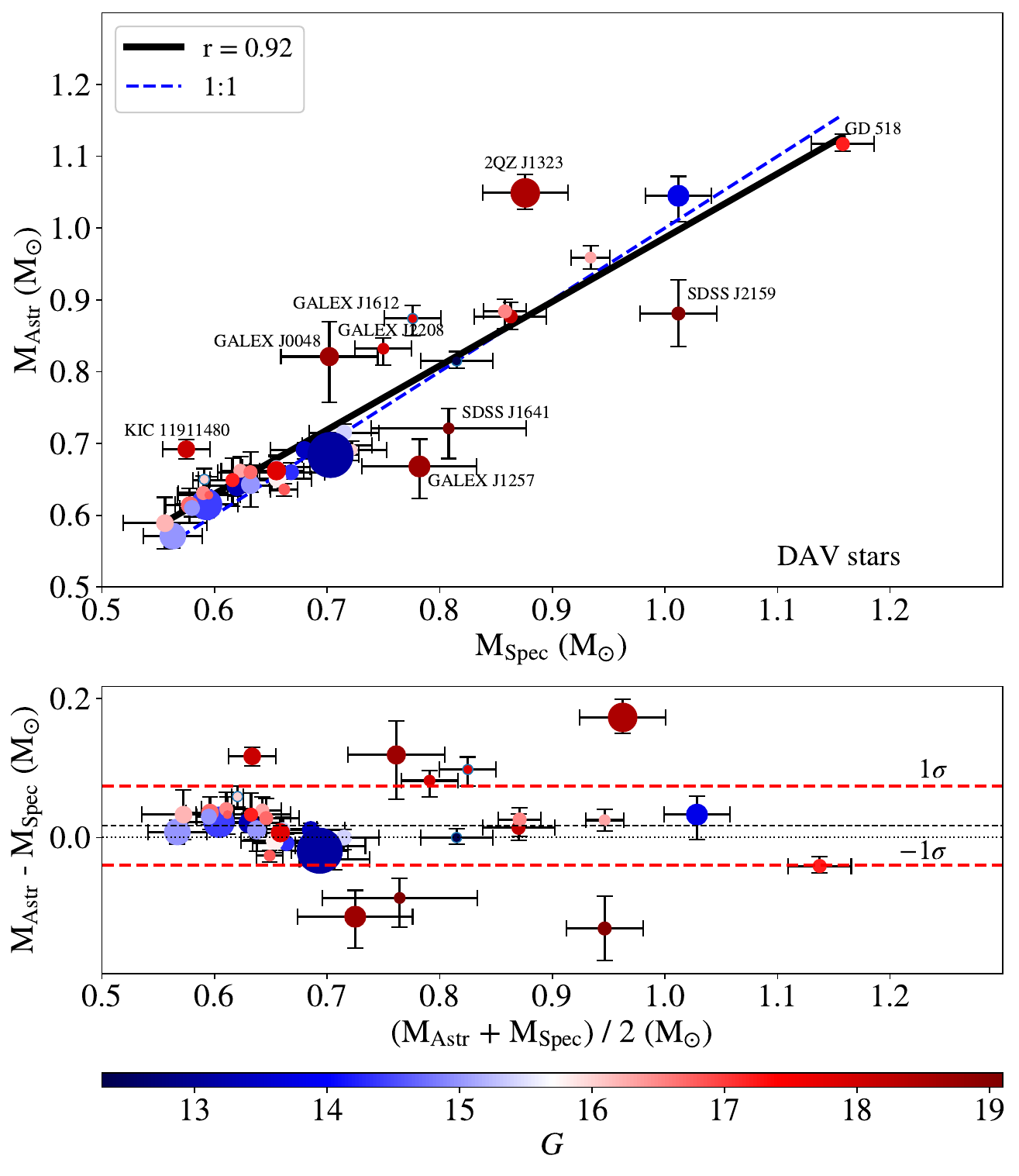}
	\caption{Similar to Fig. \ref{fig:Mspec_vs_Mseis_DA}, but for the comparison between spectroscopic  and astrometric masses. }
	\label{fig:Mspec_vs_Mastr_DA}
\end{figure}

\begin{figure}
\centering
 \includegraphics[clip,width=1.0\linewidth]{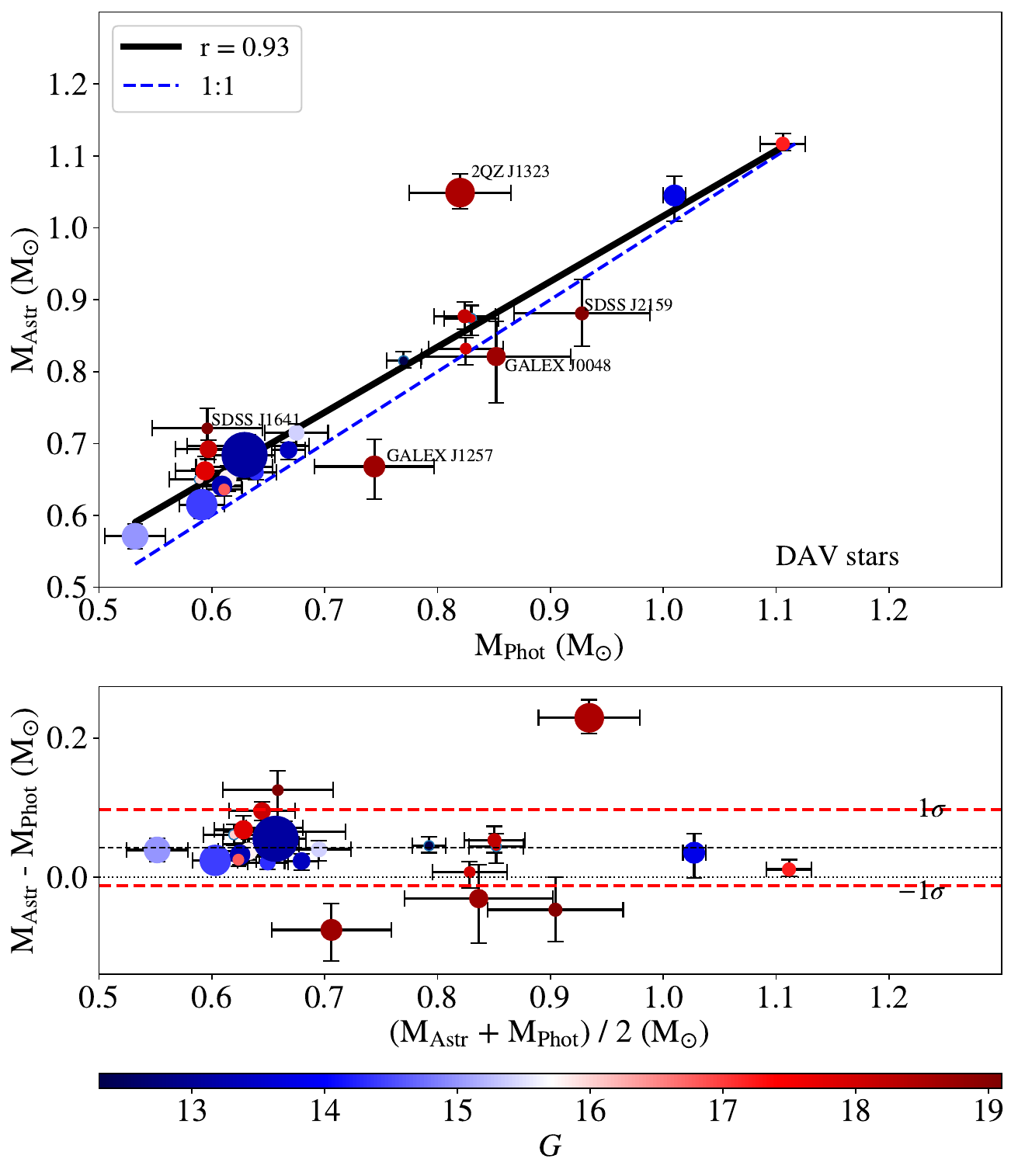}
	\caption{Similar to Fig. \ref{fig:Mspec_vs_Mseis_DA}, but for the comparison between photometric and astrometric masses.}
	\label{fig:Mphot_vs_Mastr_DA}
\end{figure}

\begin{figure}
\centering
 \includegraphics[clip,width=1.0\linewidth]{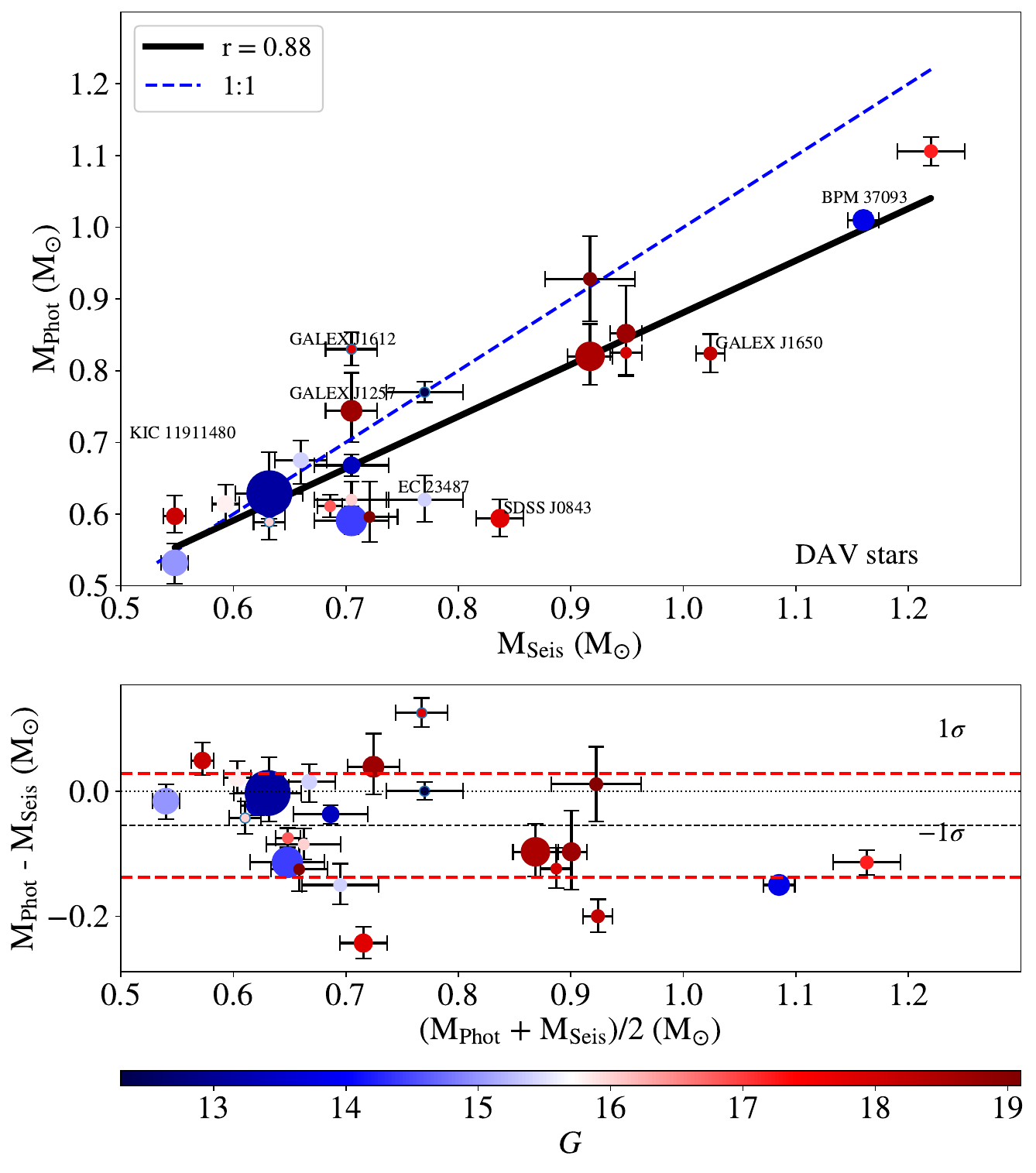}
	\caption{Similar to Fig. \ref{fig:Mspec_vs_Mseis_DA}, but for the comparison between seismological and photometric masses.}
	\label{fig:Mseis_vs_Mphot_DA}
\end{figure}

\begin{figure}
\centering
 \includegraphics[clip,width=1.0\linewidth]{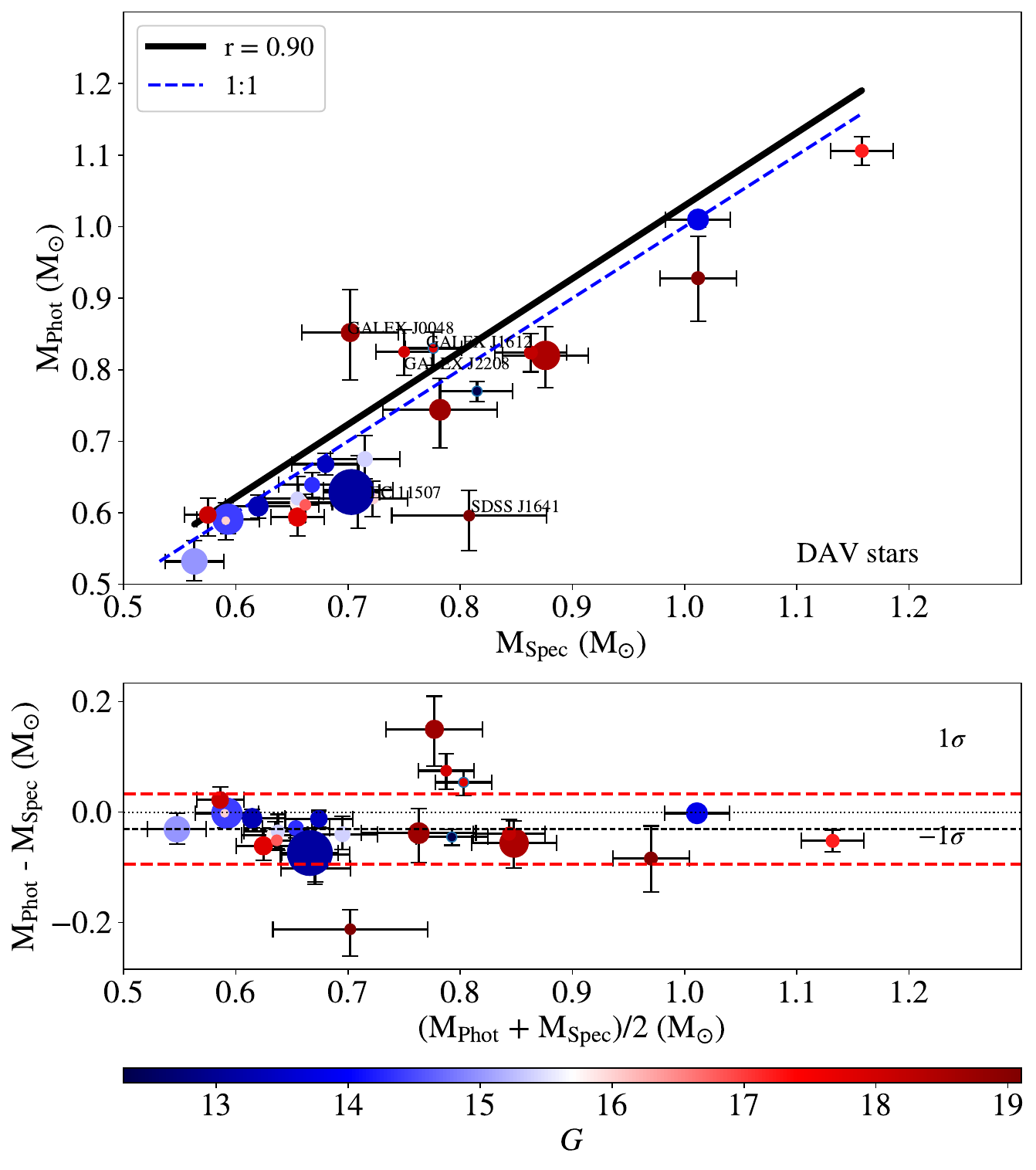}
	\caption{Similar to Fig. \ref{fig:Mspec_vs_Mseis_DA}, but for the comparison between spectroscopic and photometric masses.}
	\label{fig:Mspec_vs_Mphot_DA}
\end{figure}

We compare $M_{\rm Spec}$ and $M_{\rm Seis}$, $M_{\rm Seis}$ and $M_{\rm Astr}$, $M_{\rm Spec}$ and $M_{\rm Astr}$, $M_{\rm Phot}$ and $M_{\rm Astr}$, $M_{\rm Seis}$ and $M_{\rm Phot}$, and $M_{\rm Spec}$ and $M_{\rm Phot}$ for DAV stars in Figs. \ref{fig:Mspec_vs_Mseis_DA} to \ref{fig:Mspec_vs_Mphot_DA}.  Upon inspection of the figures, significant agreement is observed among the different sets of masses, particularly for masses below $\sim 0.75 M_{\sun}$, a trend that becomes apparent upon examination of the Pearson linear correlation coefficient. It is evident that in all cases this coefficient exceeds $\sim +0.8$, indicating a strong correlation. However, we are also interested in knowing how closely the sets of mass values agree, discovering whether there are global biases or not, and finding outlier stars whose masses should be re-evaluated in future analyses.

When comparing spectroscopic and seismological masses in a dispersion diagram (upper panel of Fig. \ref{fig:Mspec_vs_Mseis_DA}), a high level of correlation is found ($r= +0.83$). $M_{\rm Spec}$ and $M_{\rm Seis}$ show good agreement, which is somewhat anticipated, as seismological models are typically selected to satisfy the spectroscopic constraints of $T_{\rm eff}$ and $\log g$. This selection is made to mitigate the intrinsic degeneracy of solutions often encountered as a result of period-to-period fits \citep[see, for instance,][]{2012MNRAS.420.1462R, 2013ApJ...779...58R}. However, there are significant discrepancies for some stars more massive than $\sim 0.75 M_{\sun}$,  as evidenced by their deviation from the 1:1 line of correspondence.
The lower panel of the figure depicts the Bland-Altman diagram, where the mass differences, $\Delta M_{\star}= M_{\rm Seis}-M_{\rm Spec}$, are plotted against the mean mass value, $\langle M_{\star}\rangle= (M_{\rm Seis} + M_{\rm Spec})/2$, for each star. The mean difference is $\langle \Delta M_{\star}\rangle= 0.023$, indicating a small positive bias represented by the gap between the $x$ axis (corresponding to zero differences, marked by the dotted line) and the short-dashed line parallel to the $x$ axis at 0.023 units. This suggests that on average, the seismological masses are somewhat higher than the spectroscopic ones. The limits of agreement, defined as the $\pm 1\sigma$ departure from the mean difference, are displayed with red dashed  lines at $\langle\Delta M_{\star}\rangle \pm 0.089 M_{\sun}$. From this diagram, it is clear that there are outlier stars for which the mass difference is beyond the agreement limits, in accordance with what the dispersion diagram (upper panel) indicates. In these diagrams, the size of the symbols is directly proportional to the number of periods used to obtain the seismological models (fourth column of Table \ref{table:masses-dav-sample}), while the colour of each symbol is related to the brightness of the star (the apparent {\it Gaia} magnitude $G$ in the bottom palette of colours). Clearly, most outlier stars are dim (high $G$ value). One exception is BPM~37093, which is bright but, however, exhibits a large mass discrepancy. Furthermore, there is no clear indication that the number of periods is crucial for the star to have discrepant spectroscopic and seismological masses. 

The comparison between seismological and astrometric masses (upper panel of Fig. \ref{fig:Mseis_vs_Mastr_DA}) closely resembles the situation analysed above for spectroscopic and seismological masses (Fig. \ref{fig:Mspec_vs_Mseis_DA}). It is evident that there is strong agreement between $M_{\rm Seis}$ and $M_{\rm Astr}$, particularly for $M_{\star} \lesssim 0.75 M_{\sun}$. However, discrepancies become more apparent for larger masses, with data points deviating further from the 1:1 identity line. The Pearson coefficient for this comparison is $r= +0.86$, indicating a robust correlation. In the Bland-Altman diagram (lower panel of Fig. \ref{fig:Mseis_vs_Mastr_DA}), we observe a small negative mean mass difference of $\langle \Delta M_{\star}\rangle= \langle M_{\rm Astr}-M_{\rm Seis}\rangle= -0.005 M_{\sun}$, suggesting a slight negative bias between $M_{\rm Seis}$ and $M_{\rm Astr}$ masses (with $M_{\rm Seis}$ tending to be slightly larger than $M_{\rm Astr}$ on average). The limits of agreement, $\langle\Delta M_{\star}\rangle \pm 0.083 M_{\sun}$, are very similar to those observed in the comparison between spectroscopic and seismological masses (Fig. \ref{fig:Mspec_vs_Mseis_DA}). Once again, BPM~37093 stands out among outliers, exhibiting a significant discrepancy between seismological and astrometric masses despite being a bright object.

When comparing spectroscopic masses with astrometric masses, a very strong correlation is observed ($r= 0.92 $, upper panel of Fig. \ref{fig:Mspec_vs_Mastr_DA}). In particular, we can see a very dense crowding of points close to the 1:1 correspondence line for objects with masses less than $\sim 0.75 M_{\sun}$, while there are several more massive objects that deviate from the line of identity, reflecting considerable discrepancies in the value of the stellar mass depending on the method used to derive it. In the corresponding Bland-Altman diagram (lower panel), a slight bias towards larger astrometric masses than spectroscopic ones is observed, with a positive mean difference of $\langle \Delta M_{\star}\rangle= \langle M_{\rm Astr}-M_{\rm Spec}\rangle= +0.017 M_{\sun}$. The limits of agreement, $\langle\Delta M_{\star}\rangle \pm 0.057 M_{\sun}$, suggest that the mass differences are less scattered compared to the previous analyses. Notably, all outlier stars in this case are dim.

Upon analyzing the relationship between photometric and astrometric masses shown in Fig. \ref{fig:Mphot_vs_Mastr_DA}, we find a very strong correlation, as indicated by the $r= +0.93$ value in the upper panel. $M_{\rm Phot}$ and $M_{\rm Astr}$ show a significant agreement, in general, which is not surprising given that, although determined through very different procedures, the two methods employ photometry and astrometry from {\it Gaia}. The agreement between photometric and
astrometric masses is particularly true for stars with $M_{\star} \lesssim 0.75 M_{\sun}$, while for some more massive stars, a deviation from the 1:1 line of correspondence is found. The Bland-Altman diagram (lower panel) reveals a positive mean mass difference of $\langle \Delta M_{\star}\rangle= \langle M_{\rm Astr}-M_{\rm Phot}\rangle= + 0.042\ M_{\sun}$, indicating a noticeable bias where astrometric masses exceed photometric masses. The limits of agreement, $\langle\Delta M_{\star}\rangle \pm 0.055 M_{\sun}$, similarly to the previous case, are also less scattered than the first two cases ($M_{\rm Spec}$ vs $M_{\rm Seis}$ and $M_{\rm Seis}$ vs $M_{\rm Astr}$). We also note that all the outlier stars in this case are dim. 

In assessing the comparison between seismological and photometric masses, we find a similar dispersion diagram (upper panel in Fig. \ref{fig:Mseis_vs_Mphot_DA}) as the one shown in the comparison between astrometric and seismological masses. In this case, $r= +0.88$, indicating again a strong correlation. It is clear that there is close agreement between the seismological and photometric masses, although some discrepancies are found, particularly for stars between, roughly, $0.7$ and $1\ M_{\sun}$. In the Bland-Altman diagram (lower panel in Fig. \ref{fig:Mseis_vs_Mphot_DA}), there is a negative mean mass difference of $\langle \Delta M_{\star}\rangle= \langle M_{\rm Phot}-M_{\rm Seis}\rangle= - 0.055\ M_{\sun}$, indicating that photometric masses are, on average, larger than seismological masses. The limits of agreement, $\langle\Delta M_{\star}\rangle \pm 0.083 M_{\sun}$ are also very similar to those found in the comparison between astrometric and seismological masses. In this case, as before, we once again find BPM~37093 as an outlier star.

Finally, when comparing spectroscopic and photometric masses, illustrated in Fig. \ref{fig:Mspec_vs_Mphot_DA}, we observe a strong correlation as reflected by the $r= +0.90$ value shown in the upper panel. The Bland-Altman diagram (lower panel) indicates that there is a negative mean mass difference of $\langle \Delta M_{\star}\rangle= \langle M_{\rm Phot}-M_{\rm Spec}\rangle= - 0.031\ M_{\sun}$, implying that photometric masses are generally lower than spectroscopic masses. In this case, the limits of agreement are $\langle\Delta M_{\star}\rangle \pm 0.063 M_{\sun}$, and it is clear that all outlier stars are dim.

In summary, the masses of DAV stars derived from the four methods considered generally show good agreement, particularly for masses below approximately $0.75 M_{\sun}$. However, there are notable discrepancies for certain DAV stars. Table \ref{table:comparisons-DAVs} lists the objects classified as outliers, where there are discrepancies between the mass values derived from different methods. An outlier is identified when 
$\langle M_{\star} \rangle-\sigma  \geq  \Delta M_{\star} \geq \langle  M_{\star} \rangle +\sigma$.
For outlier DAV stars BPM~37093, GALEX~J1650+3010, GALEX~J2208+0654, GALEX~J0048+1521, SDSS~J0843+0431, and EC~23487$-$2424, we observe $M_{\rm Seis} > M_{\rm Astr} \gtrsim M_{\rm Spec}$. This suggests that spectroscopic masses for these stars might be slightly underestimated, while seismological masses could be overestimated and require reassessment. Regarding BPM~37093, we note that the mass of this star would vary slightly if we were to use CO-core WD models. However, the seismological analysis of this star was performed using ONe core WD models \citep{2019A&A...632A.119C}. Therefore, for consistency, we have used ONe-core WD evolutionary tracks in this paper to derive both spectroscopic and astrometric masses. It is interesting to note that the photometric mass of this star is very similar to the spectroscopic mass. Conversely, for SDSS~J2159+1322 and GALEX~J1257+0124, we find $M_{\rm Spec} > M_{\rm Phot} \gtrsim M_{\rm Seis} \gtrsim M_{\rm Astr}$, which indicates the potential overestimation of spectroscopic masses. TIC~167486543 and GALEX~J1612+0830 exhibit $M_{\rm Astr} > M_{\rm Spec} > M_{\rm Seis}$, suggesting that both seismological and spectroscopic masses might be underestimated. Similarly, for the pair 2QZ~J1323+0103 and KIC~11911480, we observe $M_{\rm Astr} > M_{\rm Seis} \gtrsim M_{\rm Spec}$, indicating the potential underestimation of both seismological and spectroscopic masses. The case of Ross~808  shows $M_{\rm Seis} > M_{\rm Astr} \simeq M_{\rm Spec} \simeq M_{\rm Phot}$, suggesting a possible overestimation of the seismological mass. Finally, for SDSS~J1641+3521, we find $M_{\rm Spec} > M_{\rm Seis} = M_{\rm Astr} > M_{\rm Phot}$, indicating a potential overestimation of spectroscopic mass, necessitating a review of $T_{\rm eff}$ and $\log g$, and a potential underestimation of photometric mass.

\begin{table*}[h!]
\centering
\caption{List of the outlier DAV stars including the mass differences, the number of $g$-mode periods employed to derive the seismological model, and their apparent brightness.}
\begin{tabular}{llrcc}
\hline
\noalign{\smallskip}
Comparison      & Outlier   &   $\Delta M_{\star}$  &  $N_{\Pi}$ & $G$ \\
                &           &                  $[M_{\sun}]$                 &            &  [mag] \\  
\hline
\noalign{\smallskip}
$M_{\rm Spec}$ vs $M_{\rm Seis}$ & BPM~37093     & 0.148  &  8  & 13.793 \\
($\Delta M_{\star}\equiv M_{\rm Seis} - M_{\rm Spec}$)  & GALEX~J1650+3010  & 0.161 &  3  & 18.151 \\   
                                 & GALEX~J2208+0654   &  0.199  & 2  & 17.972 \\  
                                 & GALEX~J0048+1521   &  0.247  & 6  & 18.711 \\  
                                 & SDSS~J0843+0431    &  0.182  & 6  & 17.835 \\  
                                 & EC~23487$-$2424    &  0.115  & 3  & 15.355 \\  
                                 & Ross~808           &  0.112  & 17 & 14.408 \\
                                 & SDSS~J2159+1322    & $-0.095$  & 3  & 18.959 \\
                                 & TIC~167486543      & $-0.114$  & 2  & 16.233\\
                                 & SDSS~J1641+3521    & $-0.087$  & 2  & 19.092 \\
                                 & GALEX~J1612+0830   & $-0.071$  & 1  & 17.806 \\  
                                 & GALEX~J1257+0124   & $-0.077$  & 8  & 18.665 \\  
                                 & G~29$-$38          & $-0.071$  & 38 & 13.062 \\
\noalign{\smallskip}
\hline
\noalign{\smallskip}
$M_{\rm Seis}$ vs $M_{\rm Astr}$ & 2QZ~J1323+0103     & 0.132 & 15  & 18.549\\
($\Delta M_{\star}\equiv M_{\rm Astr} - M_{\rm Seis}$)                            
                                 & TIC~167486543      & 0.139 &  2  & 16.233 \\
                                 & GALEX~J1612+0830   & 0.169 & 3 & 17.806 \\
                                 & KIC~11911480       & 0.144 &  5  & 18.064 \\         
                                 & GD~518             & $-0.103$ & 3 & 17.244 \\  
                                 & BPM~37093          & $-0.115$ & 3 & 13.793 \\ 
                                 & GALEX~J1650+3010   & $-0.147$ & 3 & 18.151 \\   
                                 & GALEX~J2208+0654   & $-0.117$ & 2 & 17.972 \\
                                 & GALEX~J0048+1521   & $-0.128$ & 6 & 18.711 \\
                                 & SDSS~J0843+0431    & $-0.175$ & 6 & 17.835 \\
                                 & EC~23487$-$2424    & $-0.103$ & 3 & 15.355 \\
                                 & Ross~808           & $-0.090$ & 17 & 14.408 \\
\noalign{\smallskip}
\hline
\noalign{\smallskip}
$M_{\rm Spec}$ vs $M_{\rm Astr}$ & 2QZ~J1323+0103     & 0.173 &  15 & 18.549 \\
($\Delta M_{\star}\equiv M_{\rm Astr} - M_{\rm Spec}$)     
                                 & GALEX~J1612+0830   & 0.098 &   1 & 17.806 \\
                                 & GALEX~J2208+0654   & 0.082 &   2 & 17.972 \\
                                 & GALEX~J0048+1521   & 0.119 &   6 & 18.711 \\
                                 & KIC~11911480       & 0.117 &   5 & 18.064 \\
                                 & GD~518             & $-0.041$ &   3 & 17.244 \\
                                 & SDSS~J2159+1322    & $-0.131$ &   3 & 18.959 \\        
                                 & SDSS~J1641+3521    & $-0.087$ &   2 & 19.092 \\ 
                                 & GALEX~J1257+0124   & $-0.114$ &   8 & 18.665 \\
\hline
\noalign{\smallskip}
$M_{\rm Astr}$  vs $\bm {M_{\rm Phot}}$ & GALEX~J0048+1521  & $-0.031$ &  6 & 18.711 \\
($\Delta M_{\star}\equiv M_{\rm Astr} - M_{\rm Phot}$)     
                                 & GALEX~J1257+0124   & $-0.076$ &    1 &  18.665 \\
                                 & 2QZ~J1323+0103     & 0.229    &   15 &  18.549 \\
                                 & SDSS~J1641+3521    & 0.125    &    2 &  19.092 \\ 
                                 & SDSS~J2159+1322    & $-0.047$ &    3 &  18.959 \\ 
\hline
\noalign{\smallskip}
$M_{\rm Phot}$  vs $M_{\rm Seis}$ &  KIC~11911480       &  0.049 &    5 &  18.064 \\
($\Delta M_{\star}\equiv M_{\rm Phot} - M_{\rm Seis}$)     
                                 & EC~23487$-$2424    & $-0.150$ &   3 &  15.355 \\
                                 & SDSS~J0843+0431    & $-0.243$ &   6 &  17.835 \\
                                 & GALEX~J1257+0124   &  0.039   &   1 &  18.665 \\
                                 & GALEX~J1612+0830   &  0.125   &   1 &  17.806 \\
                                 & GALEX~J1650+3010   & $-0.200$ &   3 &  18.151 \\
                                 & BPM~37093          & $-0.150$ &   3 &  13.793 \\ 
\hline
\noalign{\smallskip}
$M_{\rm Phot}$ vs$M_{\rm Spec}$ &  EC~11507$-$1519 & $-0.102$ &   2 &  16.031 \\
($\Delta M_{\star}\equiv M_{\rm Phot} - M_{\rm Spec}$)    
                                 &  GALEX~J0048+1521   &  0.150   &  6 &  18.711 \\
                                 &  GALEX~J1612+0830   &  0.054   &  1 &  17.806 \\
                                 &  SDSS~J1641+3521    & $-0.212$ &  2 &  19.092 \\ 
                                 &  GALEX~J2208+0654   &  0.075   &  2 &  17.972 \\
\noalign{\smallskip}
\hline
\noalign{\smallskip}
\end{tabular}
%\tablefoot{xxx.}
\label{table:comparisons-DAVs}
\end{table*}

\subsection{DBV stars}
\label{sec:DBVs}

\begin{figure}
\centering
 \includegraphics[clip,width=1.0\linewidth]{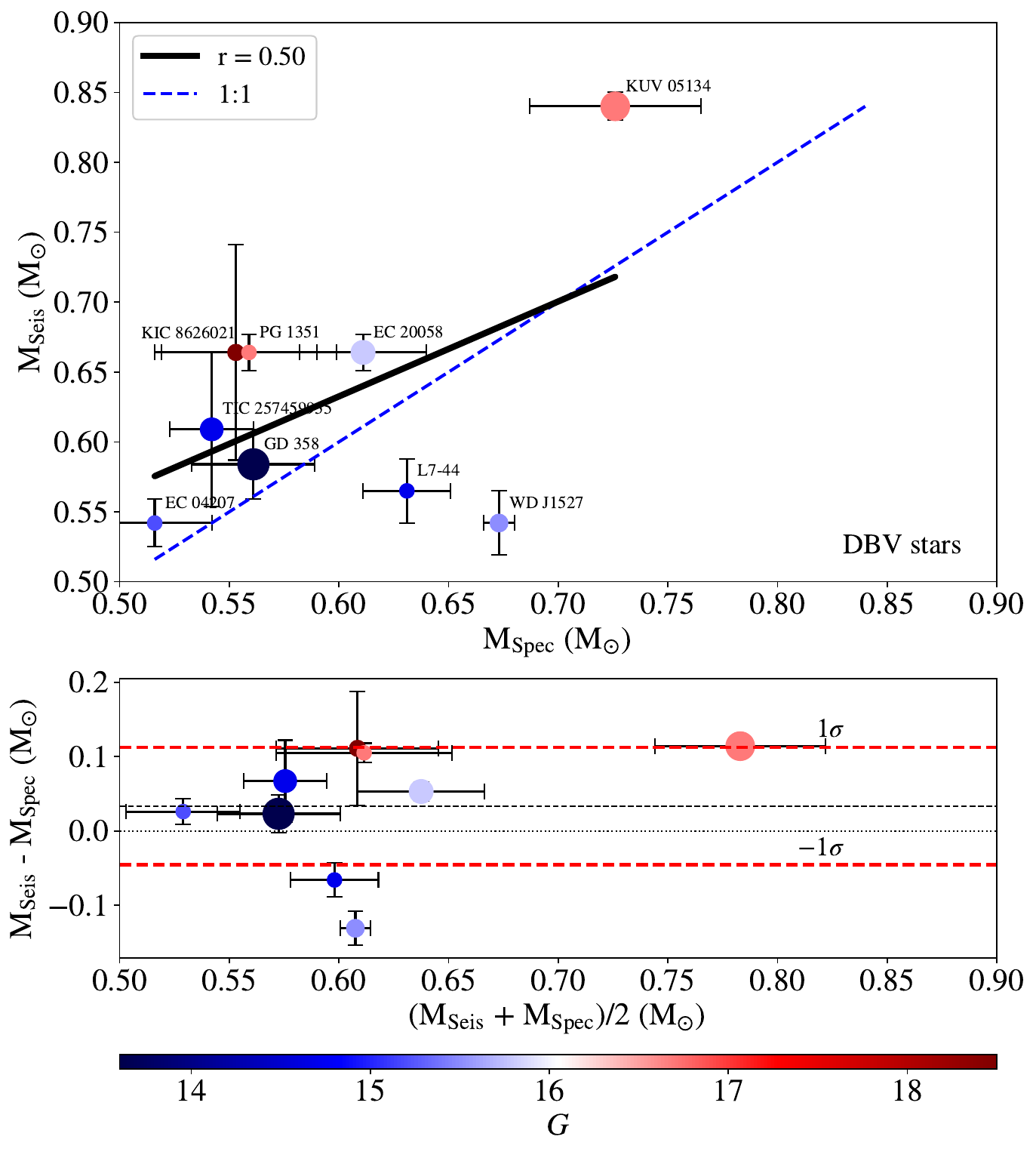}
	\caption{Upper panel: dispersion diagram displaying the comparison between the stellar masses of DBVs obtained from spectroscopy and asteroseismology (see Table \ref{table:masses-dbv-sample}). All the stars are identified by their names. Bottom panel: the corresponding Bland-Pearson diagram. The meaning of the different lines in both panels is the same as in Fig. \ref{fig:Mspec_vs_Mseis_DA}.}
	\label{fig:Mspec_vs_Mseis_DB}
\end{figure}

\begin{figure}
\centering
 \includegraphics[clip,width=1.0\linewidth]{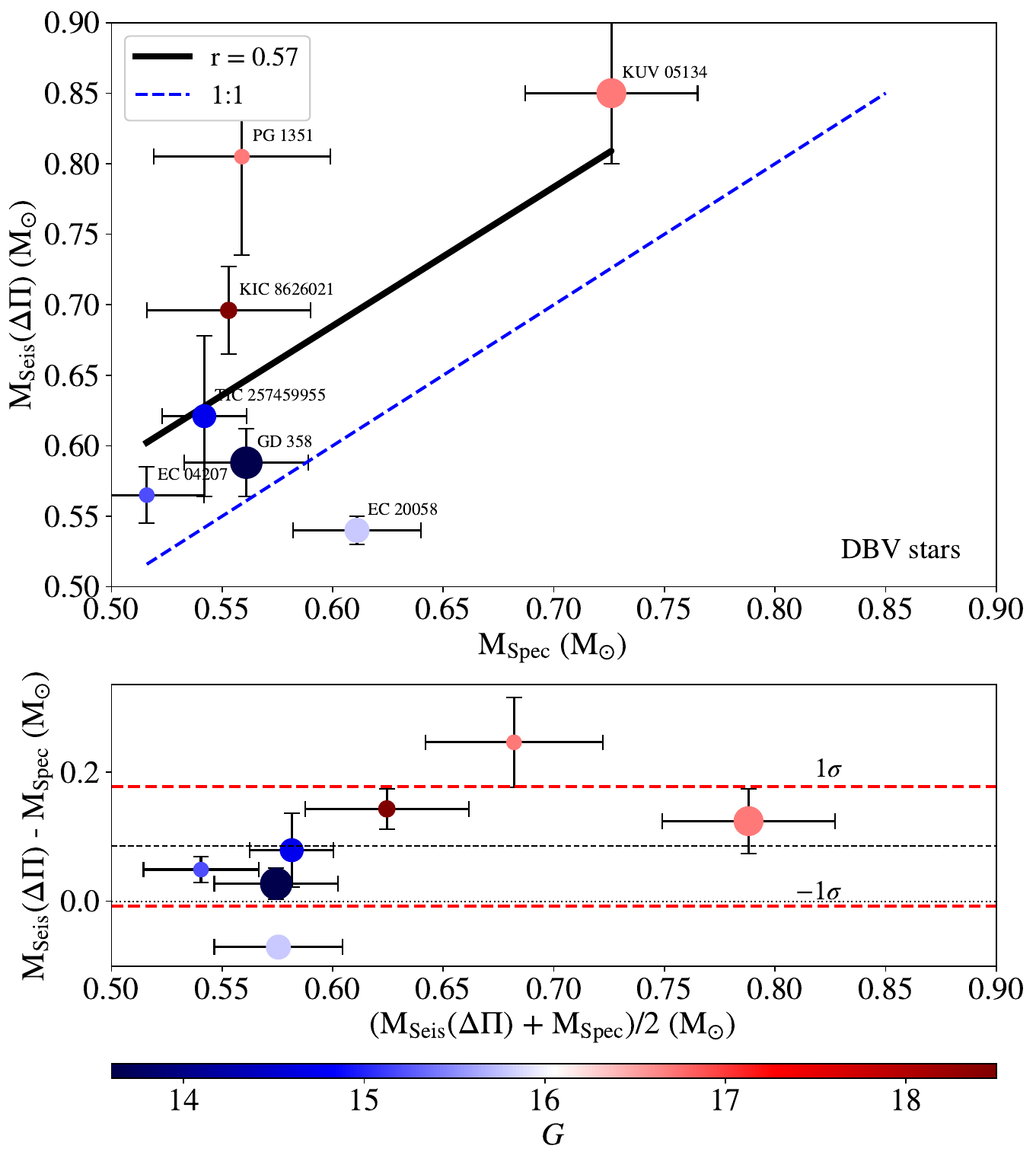}
	\caption{Similar to Fig. \ref{fig:Mspec_vs_Mseis_DB}, this figure illustrates the comparison between 
 spectroscopic and seismological masses, with the seismological mass derived from the period spacing ($\Delta \Pi$).  It is worth noting that only seven out of the total nine DBV stars analyzed have an estimate of seismological mass from the period spacing (see Table \ref{table:masses-dbv-sample}).}
	\label{fig:Mspec_vs_Mseis_PS_DB}
\end{figure}

\begin{figure}
\centering
 \includegraphics[clip,width=1.0\linewidth]{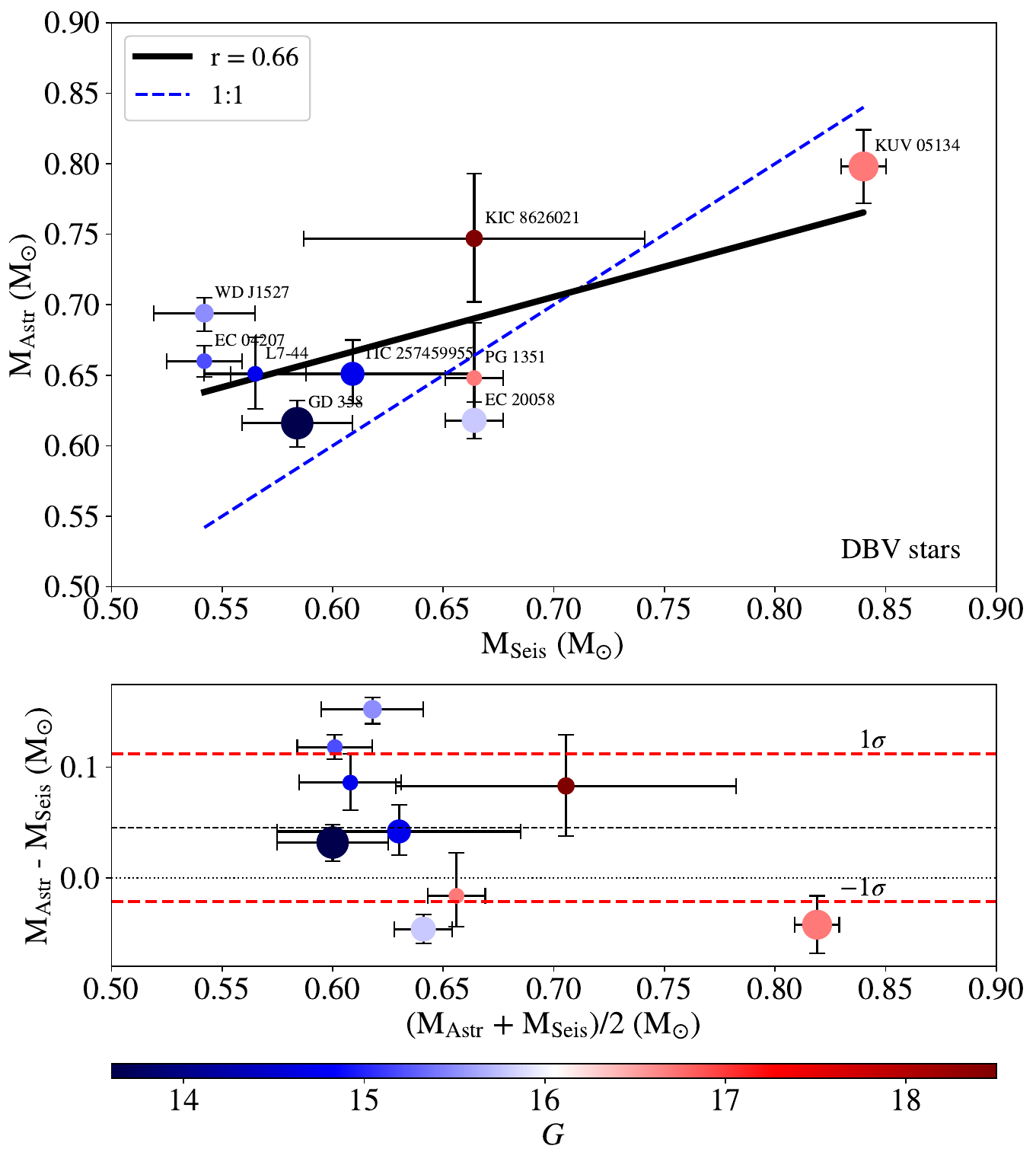}
	\caption{Similar to Fig. \ref{fig:Mspec_vs_Mseis_DB}, but for the comparison between the seismological and astrometric masses.}
	\label{fig:Mseis_vs_Mastr_DB}
\end{figure}

\begin{figure}
\centering
 \includegraphics[clip,width=1.0\linewidth]{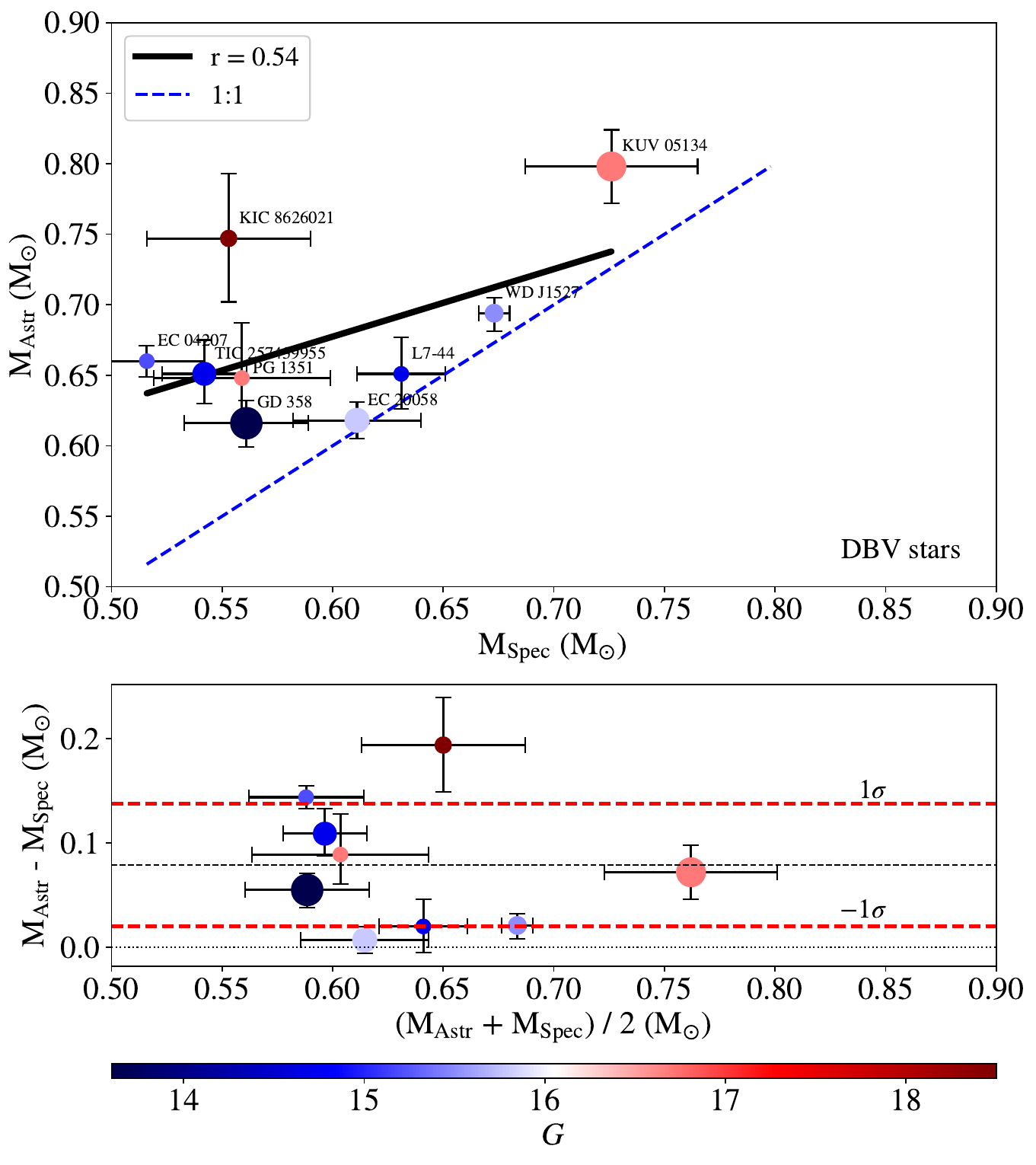}
	\caption{Similar to Fig. \ref{fig:Mspec_vs_Mseis_DB}, but for the comparison between spectroscopic  
 and astrometric masses.}
	\label{fig:Mspec_vs_Mastr_DB}
\end{figure}

\begin{figure}
\centering
 \includegraphics[clip,width=1.0\linewidth]{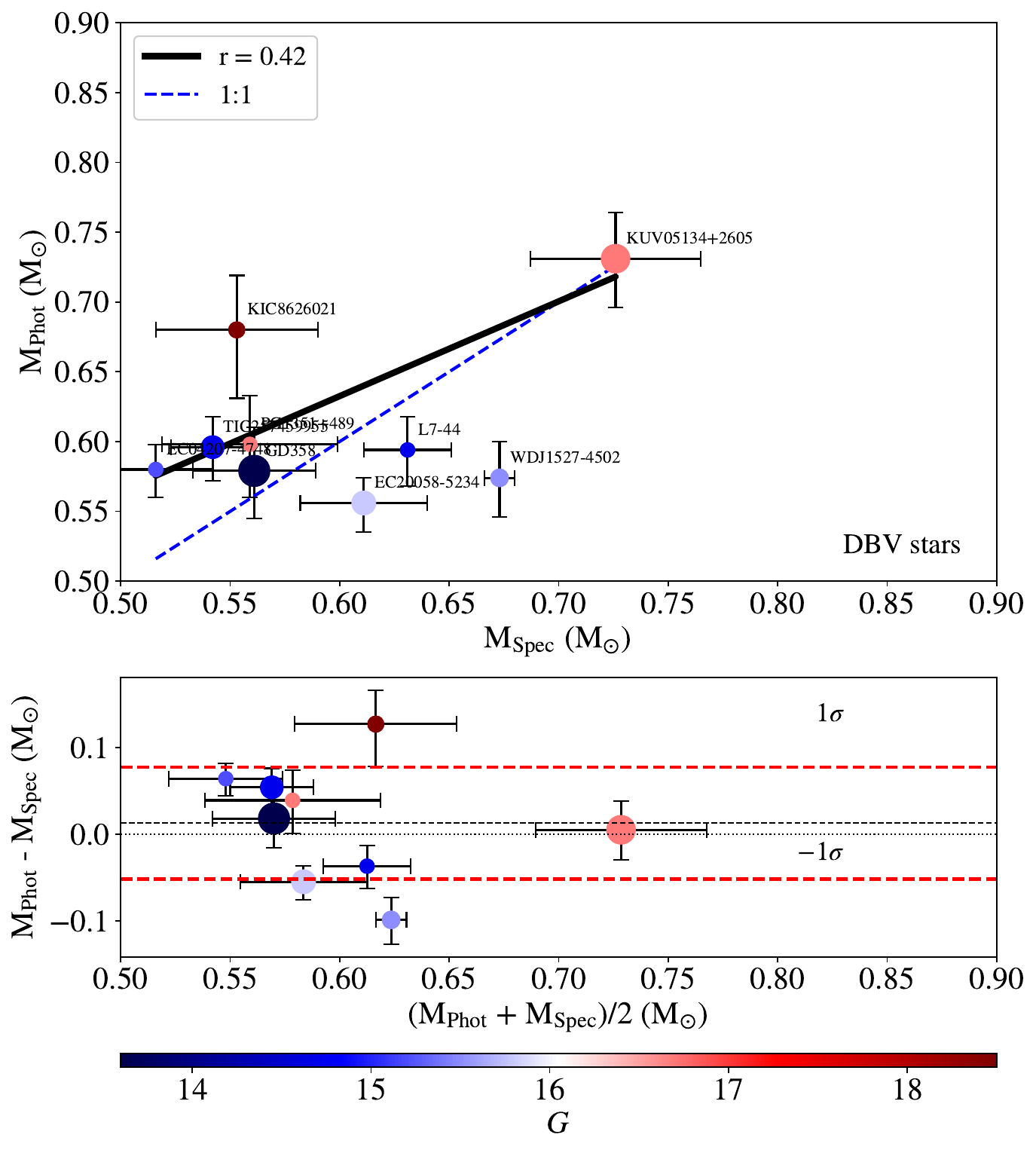}
	\caption{Similar to Fig. \ref{fig:Mspec_vs_Mseis_DB}, but for the comparison between spectroscopic and photometric masses.}
	\label{fig:Mspec_vs_Mphot_DB}
\end{figure}

\begin{figure}
\centering
 \includegraphics[clip,width=1.0\linewidth]{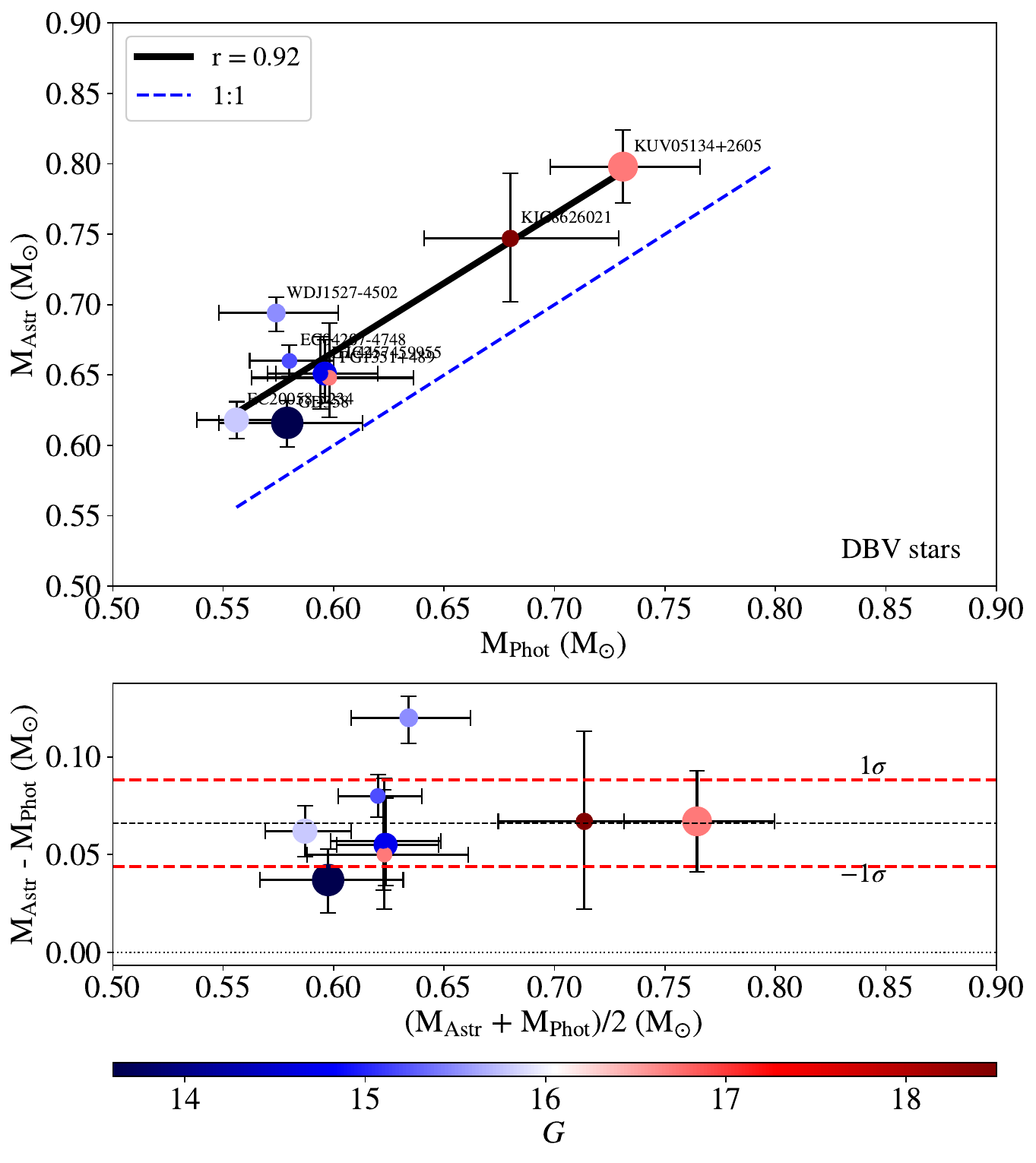}
	\caption{Similar to Fig. \ref{fig:Mspec_vs_Mseis_DB}, but for the comparison between photometric and astrometric masses.}
	\label{fig:Mphot_vs_Mastr_DB}
\end{figure}

\begin{figure}
\centering
 \includegraphics[clip,width=1.0\linewidth]{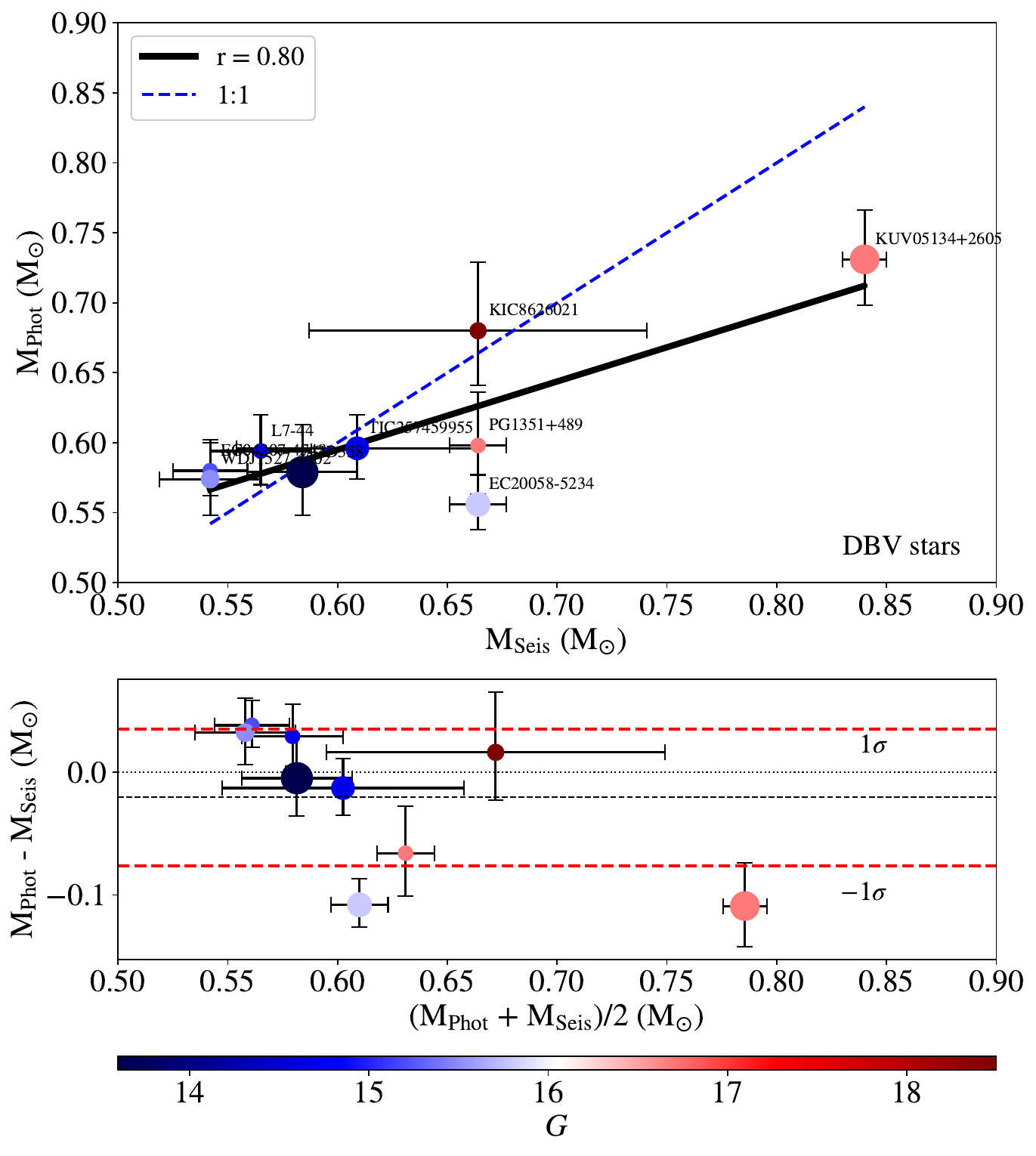}
	\caption{Similar to Fig. \ref{fig:Mspec_vs_Mseis_DB}, but for the comparison between seismological and photometric masses.}
	\label{fig:Mseis_vs_Mphot_DB}
\end{figure}

\begin{figure}
\centering
 \includegraphics[clip,width=1.0\linewidth]{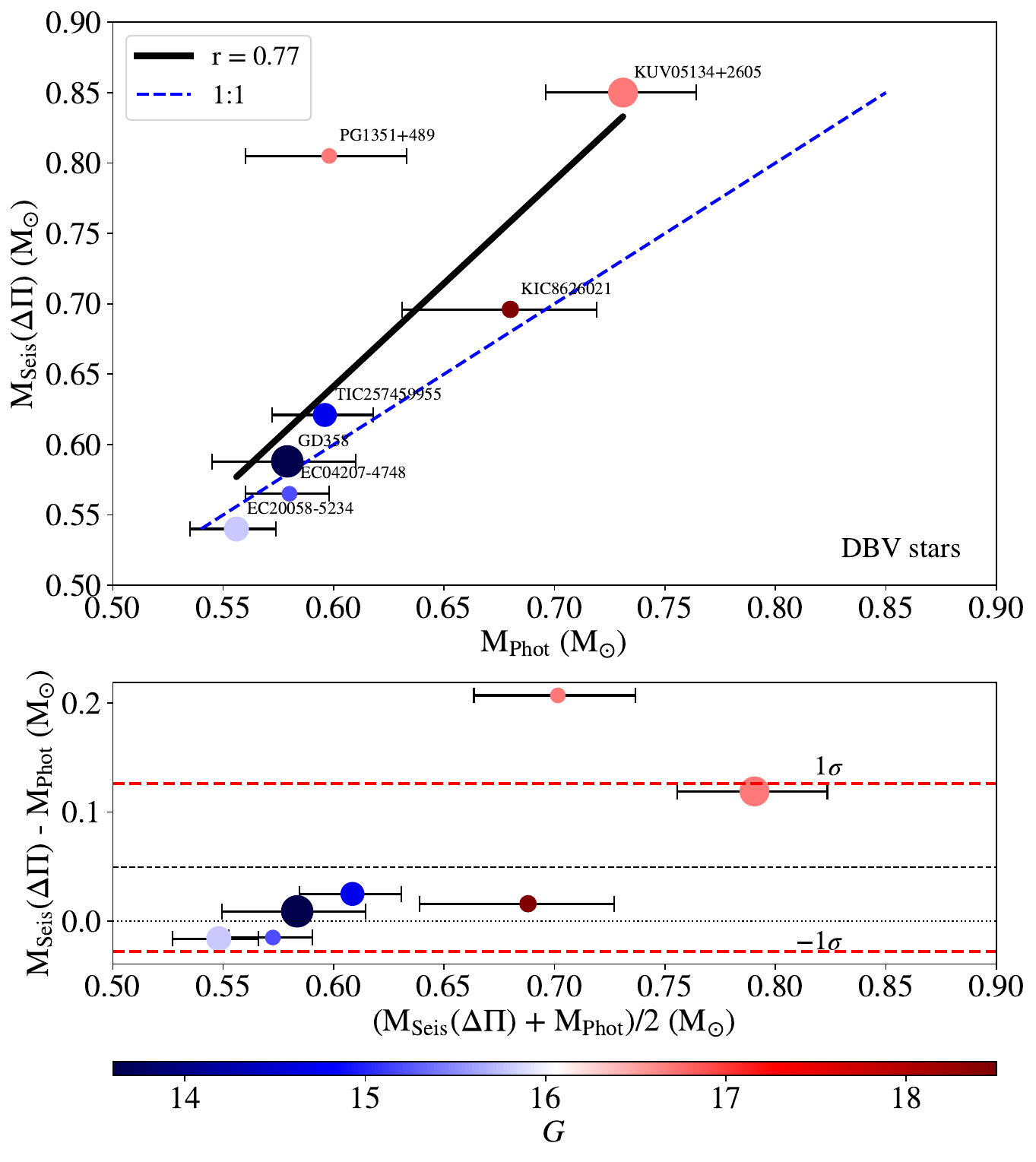}
	\caption{Similar to Fig. \ref{fig:Mspec_vs_Mseis_PS_DB}, but for the comparison between photometric and seismological masses derived from the period spacing.}
	\label{fig:Mphot_vs_Mseis_PS_DB}
\end{figure}

We present comparisons between $M_{\rm Spec}$ and $M_{\rm Seis}$, $M_{\rm Spec}$ and $M_{\rm Seis}(\Delta \Pi)$, $M_{\rm Seis}$ and $M_{\rm Astr}$, $M_{\rm Spec}$ and $M_{\rm Astr}$,  $M_{\rm Spec}$ and $M_{\rm Phot}$, $M_{\rm Phot}$ and $M_{\rm Astr}$, $M_{\rm Seis}$ and $M_{\rm Phot}$, and $M_{\rm Phot}$ and $M_{\rm Seis(\Delta \Pi)}$ for DBVs in Figs. \ref{fig:Mspec_vs_Mseis_DB} to \ref{fig:Mphot_vs_Mseis_PS_DB}. $M_{\rm Seis}(\Delta \Pi)$ represents the seismological stellar mass derived on the basis of uniform period spacing. Among these figures, we observe two different levels of correlation. Five out of eight figures (\ref{fig:Mspec_vs_Mseis_DB} to \ref{fig:Mspec_vs_Mphot_DB}, displaying $M_{\rm Spec}$ and $M_{\rm Seis}$, $M_{\rm Spec}$ and $M_{\rm Seis}(\Delta \Pi)$, $M_{\rm Seis}$ and $M_{\rm Astr}$, $M_{\rm Spec}$ and $M_{\rm Astr}$, and $M_{\rm Spec}$ and $M_{\rm Phot}$) indicate a moderate linear correlation, with Pearson coefficients that do not exceed $\sim +0.66$, while the remaining three (figures \ref{fig:Mphot_vs_Mastr_DB} to \ref{fig:Mphot_vs_Mseis_PS_DB}, showing $M_{\rm Phot}$ and $M_{\rm Astr}$, $M_{\rm Seis}$ and $M_{\rm Phot}$, and $M_{\rm Phot}$ and $M_{\rm Seis}(\Delta \Pi)$) exhibit a stronger correlation, with Pearson coefficients that range from $+ 0.77$ to $+ 0.92$. Notably, of the four figures that involve comparisons with photometric masses, three of them are the ones showing a higher degree of correlation. This suggests a potential trend where comparisons involving photometric masses tend to exhibit a stronger correlation.

We first turn our attention to the cases with moderate correlation (figures \ref{fig:Mspec_vs_Mseis_DB} to \ref{fig:Mspec_vs_Mphot_DB}). The Bland-Pearson diagrams reveal significant biases among the methods of stellar mass derivation, indicating poor agreement among the various mass sets. Identifying outlier stars based on agreement limits becomes a challenge. In cases where there is a notable bias between the two mass sets, characterised by a mean difference considerably different from zero, a star may fall within admissibility limits because its distance from the mean difference, $\langle \Delta M_{\star} \rangle$, is within $\pm \sigma$. However, it could still be located far from the 1:1 correspondence line, indicating significant mass discrepancies. Conversely, an object may appear close to the 1:1 correspondence line, suggesting good mass agreement, but fall outside admissibility limits in the Bland-Altman diagram, leading to erroneous outlier classification.

Notably, a consistent trend is observed in most comparisons: spectroscopic masses are systematically smaller than seismological and astrometric masses. For example, in Fig. \ref{fig:Mspec_vs_Mseis_DB}, where seismological masses are compared with spectroscopic masses, seven out of nine DBV stars exhibit spectroscopic masses smaller than seismological masses. This trend is evident in the mean mass difference of $\langle \Delta M_{\star} \rangle= 0.034 M_{\sun}$, indicating a notable bias towards seismological masses being greater than spectroscopic masses. It is important to note that the limits of agreement, $\langle\Delta M_{\star}\rangle \pm 0.079 M_{\sun}$, prove to be inappropriate to identify outliers. In fact, KUV~05134+2605, KIC~8626021, and PG~1351+489 clearly stand out as outliers, positioned significantly away from the 1:1 correspondence line, despite falling at the upper limit of agreement. TIC~257459955 is also evidently an outlier, despite falling within the agreement limits.

The discrepancy between different sets of masses becomes more apparent when comparing spectroscopic masses with seismological masses derived from period spacing (Fig. \ref{fig:Mspec_vs_Mseis_PS_DB}). In this comparison,  seven out of nine stars exhibit spectroscopic masses smaller than their seismological counterparts. It is important to note that for two stars (PG~1351+489 and EC~20058$-$5234), the seismological mass is defined within a range of values (see Table \ref{table:masses-dbv-sample}), where we have taken the average value of both extremes as the seismological mass. The bias towards larger seismological masses compared to spectroscopic ones is more pronounced here, with an average mass difference of $\langle \Delta M_{\star} \rangle= 0.085 M_{\sun}$. Unfortunately, the Bland-Altman diagram does not effectively identify outliers. For instance,  despite KUV~05134+2605 and KIC~8626021 being outliers according to the scatter plot, they fall well within the limits of agreement in the Bland-Altman diagram.

In the comparison between seismological and astrometric masses (Fig. \ref{fig:Mseis_vs_Mastr_DB}), it is clear that astrometric masses generally exceed seismological masses, with an average mass difference of $\langle \Delta M_{\star} \rangle= 0.045 M_{\sun}$. Furthermore, when comparing spectroscopic masses with astrometric masses (Fig. \ref{fig:Mspec_vs_Mastr_DB}), it is noteworthy that \textit{all} spectroscopic masses are below the astrometric masses. Here, the average mass difference amounts to $\langle \Delta M_{\star} \rangle= 0.079 M_{\sun}$. In fact, spectroscopic masses are approximately $14 \%$ smaller than astrometric masses on average. Validating this trend would require analysing a broader sample of DB WD stars, a task planned for a future paper.

The observed trend of spectroscopic masses being systematically smaller than the other mass estimates does not hold, however, when comparing them to photometric masses, as shown in Fig. \ref{fig:Mspec_vs_Mphot_DB}. Indeed, we find a slight bias towards larger photometric masses, reflected in the mean mass difference $\langle \Delta M_{\star} \rangle= 0.013 M_{\sun}$. The lower panel reveals that the limits of agreement, $\langle\Delta M_{\star}\rangle \pm 0.065 M_{\sun}$, are once again inadequate for identifying outliers. This is evident from L7$-$44 and EC~20058$-$5234, which, despite being clear outliers in the upper panel, fall within or almost exactly on the limits of agreement in the lower panel.

Now we focus on the cases exhibiting strong correlations, specifically between $M_{\rm Phot}$ and $M_{\rm Astr}$, $M_{\rm Phot}$ and $M_{\rm Seis}(\Delta \Pi)$, and $M_{\rm Seis}$ and $M_{\rm Phot}$. In the first case, where we compare photometric and astrometric masses, as illustrated in Fig. \ref{fig:Mphot_vs_Mastr_DB}, the Pearson coefficient is high, $r= +0.92$. The upper panel clearly shows that \textit{all} astrometric masses surpass the photometric ones, similar to what is found in the comparison between spectroscopic and astrometric masses. Here, the mean mass difference is $\langle \Delta M_{\star} \rangle= 0.045 M_{\sun}$, and we find that astrometric masses are approximately $11\%$ greater than their photometric counterparts. Confirming this trend between astrometric and photometric masses will necessitate further analysis with a larger sample of DB WDs, as already mentioned and planned for future work.
%  The limits of agreement in this case are $\langle\Delta M_{\star}\rangle \pm 0.066 M_{\sun}$.}

When comparing seismological and photometric masses, a strong correlation is evident, with a Pearson coefficient of  $r= +0.80$, as shown in the upper panel of Fig. \ref{fig:Mseis_vs_Mphot_DB}. This is particularly true for stars with masses $\lesssim 0.6\ M_{\sun}$. In the lower panel, the mean mass difference of $\langle \Delta M_{\star} \rangle= -0.021 M_{\sun}$ indicates that seismological masses slightly exceed photometric ones. %The limits of agreement in this case are $\langle\Delta M_{\star}\rangle \pm 0.055 M_{\sun}$.
Lastly, Fig. \ref{fig:Mphot_vs_Mseis_PS_DB} presents the comparison between $M_{\rm Phot}$ and $M_{\rm Seis}(\Delta \Pi)$. The correlation in this case is $r= +0.77$, slightly lower than in the previous case. However, it remains evident that objects with masses $\lesssim 0.6\ M_{\sun}$ are close to the 1:1 correspondence line. The mean mass difference of $\langle \Delta M_{\star} \rangle= 0.031 M_{\sun}$ indicates a tendency for seismological masses derived from the period spacing to be slightly larger than photometric ones.

In the following, we will delve into describing the significant discrepancies observed in some notable cases of disagreement.

\begin{itemize}

\item[-] KUV~05134+2605: With an extensive dataset of 16 detected periods, but not being very bright ($G= 16.749$), this star exhibits the trend $M_{\rm Seis} \simeq M_{\rm Seis}(\Delta \Pi) > M_{\rm Astr} \gtrsim M_{\rm Phot} \simeq M_{\rm Spec}$. The masses derived from asteroseismology ($M_{\rm Seis}, M_{\rm Seis}(\Delta \Pi)$) appear notably high among the other estimates, in clear contrast to the low value of the spectroscopic mass ($M_{\rm Spec}$). Given the close agreement between $M_{\rm Phot}$, $M_{\rm Spec}$, and $M_{\rm Astr}$, our results might suggest that the higher masses obtained from asteroseismology warrant a re-evaluation.

\item[-] KIC~8626021: Here, we observe $M_{\rm Astr} > M_{\rm Seis}(\Delta \Pi) \simeq M_{\rm Phot} \simeq M_{\rm Seis} > M_{\rm Spec}$. This star, with a faint brightness ($G= 18.500$) and a limited number of periods (5), exhibits a notably small spectroscopic mass, highlighting the need for a thorough revision of the spectroscopic parameters ($T_{\rm eff}$ and $\log g$).

\item[-] PG~1351+489: In this case, we find $M_{\rm Seis}(\Delta \Pi) > M_{\rm Seis} \simeq M_{\rm Astr} > M_{\rm Phot}  > M_{\rm Spec}$. The seismological mass derived from the period spacing appears to be notably high ($0.740 \lesssim M_{\star}/M_{\sun} \lesssim 0.870$). In contrast, the spectroscopic mass is markedly low, indicating a need for refinement to determine $T_{\rm eff}$ and $\log g$.

\item[-] WD~J1527$-$4502: For this star, we get $M_{\rm Astr} \gtrsim M_{\rm Spec} > M_{\rm Phot} > M_{\rm Seis}$, suggesting that the spectroscopic mass is well determined, but the seismological mass is low in excess. The photometric mass is also lower compared to $ M_{\rm Astr}$ and $M_{\rm Spec}$, but to a lesser extent than the seismological mass. For this star, there is no assessment of the seismological mass based on the period spacing. Our results suggest a review of the seismological mass of WD~J1527$-$4502. 
  
\end{itemize}

It is worth highlighting that, beyond these four troublesome stars, the remaining DBV stars in the small sample show varying degrees of inconsistency in their mass values depending on the method used, as we show below: 

\begin{itemize}

  \item[-] EC~20058$-$5234: In this case, we observe $M_{\rm Seis} > M_{\rm Spec} \simeq M_{\rm Astr} > M_{\rm Phot}  \simeq M_{\rm Seis}(\Delta \Pi)$. This suggests the reliability of the spectroscopic mass, which aligns with the astrometric mass. However, the seismological mass derived from individual periods appears to be overstated. Interestingly, the photometric mass aligns well with the seismological mass based on period spacing, although both of these are lower compared with the spectroscopic and astrometric values. 
Consequently, a reassessment of the seismological determinations of EC20058$-$5234's mass seems warranted.

\item[-] TIC~257459955: In this case, $M_{\rm Astr} \simeq M_{\rm Seis}(\Delta \Pi) \simeq M_{\rm Seis} \simeq M_{\rm Phot} > M_{\rm Spec}$, indicating a notably low spectroscopic mass. Therefore, a reevaluation of the parameters $T_{\rm eff}$ and $\log g$, utilised to derive the spectroscopic mass, appears to be necessary.

\item[-]  EC~04207$-$4748: Here, $M_{\rm Astr} > M_{\rm Phot} \simeq M_{\rm Seis} \simeq M_{\rm Seis}(\Delta \Pi) > M_{\rm Spec}$. This suggests an underestimation of the spectroscopic mass, implying a need for a reassessment of the spectroscopic parameters for this star. 

\item[-] L~7$-$44: Here, we find $M_{\rm Astr} \gtrsim M_{\rm Spec} > M_{\rm Phot} > M_{\rm Seis}$, suggesting an underestimation of the seismological mass, which warrants a re-assessment.

\item[-] GD~358: As the prototype of DBV stars (also known as V777 Her stars), we observe $M_{\rm Astr} \gtrsim M_{\rm Seis} \simeq M_{\rm Seis}(\Delta \Pi) \simeq M_{\rm Phot} > M_{\rm Spec}$. These findings imply a possible underestimation of the spectroscopic mass of GD~358. Therefore, it is advisable to reevaluate the spectroscopic parameters ($T_{\rm eff}$ and $\log g$) utilised to calculate $M_{\rm Spec}$.

\end{itemize}

We conclude our analysis of the DBV stars by highlighting a trend that is consistent across most cases: the spectroscopic mass values appear to be systematically underestimated. This observation raises the possibility of encountering challenges in precisely determining $T_{\rm eff}$ and $\log g$ through spectroscopic methods for DBV stars (see Sect. \ref{discussion}). Similar to DAVs, we do not observe a direct correlation between the reliability of seismological mass determination and either the star's brightness or the number of periods used in determining the seismological mass.

\subsection{GW Vir stars}
\label{sec:GWVir}

\begin{figure}
\centering
 \includegraphics[clip,width=1.0\linewidth]{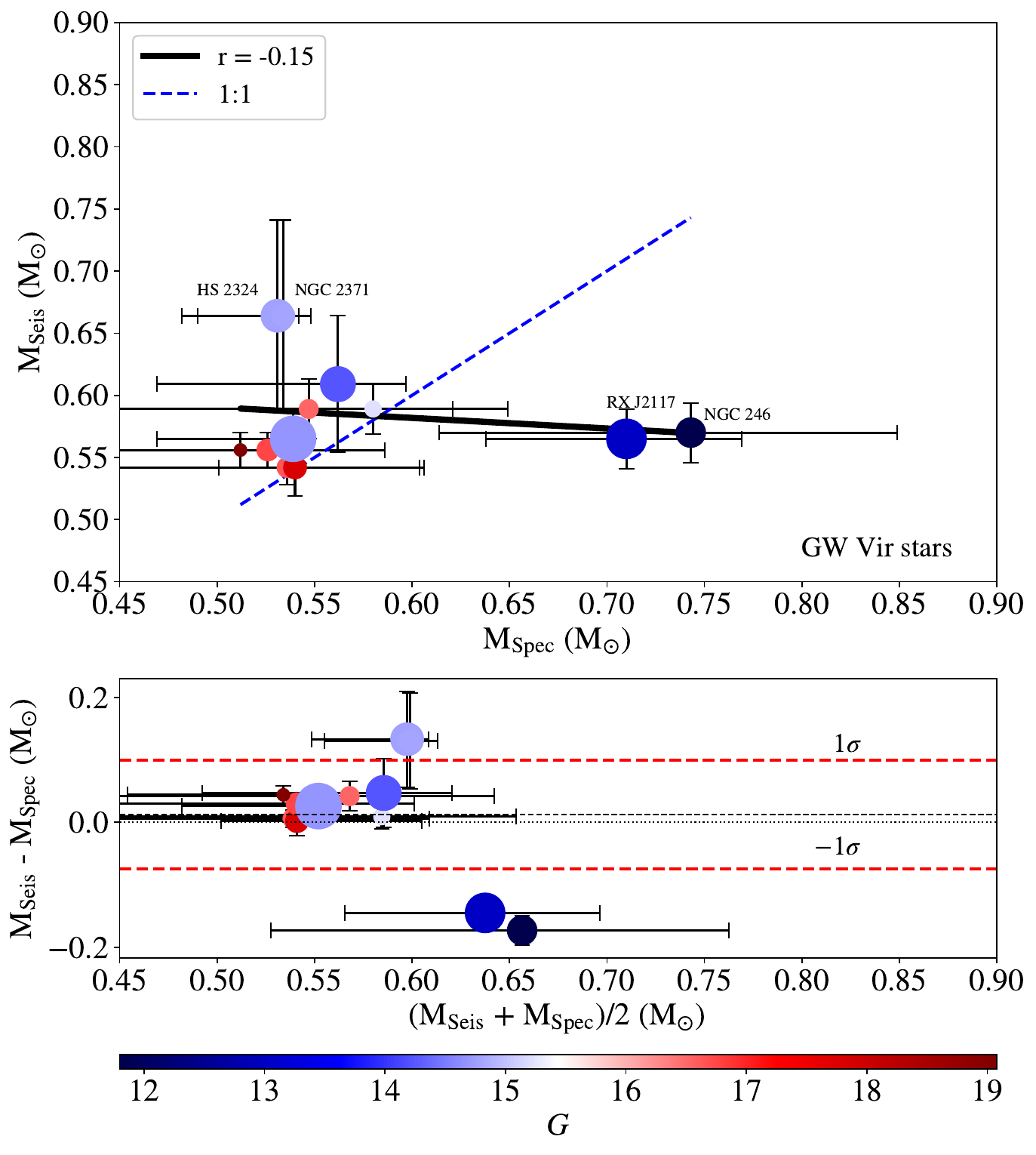}
	\caption{Upper panel: dispersion diagram showing the comparison of stellar masses for GW Vir stars derived from spectroscopy and asteroseismology (refer to Table \ref{table:masses-gwvir-sample}). Stars with mass estimates that significantly disagree are annotated with their names. Bottom panel: the corresponding Bland-Pearson diagram. The meaning
of the different lines in both panels is the same as in Fig. \ref{fig:Mspec_vs_Mseis_DA}.}
 	\label{fig:Mspec_vs_Mseis_GWVIR}
\end{figure}

\begin{figure}
\centering
 \includegraphics[clip,width=1.0\linewidth]{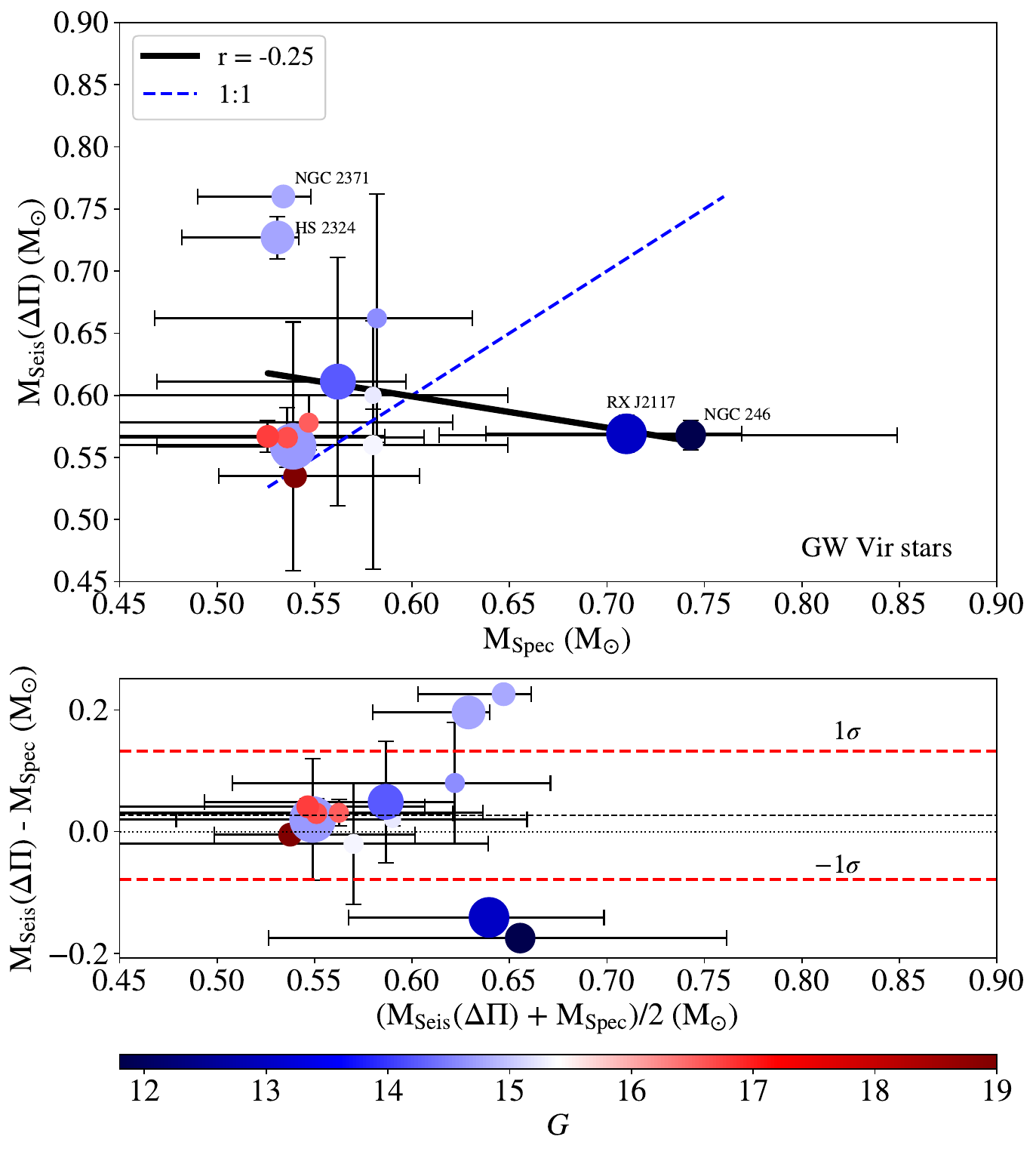}
	\caption{Similar to Fig. \ref{fig:Mspec_vs_Mseis_GWVIR}, this figure presents the comparison between spectroscopic and seismological masses, where the seismological masses are derived from the period spacing ($\Delta \Pi$).}
	\label{fig:Mspec_vs_Mseis_PS_GWVIR}
\end{figure}

\begin{figure}
\centering
 \includegraphics[clip,width=1.0\linewidth]{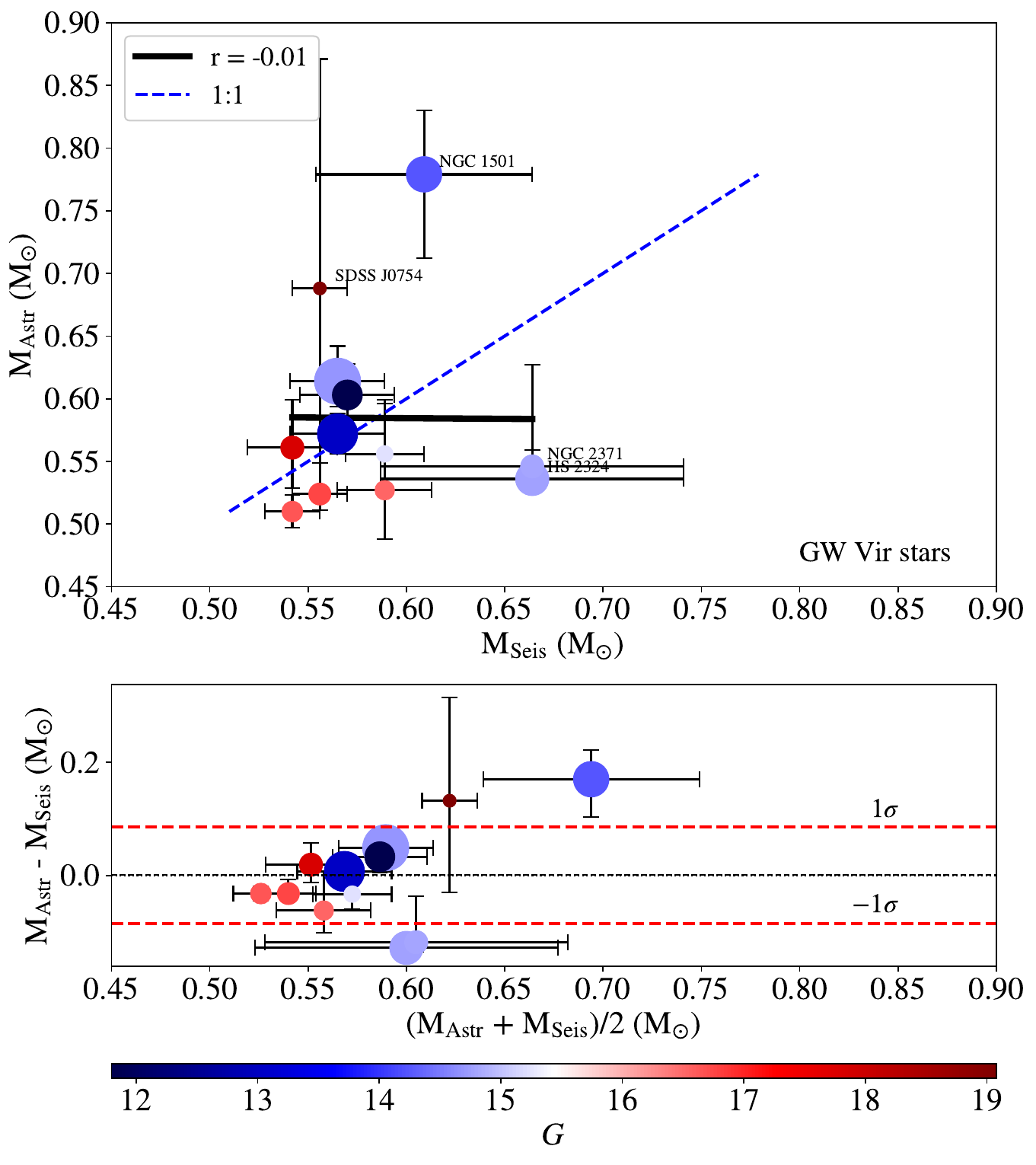}
	\caption{Similar to Fig. \ref{fig:Mspec_vs_Mseis_GWVIR}, this figure displays the comparison between seismological and astrometric masses.}
	\label{fig:Mseis_vs_Mastr_GWVIR}
\end{figure}

\begin{figure}
\centering
 \includegraphics[clip,width=1.0\linewidth]{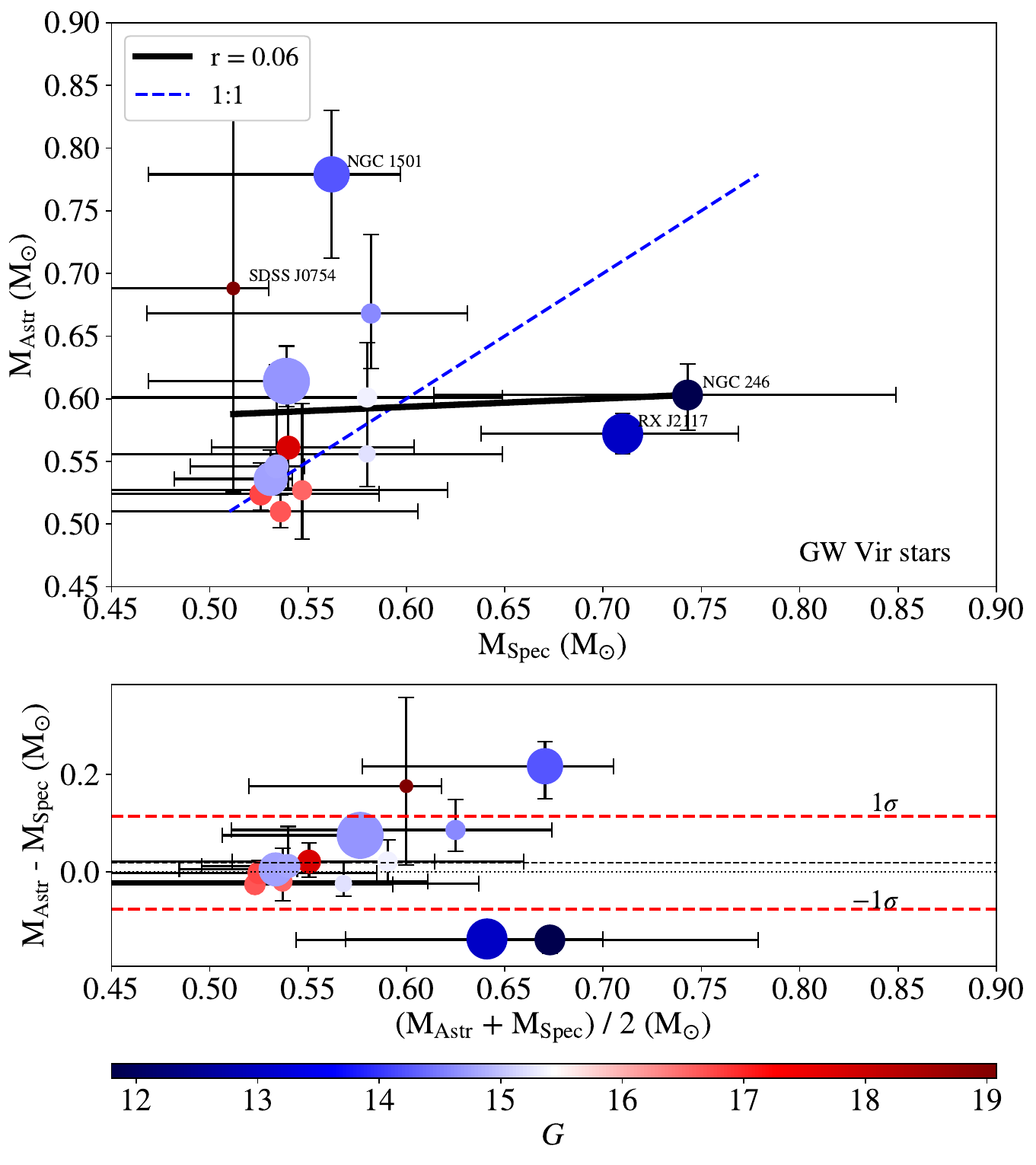}
	\caption{In analogy to Fig. \ref{fig:Mspec_vs_Mseis_GWVIR}, this figure illustrates the comparison between spectroscopic and astrometric masses.}
	\label{fig:Mspec_vs_Mastr_GWVIR}
\end{figure}

In this section, we present comparisons between $M_{\rm Spec}$ and $M_{\rm Seis}$, $M_{\rm Spec}$ and $M_{\rm Seis}(\Delta \Pi)$, $M_{\rm Seis}$ and $M_{\rm Astr}$, and $M_{\rm Spec}$ and $M_{\rm Astr}$ for GW Vir stars in Figures \ref{fig:Mspec_vs_Mseis_GWVIR}, \ref{fig:Mspec_vs_Mseis_PS_GWVIR}, \ref{fig:Mseis_vs_Mastr_GWVIR}, and \ref{fig:Mspec_vs_Mastr_GWVIR}, respectively. As we established before, for this class of pulsating stars we do not have available photometric radii or masses, which prevents us from considering comparisons with $M_{\rm Phot}$.

Let us begin by examining the comparison between the spectroscopic and seismological masses of GW Vir stars. From the upper panel of Fig. \ref{fig:Mspec_vs_Mseis_GWVIR}, it is evident that there is generally good agreement between the two sets of masses, as indicated by their proximity to the 1:1 identity line. However, certain stars exhibit significant discrepancies. For example, the seismological masses are significantly larger (by approximately 20\%) than the spectroscopic masses for NGC~2371 and HS~2324+3944, while for others, such as RX~J2117+3412 and NGC~246, the spectroscopic mass exceeds the seismological mass by approximately 30\%. In this case, the Pearson coefficient is small and negative ($r= -0.15$), suggesting a weak and meaningless anticorrelation between the two sets of masses. Pearson's linear correlation analysis loses meaning in this case.
On the other hand, the Bland-Altman diagram (bottom panel) illustrates that most stars cluster near the mean value of the mass differences, which is very small ($\langle \Delta M_{\star} \rangle= 0.013 M_{\sun}$). This indicates a negligible bias between the two mass sets, with only four outlier stars deviating beyond the limits of agreement ($\langle\Delta M_{\star}\rangle \pm 0.087 M_{\sun}$). These outliers include RX~J2117+2117 and NGC~246 (where $M_{\rm Spec} > M_{\rm Seis}$), and HS~2324+3944 and NGC~2371 (where $M_{\rm Spec} < M_{\rm Seis}$).

Something completely analogous happens when comparing the spectroscopic masses with the seismological masses derived from the period spacing (Fig. \ref{fig:Mspec_vs_Mseis_PS_GWVIR}). Once again, significant disparities are evident between the spectroscopic and seismological masses of the same stars NGC~2371, HS~2324+3944, RX~J2117+3412, and NGC~246. In this case, the Pearson coefficient is $r= -0.25$, indicating a weak anticorrelation between $M_{\rm Seis}(\Delta \Pi)$ and $M_{\rm Spec}$. A feature worth highlighting is that the uncertainties of the spectroscopic masses are by far greater than the seismological ones. This is due to the enormous uncertainties in the spectroscopic parameters of GW Vir stars, in particular in surface gravity (see Table \ref{table:sample_GWVIR}). This is a clear symptom of the difficulty in modelling the atmospheres of these extremely hot stars (see Sect. \ref{discussion}). 

The comparison between seismological and astrometric masses of GW Vir stars is depicted in Fig. \ref{fig:Mseis_vs_Mastr_GWVIR}. Here, the Pearson coefficient is extremely small ($r= -0.01$), indicating a negligible correlation between $M_{\rm Seis}$ and $M_{\rm Astr}$. The Bland-Altman diagram reveals an almost negligible bias between the seismological and astrometric masses, with an average difference of $\langle \Delta M_{\star} \rangle= 0.0004 M_{\sun}$. This suggests a very close agreement between both methods for deriving the masses of GW Vir stars on average.
Moreover, the diagram highlights the presence of some outliers, whose mass differences exceed the agreement limits. These limits, set at $\langle\Delta M_{\star}\rangle \pm 0.085 M_{\sun}$, are surpassed by SDSS~J0754+0852 and NGC~1501, where $M_{\rm Astr} > M_{\rm Seis}$, as well as HS~2324+3944 and NGC~3271, which exhibit $M_{\rm Astr} < M_{\rm Seis}$.

Finally, in Fig. \ref{fig:Mspec_vs_Mastr_GWVIR}, we explore the compatibility between spectroscopic and astrometric masses. Once again, the large size of the error bars associated with the spectroscopic masses is noteworthy, indicating the substantial uncertainties in deriving $\log g$ through spectroscopy. In this case, the Pearson coefficient is $r= +0.06$, indicating again a negligible linear correlation. The average of the mass differences is in this case $\langle \Delta M_{\star} \rangle= 0.013 M_{\sun}$ , and the agreement limits are $\langle\Delta M_{\star}\rangle \pm 0.101 M_{\sun}$. Also notable in this case is the strong discrepancy between the spectroscopic mass and the astrometric mass in several objects, such as NGC~1501, SDSS~J0754+0852, NGC~246, and RX~J2117+3412.

In summary, there is consensus among the sets of stellar masses for some GW Vir stars derived using three (or four) different methods, although there are outliers where mass determinations by various methods display significant discrepancies. In the following, we delve into each of these cases.

\begin{itemize}
\item[-] HS~2324+3944: For this star, we find $M_{\rm Seis}(\Delta \Pi) > M_{\rm Seis} > M_{\rm Astr} \simeq M_{\rm Spec}$. This suggests that the seismological masses are overestimated, particularly the one based on period spacing, while the spectroscopic mass, which closely matches the astrometric mass, appears to be well-determined.

\item[-] NGC~2371: The situation for this star is analogous to the case of HS~2324+3944, where the different estimates of the stellar mass verify the inequalities $M_{\rm Seis}(\Delta \Pi) > M_{\rm Seis} > M_{\rm Astr} \gtrsim M_{\rm Spec}$. So, again, we conclude that the seismological masses are overestimated.

\item[-] RX~J2117+3412: This star presents a notable scenario where $M_{\rm Spec} > M_{\rm Astr} \gtrsim M_{\rm Seis}(\Delta \Pi) \gtrsim M_{\rm Astr}$. Unlike previous cases, the spectroscopic mass of RX~J2117+3412 appears to be overstated, suggesting a potential need to re-assess the parameters $T_{\rm eff}$ and $\log g$ derived from spectroscopic analysis.

\item[-] NGC~246: This is a twin case to the RX~J2117+3412 case ($M_{\rm Spec} > M_{\rm Astr} \gtrsim M_{\rm Seis}(\Delta \Pi) \gtrsim M_{\rm Astr}$). We suggest that the spectroscopic parameters of NGC~246, and thus the evaluation of its spectroscopic mass, need to be revised. Both NGC~246 and RX~J2117+3412 exhibit P~Cygni line profiles in the ultraviolet wavelength region that allowed to measure their mass-loss rates \citep{WernerKoesterke1998}. It would be worthwhile to repeat the analysis of their optical spectra using expanding model atmospheres instead of hydrostatic ones which were employed to determine the spectroscopic parameters.

\item[-] NGC~1501: In this case, we find $M_{\rm Astr} > M_{\rm Seis} \simeq M_{\rm Seis}(\Delta \Pi) \gtrsim M_{\rm Spec}$. In this way, we conclude that the seismological and spectroscopic masses of NGC~1501 are quite low, and then $T_{\rm eff}$ and $\log g$ need to be revised. We remark that it is not trivial to determine effective temperature and surface gravity of [WR] stars, because they have extended atmospheres. One
has to define precisely the stellar radius, as this affects the values of temperature and gravity. 
In addition, using such atmosphere models as boundary conditions for evolution calculations could be significant for the location of the tracks in the Hertzsprung-Russell Diagram (HRD).

\item[-] SDSS~J0754+0852: For this star, we have $M_{\rm Astr} > M_{\rm Seis} > M_{\rm Spec}$, 
the situation being completely analogous to the case of NGC~1501: the seismological and spectroscopic masses are quite small. We suggest revising the seismological and spectroscopic masses and reevaluating the spectroscopic parameters of SDSS~J0754+0852. We note, however, that the 
astrometric mass error is the largest of all GW~Vir stars because it is very faint.

\end{itemize} 

\section{Discussion}
\label{discussion}

The results of the previous section point to several cases of
important discrepancies between the masses derived through different
  methods for the three classes
  of pulsating WDs examined. In particular, we have found
situations in which the spectroscopic mass is in significant
disagreement with the other mass determinations. In other cases, it is
the seismological mass that is dissonant. In this section, we discuss
some of the possible reasons that could explain, at least in part, the
discrepancies found.

When considering stars with spectroscopic masses that disagree with other determinations, it becomes apparent that the uncertainties in determining $M_{\rm Spec}$ for WDs are closely related to uncertainties in determining the effective temperature and surface gravity. The accuracy of spectroscopic determinations depends heavily on the input physics incorporated into atmosphere models, particularly with respect to the reliability of modelled line profiles, especially in ultraviolet and optical spectra. Factors such as line broadening and convective energy transport significantly influence the shape and intensity of spectral lines \citep{2022PhR...988....1S}. 

Various systematic effects impact the assessment of $T_{\rm eff}$ and $\log g$ for WDs, particularly pulsating WDs. \cite{2017PhDT........20F} spectroscopically observed 122 DA WDs that either pulsate or are close to the DAV instability strip and estimated $T_{\rm eff}$ and $\log g$ for each WD based on Balmer line profile shapes. They conducted a meticulous study of several systematics involved in data reduction and spectral fitting procedures, including extinction correction, flux calibration, and signal-to-noise ratios of the spectra, which can affect the final atmospheric parameters $T_{\rm eff}$ and $\log g$. They concluded that neglecting these systematic effects could introduce errors in atmospheric parameter determinations, potentially affecting the determination of spectroscopic masses for DAVs.

Similarly to DAVs, the determination of $T_{\rm eff}$ and $\log g$ for DBVs is subject to uncertainties arising from various sources, such as the input physics of atmosphere models, signal-to-noise ratio of spectra, use of different data sets, and accuracy of flux calibrations \citep[see, e.g.,][]{2023MNRAS.520.2843I}. Specifically, in the case of the model atmosphere fit, precise values of $T_{\rm eff}$ are difficult to obtain within the range of $\sim 21,000 - 31,000$ K. Here, a plateau in the strength of the HeI absorption lines results in hot and cold solutions for $T_{\rm eff}$ due to the insensitivity of these lines to changes in temperature \citep{2011ApJ...737...28B}. This overlap with the DBV instability strip complicates their characterisation \citep{2022ApJ...927..158V}.

Notably, the photometric method provides an alternative value of the effective temperature as part of the fitting process itself, as previously mentioned. In fact, the photometric technique in DA WDs appears to be more accurate than the spectroscopic technique for the $T_{\rm eff}$ determination, as \cite{2019ApJ...871..169G} show. However, these authors also state that both techniques have a similar accuracy at determining the stellar masses of DA WDs. For DB WDs, \cite{2019ApJ...882..106G} report that both techniques yield the effective temperature with comparable accuracy, but the photometric technique is a superior option for the estimation of WD masses. Overall, this technique is generally more advantageous because broadband fluxes are significantly less affected by the complexities of the atomic physics and the equation-of-state compared to line profiles. Nonetheless, its accuracy relies on the photometric calibration \citep{2021A&ARv..29....4S}.

Finally, both $T_{\rm eff}$ and $\log g$ in GW Vir stars exhibit considerable uncertainties. As shown in Table~\ref{table:sample_GWVIR}, $T_{\rm eff}$ uncertainties are generally manageable, typically being within $10\%$. When UV spectra or high-resolution and high S/N ratio optical spectra are available, the error can decrease to approximately $5\%$. In particular, the presence of metal lines, often found in UV or optical spectra, serves as a valuable tool to constrain $T_{\rm eff}$ \citep{2014A&A...569A..99W}. However, $\log g$ uncertainties pose a significant challenge. Although errors for DAs and DBs typically hover around 0.05 dex, they can escalate to approximately 0.5 dex for GW Vir stars, with only a few exceptions as low as 0.3 or 0.2 dex, as detailed in Table~\ref{table:sample_GWVIR}. The main problem is that the primary gravity indicator, the wings of the HeII line, exhibits weak sensitivity to variations of $\log g$, which directly affects the determination of gravity. Existent uncertainties in line-broadening theory also affect this determination. Moreover, the normalisation of optical spectra introduces systematic errors due to challenges in determining the true continuum, particularly given the width of the HeII (and CIV) lines. It is important to note that in the determinations of $\log g$ of GW Vir stars from the T\"ubingen group \citep[see e.g.,][]{2006PASP..118..183W} not only are internal errors taken into account, coming from spectrum fits, but systematic errors are also accounted for, an aspect that may explain the relatively higher values of the uncertainty of $\log g$ compared to the other classes of pulsating WD stars studied in this work. 

When considering stars with notably discrepant seismological masses, it becomes apparent that revisions may be necessary in the derivation of seismological models. Our analysis suggests that neither the number of pulsation periods nor the apparent brightness of pulsating WDs can be directly linked to the reliability of $M_{\rm Seis}$ determination. Specifically, having numerous periods available for asteroseismology does not guarantee the identification of a single and robust seismological solution. Instead, it is the distribution of these periods, particularly in terms of radial order, that significantly influences solution uniqueness or degeneracy. This issue has been exemplified in the case of DAVs by \cite{2017A&A...598A.109G}, who demonstrated that even with a limited number of modes, precise determination of core chemical stratification can be achieved due to the considerable sensitivity of certain confined modes to partial mode trapping effects. Moreover, they highlighted that the ability to unravel the core structure and obtain a unique seismological model depends on the information content of available seismological data in terms of the weight functions\footnote{Weight functions pinpoint the internal regions of the star that influence the period of the mode, providing insight into its sensitivity to specific parts \citep{1985ApJ...295..547K,2023RNAAS...7..166T}.} of the observed $g$-modes. In some cases, the isolation of a unique and well-defined seismological solution can prove challenging, leaving the problem degenerate. An additional factor contributing to potential errors in seismological model derivation is the inherent symmetry observed in high-overtone stellar pulsations as they probe both the core and envelope of pulsating WDs \citep{2003MNRAS.344..657M}. This symmetry has the potential to introduce ambiguity in the derived internal structure locations, as well as in the seismological models themselves. Finally, the choice of the asteroseismological approach and the criteria for selecting seismological models, when faced with a range of potential solutions, may influence the determination of the seismological mass, eventually leading to discrepancies compared to masses derived from other methods. In other cases, for stars with a limited number of periods, seismological analyses may resort to models constrained by external parameters such as spectroscopic $T_{\rm eff}$ and $\log g$. In these instances, we observe an obvious close alignment between the seismological and spectroscopic masses.

We close this section by pointing out that there could be possible systematic uncertainties in the WD evolutionary tracks that would affect the determination of the four types of mass determination, including the astrometric mass because this method uses the luminosity versus the effective temperature extracted from the evolutionary tracks, as well as the photometric mass, because of the use of the mass-radius relationships. It is important for future studies of this nature to employ alternative sets of WD evolutionary tracks, distinct from those used in this paper generated using the {\tt LPCODE} evolutionary code. Concerning our results regarding the masses of the DAVs, the observed dispersion across various methods becomes particularly notable for masses exceeding approximately $0.75 M_{\sun}$ (Figs. \ref{fig:Mspec_vs_Mseis_DA}, \ref{fig:Mseis_vs_Mastr_DA}, \ref{fig:Mspec_vs_Mastr_DA}, and \ref{fig:Mphot_vs_Mastr_DA}). Assessing whether this phenomenon reflects genuine discrepancies rooted in unexplored aspects of WD structure and evolution requires the examination of a substantially larger sample of objects. Investigating such a broader dataset will be the primary focus of future research.

\section{Conclusions}
\label{sec:conclusions}

In this paper, we conducted a comparative analysis to determine the level of agreement between various methodologies used in determining the stellar mass of isolated pulsating WDs. We computed the stellar mass for a sample of selected DAV, DBV, and GW Vir stars using four distinct approaches: spectroscopy, asteroseismology, astrometry,  and photometry (although the latter was only applied for DAVs and DBVs). Specifically, we used the spectroscopic measurements of $T_{\rm eff}$ and $\log g$, along with WD evolutionary tracks, to estimate spectroscopic masses. Seismological mass values were sourced from the existing literature. Additionally, we derived astrometric masses using evolutionary tracks, spectroscopic effective temperatures, apparent magnitudes, and geometric distances obtained from {\it Gaia} parallaxes, as well as bolometric corrections from model atmospheres.  We also utilized photometric masses and photometric effective temperatures from the literature, which have been determined using {\it Gaia} parallaxes and photometry combined with synthetic fluxes from model atmospheres. These values are then integrated with our evolutionary tracks to derive our own photometric masses. Our methodology involved employing identical evolutionary tracks and WD models in all methods. In particular, for assessing spectroscopic, astrometric, and photometric masses, we applied the same evolutionary tracks linked to sets of WD stellar models used in the asteroseismological analyses to derive seismological masses. This approach ensures coherence and reliability in comparing the four estimates of stellar mass.

The results of our analysis vary according to the category of pulsating WD considered. For DAVs, there is broad consensus among the  four methods for stars with masses up to around $\sim 0.75 M_{\sun}$, but significant inconsistencies emerge for more massive DAVs (Figs. \ref{fig:Mspec_vs_Mseis_DA}, \ref{fig:Mseis_vs_Mastr_DA}, \ref{fig:Mspec_vs_Mastr_DA},  and \ref{fig:Mphot_vs_Mastr_DA}). Assessing whether this phenomenon reflects genuine discrepancies rooted in unexplored aspects of WD structure and evolution requires the examination of a substantially larger sample of objects. Investigating such a broader dataset will be the primary focus of future research. Regarding the examined DBVs, all objects in the sample exhibit astrometric masses that surpass their seismological, spectroscopic, and photometric counterparts (Figs. \ref{fig:Mspec_vs_Mseis_DB}, \ref{fig:Mspec_vs_Mseis_PS_DB}, \ref{fig:Mseis_vs_Mastr_DB}, \ref{fig:Mspec_vs_Mastr_DB}, and \ref{fig:Mphot_vs_Mastr_DB}). Finally, for GW~Vir stars, while some display strong agreement among $M_{\rm Spec}$, $M_{\rm Seis}$, and $M_{\rm Astr}$, others reveal substantial disparities (Figs. \ref{fig:Mspec_vs_Mseis_GWVIR}, \ref{fig:Mspec_vs_Mseis_PS_GWVIR}, \ref{fig:Mseis_vs_Mastr_GWVIR} and \ref{fig:Mspec_vs_Mastr_GWVIR}). The dispersion of $M_{\star}$ values of the three classes of pulsating WDs considered in this paper, depending on the method used, suggests the need to re-assess the derivation of spectroscopic parameters ($T_{\rm eff}$ and $\log g$), as well as to revise the seismological models for some stars. Also, there is a need to continue improvements in the parallax measurements to exclude any possible astrometric error.

Future spectroscopic and photometric observations of pulsating WDs, particularly those deficient in H, are crucial for a more thorough comparison of these stars. We emphasise the importance of discovering pulsating WDs in eclipsing binaries, as this provides an opportunity to independently test seismological models. Ongoing and forthcoming large-scale spectroscopic surveys such as the Large Sky Area Multi-Object Fiber Spectroscopic Telescope \citep[{\it LAMOST};][]{2012RAA....12.1197C}, Sloan Digital Sky Survey V  \citep[{\it SDSS-V};][]{2017arXiv171103234K}, WEAVE \citep[][]{2012SPIE.8446E..0PD}, and 4MOST \citep{2019Msngr.175....3D} will play a significant role in confirming the spectroscopic classifications of pulsating WDs and expanding the sample size.  Furthermore, ongoing and upcoming photometric observations from space missions such as the Transiting Exoplanet Survey Satellite \citep[{\it TESS};][]{2015JATIS...1a4003R} and \citep[{\it PLATO};][]{2014ExA....38..249R}, respectively, as well as ground-based initiatives such as the Large Synoptic Survey Telescope \citep[{\it LSST};][]{2019ApJ...873..111I} and {\it BlackGEM} \citep[][]{2016SPIE.9906E..64B}, will offer valuable insights into the nature of variability among pulsating WDs and allow the discovery of more pulsation periods for better modelling.

\begin{acknowledgements}

  We would like to express our sincere gratitude to our referee,
    Prof. Pierre Bergeron, for his generous support and assistance,
    which significantly enhanced the scientific content of this
    paper. Part of this work was supported by AGENCIA through the
  Programa de Modernizaci\'on Tecnol\'ogica BID 1728/OC-AR, and by the
  PIP 112-200801-00940 grant from CONICET.  M. U. gratefully
  acknowledges funding from the Research Foundation Flanders (FWO)
  through a junior postdoctoral fellowship (grant agreement
  No. 1247624N).  This research has used NASA Astrophysics Data System
  Bibliographic Services and the SIMBAD and VizieR databases, operated
  at CDS, Strasbourg, France.

\end{acknowledgements}

\bibliographystyle{aa}
\bibliography{biblio}

\appendix
\section{Lists of selected objects}
\label{sec:appendix}

In this appendix, we provide tables containing basic information about all the objects analysed in this paper. Table \ref{table:sample_DAVS} corresponds to DAV stars, Table \ref{table:sample_DBVS} is associated with DBV stars, and Table \ref{table:sample_GWVIR} is for GW Vir stars.

\onecolumn

\begin{landscape}

\small\tabcolsep=4pt

\centering

\begin{longtable}{lccccccccccccc}

%\caption{Input catalogue of DA white dwarfs studied in this work.} \\
\caption{Sample of the 37 DAV stars analyzed in this study. In column (a), we show the star names, while columns (b) and (c) correspond to the equatorial coordinates. Column (d) corresponds to the $V$ apparent magnitude, and columns (e), (f), and (g) display the DR3 {\it Gaia} apparent magnitudes. Column (h) depicts the spectral type, while columns (i) and (j) show the spectroscopic effective temperature and surface gravity, respectively, with 3D corrections \citep[following][]{2013A&A...559A.104T} applied. Finally, column (k) shows the DR3 {\it Gaia} parallax \citep{2023A&A...674A...1G}, column (l) depicts the geometric distance from \cite{2021AJ....161..147B} based on the DR3 {\it Gaia} parallax. The last column (m) corresponds to the references from which $T_{\rm eff}$ and $\log g$ have been extracted.}\\

\toprule
                    Name &     RA   &   DEC   &    $V$  &    $G$  &  $G_{\rm BP}$  & $G_{\rm RP}$  & S. Type & $T_{\rm eff}$   & $\log{g}$    &   $\pi$  & $d_{\rm BJ}$ &  Reference   \\     
                         &   (deg)  &  (deg)  &   (mag) &   (mag) &   (mag)        &  (mag)        &         & (K)             &  (cgs)       &   (mas)  &  (pc)        &               \\    
\hline                    
                    (a) &  (b) &  (c)  & (d) & (e)  & (f) & (g) & (h)  & (i)  & (j) &  (k) & (l) & (m)           \\                                          

\midrule
\endhead
\midrule
\multicolumn{9}{r}{{Continued on next page}} \\
\midrule
\endfoot

\bottomrule
\endlastfoot
GD~244 & 344.193167               & 12.880681 & $15.788\pm0.022$ & 15.761 & 15.779 & 15.762 &      DA & $11\,760\pm200$ &  $8.09\pm0.05$ &           $15.98^{+0.04}_{-0.05}$ &  $62.46^{+0.16}_{-0.17}$       & \text{\cite{2019A&ARv..27....7C}} \\
G~226$-$29 & 252.107958           & 59.054969 & $12.306\pm0.006$ & 12.279 & 12.278 & 12.296 &      DA & $12\,510\pm200$ &  $8.35\pm0.05$ &           $91.34^{+0.02}_{-0.02}$ &  $10.95^{+0.01}_{-0.01}$       & \text{\cite{2019A&ARv..27....7C}}   \\ 
HS~0507+0434B &  77.556340        &  4.648649 & $15.403\pm0.025$ & 15.376 & 15.383 & 15.380 &      DA4.6 & $12\,010\pm200$ &  $8.19\pm0.05$ &           $20.13^{+0.04}_{-0.03}$ &  $49.57^{+0.11}_{-0.08}$    & \text{\cite{2019A&ARv..27....7C}} \\
EC~11507$-$1519 & 178.313576  & $-15.610141$  & $16.058\pm0.013$ & 16.031 & 16.046 & 16.042 &   DA4.4 & $12\,440\pm200$ &  $8.20\pm0.05$ &           $14.03^{+0.06}_{-0.04}$ &  $71.05^{+0.33}_{-0.29}$       & \text{\cite{2019A&ARv..27....7C}} \\
L~19$-$2 & 218.281819 & $-81.337256$          & $13.452\pm0.008$ & 13.424 & 13.429 & 13.451 &           DA & $12\,070\pm200$ &  $8.13\pm0.05$ &           $47.88^{+0.02}_{-0.02}$ &  $20.87^{+0.01}_{-0.01}$  & \text{\cite{2019A&ARv..27....7C}} \\   
MCT~2148$-$2911 & 327.916151 & $-28.948080$   & $16.074\pm0.018$ & 16.047 & 16.061 & 16.053 &    DA & $12\,220\pm52 $ &  $7.97\pm0.01$ &           $13.51^{+0.05}_{-0.04}$ &  $73.83^{+0.30}_{-0.28}$         & \text{\cite{2020yCat.1350....0G}} \\
EC~14012$-$1446 & 210.988205 & $-15.019605$   & $15.742\pm0.055$ & 15.716 & 15.728 & 15.705 &    DA & $12\,020\pm200$ &  $8.18\pm0.05$ &           $16.82^{+0.05}_{-0.06}$ &  $59.28^{+0.17}_{-0.18}$         & \text{\cite{2019A&ARv..27....7C}} \\ 
EC~23487$-$2424 & 357.842043 & $-24.138146$   & $15.382\pm0.039$ & 15.355 & 15.391 & 15.348 &    DA & $11\,560\pm200$ &  $8.09\pm0.05$ &           $20.02^{+0.03}_{-0.02}$ &  $49.83^{+0.09}_{-0.08}$         & \text{\cite{2019A&ARv..27....7C}} \\ 
GD~165 & 216.162167 &   9.286603              & $14.371\pm0.013$ & 14.344 & 14.340 & 14.369 &               DA & $12\,220\pm200$ &  $8.11\pm0.05$ &    $29.99^{+0.03}_{-0.02}$ &  $33.30^{+0.03}_{-0.02}$     & \text{\cite{2019A&ARv..27....7C}} \\    
Ross~808 & 240.347167 &  36.807100            & $14.435\pm0.018$ & 14.408 & 14.437 & 14.387 &             DA & $11\,120\pm200$ &  $7.98\pm0.05$ &       $30.47^{+0.02}_{-0.02}$ &  $32.79^{+0.02}_{-0.02}$    & \text{\cite{2019A&ARv..27....7C}} \\
HL~Tau$-$76 &  64.735991 &  27.296754         & $15.057\pm0.027$ & 15.030 & 15.051 & 15.016 &          DA & $11\,470\pm200$ & $7.92\pm0.05 $ &           $29.99^{+0.03}_{-0.02}$ &  $33.30^{+0.03}_{-0.02}$   & \text{\cite{2019A&ARv..27....7C}} \\
GALEX~J004855.2+152149 &  12.229877 &  15.363531    & $18.738\pm0.047$ & 18.711 & 18.692 & 18.696 &     DA & $11\,280\pm131$ & $8.17\pm0.07 $ &           $30.47^{+0.02}_{-0.02}$ &  $32.79^{+0.02}_{-0.02}$        & \text{\cite{2019A&ARv..27....7C}} \\
SDSS~J084314.05+043131.6 & 130.808592 &   4.525483     & $17.863\pm0.021$ & 17.835 & 17.887 & 17.813 &    DA & $11\,220\pm71 $ &  $8.09\pm0.04$ &         $6.59^{+0.11}_{-0.10}$  & $151.42^{+2.60}_{-2.47}$           & \text{\cite{2019A&ARv..27....7C}} \\
GALEX~J125710.5+012423 & 194.293767 &   1.406382    & $18.694\pm0.052$ & 18.665 & 18.705 & 18.779 &     DA & $11\,490\pm156$ & $8.30\pm0.08 $ &           $4.38^{+0.25}_{-0.23}$  & $227.07^{+9.33}_{-7.03}$        & \text{\cite{2019A&ARv..27....7C}} \\
2QZ~J132350.3+010304 & 200.959465 &   1.051161      & $18.576\pm0.040$ & 18.549 & 18.563 & 18.569 &     DA & $11\,380\pm157$ & $8.45\pm0.06 $ &         $5.53^{+0.23}_{-0.25}$ &  $127.36^{+2.73}_{-3.4}$           & \text{\cite{2019A&ARv..27....7C}} \\
GALEX~J161218.1+083028 & 243.075313 &   8.507868    & $17.833\pm0.024$ & 17.806 & 17.841 & 17.843 &   DA & $12\,250\pm126$ & $8.29\pm0.04 $ &         $7.77^{+0.13}_{-0.14}$ & $128.36^{+2.15}_{-2.17}$             & \text{\cite{2019A&ARv..27....7C}} \\
SDSS~J164115.61+352140.6 & 250.315059 &  35.361287     & $19.119\pm0.046$ & 19.092 & 19.174 & 19.091 &    DA & $12\,025\pm185$ & $8.34\pm0.11 $ &         $3.67^{+0.17}_{-0.15}$ & $272.62^{+8.61}_{-5.95}$            & \text{\cite{2019A&ARv..27....7C}} \\
GALEX~J165020.5+301021 & 252.585514 &  30.172546    & $18.179\pm0.026$ & 18.151 & 18.204 & 18.099 &   DA & $10\,830\pm 80$ & $8.43\pm0.05 $ &         $7.74^{+0.11}_{-0.11}$ & $128.32^{+1.89}_{-1.89}$             & \text{\cite{2019A&ARv..27....7C}} \\
GALEX~J220830.0+065448 & 332.125111 &   6.913508    & $17.999\pm0.035$ & 17.972 & 18.003 & 17.974 &     DA & $11\,147\pm 81$ & $8.25\pm0.04 $ &           $7.67^{+0.13}_{-0.11}$ & $130.03^{+2.62}_{-1.13}$         & \text{\cite{2019A&ARv..27....7C}} \\
KIC~11911480 & 290.103625 &  50.289583        & $18.091\pm0.030$ & 18.064 & 18.122 & 18.079 &       DA & $12\,160\pm140$ &  $7.94\pm0.04$ &       $5.50^{+0.09}_{-0.08}$  & $177.57^{+1.26}_{-0.99}$          & \text{\cite{2019A&ARv..27....7C}} \\
GD~1212 & 354.711542 &  $-7.688981$           & $13.320\pm0.011$ & 13.293 & 13.326 & 13.257 &            DA & $10\,970\pm140$ &  $8.03\pm0.05$ &       $53.58^{+0.02}_{-0.02}$ &  $18.65^{+0.01}_{-0.01}$     & \text{\cite{2019A&ARv..27....7C}} \\
GALEX~J134550.9$-$005536 & 206.462178 &  $-0.926792$ & $16.787\pm0.018$ & 16.760 & 16.784 & 16.782 &   DA & $11\,760\pm 40$ & $8.10\pm0.02 $ &        $9.93 ^{+0.07}_{-0.08}$ & $100.82^{+0.54}_{-0.76}$             & \text{\cite{2019A&ARv..27....7C}} \\
SDSS~J215905.5+132256  & 329.773050 &  13.382184     & $18.986\pm0.065$ & 18.959 & 19.027 & 18.978 &   DA & $11\,370\pm151$ & $8.69\pm0.06 $ &         $5.10 ^{+0.24}_{-0.25}$ & $197.84^{+8.13}_{-9.25}$            & \text{\cite{2019A&ARv..27....7C}} \\
BPM~37093 & 189.707421 & $-49.800058$         & $13.820\pm0.007$ & 13.793 & 13.819 & 13.771 &          DA & $11\,620\pm500$ & $8.69\pm0.05 $ &     $67.41^{+0.02}_{-0.02}$ &  $14.83^{+0.01}_{-0.01}$         & \text{\cite{2019A&ARv..27....7C}} \\
GD~518 & 254.812917 &  66.176461              & $17.271\pm0.013$ & 17.244 & 17.270 & 17.245 &               DA & $11\,760\pm210$ &  $8.97\pm0.06$ &  $15.48^{+0.06}_{-0.05}$ &  $64.52^{+0.25}_{-0.22}$       & \text{\cite{2019A&ARv..27....7C}} \\
TIC~8445665 & 246.153382 &  32.214668         & $16.732\pm0.032$ & 16.705 & 16.744 & 16.698 &        DA & $11\,385\pm235$ & $7.95\pm0.04 $ &         $10.24^{+0.05}_{-0.06}$ &  $97.35^{+0.49}_{-0.54}$       & \text{\cite{2022MNRAS.511.1574R}} \\
TIC~46847635 & 142.319587 &  $-8.675615$      & $16.775\pm0.014$ & 16.748 & 16.768 & 16.785 &     DA & $12\,018\pm344$ &  $7.98\pm0.05$ &         $9.65 ^{+0.08}_{-0.07}$ & $103.29^{+0.81}_{-0.78}$          & \text{\cite{2022MNRAS.511.1574R}} \\
TIC~167486543 &  72.133797 & $-10.897180$     & $16.260\pm0.029$ & 16.233 & 16.245 & 16.241 &      DA & $12\,187\pm252$ &  $8.55\pm0.03$ &           $18.20^{+0.05}_{-0.06}$ &  $54.82^{+0.17}_{-0.19}$       & \text{\cite{2022MNRAS.511.1574R}} \\
TIC~441500792 &  46.701453 & $-17.392481$     & $16.709\pm0.046$ & 16.683 & 16.713 & 16.671 &    DA & $11\,393\pm273$ &  $8.05\pm0.05$ &         $10.97^{+0.07}_{-0.06}$ &  $90.97^{+0.57}_{-0.51}$           & \text{\cite{2022MNRAS.511.1574R}} \\
TIC~442962289 &  81.448483 & $-17.563872$     & $16.535\pm0.032$ & 16.508 & 16.531 & 16.512 &    DA & $11\,945\pm252$ &  $8.42\pm0.03$ &         $14.74^{+0.05}_{-0.06}$ &  $67.69^{+0.23}_{-0.25}$           & \text{\cite{2022MNRAS.511.1574R}} \\
TIC~686044219 &  65.454141 & $-35.980307$     & $17.158\pm0.045$ & 17.131 & 17.148 & 17.118 &    DA & $11\,477\pm356$ &  $8.02\pm0.06$ &         $8.71 ^{+0.06}_{-0.05}$ & $114.51^{+0.85}_{-0.57}$           & \text{\cite{2022MNRAS.511.1574R}} \\
TIC~712406809 &  99.821860 &   1.224859       & $16.246\pm0.036$ & 16.218 & 16.201 & 16.130 &      DA & $10\,669\pm298$ & $7.91\pm0.07 $ &         $13.53^{+0.09}_{-0.09}$ &  $73.73^{+0.53}_{-0.53}$         & \text{\cite{2022MNRAS.511.1574R}} \\
TIC~20979953 & 233.387357 &  $-2.115480$      & $16.540\pm0.016$ & 16.513 & 16.541 & 16.522 &     DA & $11\,859\pm236$ & $7.97\pm0.04 $ &         $10.93^{+0.09}_{-0.11}$ &  $91.24^{+0.70}_{-0.76}$          & \text{\cite{2022MNRAS.511.1574R}} \\
TIC~55650407 &  73.863635 & $-62.979060$      & $14.998\pm0.015$ & 14.971 & 14.984 & 14.984 &     DA & $11\,838\pm150$ &  $7.95\pm0.02$ &         $21.65^{+0.02}_{-0.02}$ &  $46.11^{+0.05}_{-0.05}$          & \text{\cite{2022MNRAS.511.1574R}} \\
TIC~282783760 & 198.611837 &  17.535972       & $16.306\pm0.023$ & 16.279 & 16.278 & 16.285 &        DA & $12\,111\pm268$ & $8.03\pm0.03 $ &           $12.39^{+0.05}_{-0.04}$ &  $80.58^{+0.32}_{-0.30}$     & \text{\cite{2022MNRAS.511.1574R}} \\
BPM~31594 &  55.870974 & $-45.817843$         & $15.069\pm0.035$ & 15.042 & 15.063 & 15.039 &          DA & $11\,500\pm22 $ &  $8.05\pm0.01$ &           $22.59^{+0.02}_{-0.02}$ &  $44.20^{+0.05}_{-0.05}$   & \text{\cite{2019A&ARv..27....7C}} \\
G~29$-$38 & 352.198485 &   5.248399           & $13.089\pm0.029$ & 13.062 & 13.084 & 13.053 &            DAZ & $11\,910\pm162$ & $8.17\pm0.04 $ &        $57.10^{+0.03}_{-0.03}$ &  $17.51^{+0.01}_{-0.01}$   & \text{\cite{2019A&ARv..27....7C}} \\
\label{table:sample_DAVS}
\end{longtable}

\end{landscape}

\begin{landscape}

\small\tabcolsep=2.2pt

\centering

\begin{longtable}{lccccccccccccc}

%\caption{Input catalogue of DB white dwarfs studied in this work.} \\
\caption{Sample of the nine DBV  stars considered in this work. Each column contains information similar to that presented in Table \ref{table:sample_DAVS}. In this case, column (m) gives the interstellar extinction measured in the $V$ band, computed from the reddening $E(B-V)$ for each star. The spectroscopic effective temperature and surface gravity are those tabulated by \cite{2019A&ARv..27....7C} which have been corrected by 3D effects following \cite{2018MNRAS.481.1522C}, unless another reference is indicated.}\\

\toprule
                    Name &     RA   &   DEC   &    $V$  &    $G$  &  $G_{\rm BP}$  & $G_{\rm RP}$  & S. Type & $T_{\rm eff}$   & $\log{g}$    &   $\pi$  & $d_{\rm BJ}$ & $A_V$  & Reference   \\     
                         &   (deg)  &  (deg)  &   (mag) &   (mag) &   (mag)        &  (mag)        &         & (K)             &  (cgs)       &   (mas)  &  (pc)        &  (mag)    &             \\    
\hline                    
                    (a) &  (b) &  (c)  & (d) & (e) & (f) & (g) & (h)  & (i)  & (j) &  (k) & (l) & (m) &  (n)          \\ 
                                          
\midrule
\endhead
\midrule
\multicolumn{9}{r}{{Continued on next page}} \\
\midrule
\endfoot

\bottomrule
\endlastfoot
KIC~8626021  	&  292.269496 &  44.785841		& $18.545\pm 0.033$ & 18.500 & 18.409  & 18.692  &   DB	  &  $28\,480\pm750$	&  $7.89\pm0.07$	& $2.65^{+0.11}_{-0.14}$	 &   $382.39^{+13.54}_{-16.00}$	 &   $0.152^{+0.017}_{-0.027}$    &   \text{\cite{2019A&ARv..27....7C}}    \\
KUV~05134+2605	&  79.115780  &  26.143762		& $16.795\pm0.019$ & 16.749 & 16.657	& 16.923  &   DBA	  &  $24\,680\pm1300$	&  $8.21\pm0.06$	& $6.69^{+0.07}_{-0.07}$	 &   $149.77^{+1.72}_{-1.80}$	 &   $0.063^{+0.026}_{-0.023}$    &    \text{\cite{2019A&ARv..27....7C}}   \\
TIC~257459955	&  30.236306  &  $-15.769369$	& $14.707\pm0.016$ & 14.657 & 14.554	& 14.846  &   DB	  &  $25\,500\pm1369$	&  $7.88\pm0.03$	& $14.57^{+0.04}_{-0.04}$	 &   $68.51^{+0.12}_{-0.11}$	 &   $0.025^{+0.027}_{-0.017}$    &   \text{\cite{2019A&ARv..27....7C}}    \\
GD~358       	&  251.826636 &  32.475796		& $13.630\pm0.020$ & 13.580 & 13.476	& 13.770  &   DB2.3 &  $24\,940\pm1018$	&  $7.75\pm0.05$	& $23.24^{+0.02}_{-0.03}$	 &   $42.99^{+0.04}_{-0.05}$	 &   $0.010^{+0.008}_{-0.007}$    &   \text{\cite{2019A&ARv..27....7C}}     \\
PG~1351+489  	&  208.292000 &  48.672500		& $16.728\pm0.024$ & 16.678 & 16.586	& 16.877  &   DB	  &  $26\,010\pm1536$	&  $7.91\pm0.07$	& $5.69^{+0.05}_{-0.06}$	 &   $175.73^{+1.53}_{-1.58}$	 &   $0.062^{+0.011}_{-0.019}$    &   \text{\cite{2019A&ARv..27....7C}}    \\
EC~20058$-$5234	&  302.416147 &  $-52.421304$	& $15.806\pm0.016$ & 15.756 & 15.647	& 15.938  &   DB	  &  $25\,500\pm500$	&  $8.01\pm0.05$	& $8.48^{+0.05}_{-0.06}$	 &   $117.95^{+0.68}_{-0.75}$	 &   $0.181^{+0.108}_{-0.108}$  &  \text{\cite{2019A&ARv..27....7C}}    \\
EC~04207$-$4748	&  65.547319  &  $-47.695004$	& $15.280\pm0.022$ & 15.229 & 15.128	& 15.423  &   DB	  &  $25\,970\pm545$	&  $7.79\pm0.06$	& $10.93^{+0.03}_{-0.03}$	 &   $91.48^{+0.22}_{-0.23}$	 &   $0.040^{+0.024}_{-0.024}$  &  \text{\cite{2019A&ARv..27....7C}}    \\ %HE 0420−4748
WD~J152738.37$-$450207.3 &  231.909792 &  $-45.035391$	& $15.527\pm0.018$ & 15.482 & 15.375	& 15.634  &   DB	  &  $25\,228\pm630$	&  $8.12\pm0.01$	& $10.54^{+0.04}_{-0.05}$	 &   $94.62^{+0.34}_{-0.37}$	 &   $0.927^{+0.566}_{-0.566}$  &   \text{\cite{2022A&A...668A.161C}}  \\%WD~J1527$-$4502
L~7$-$44       	&  263.228940 &  $-87.167376$	& $14.785\pm0.011$ & 14.737 & 14.639	& 14.918  &   DB	  &  $23\,980\pm1686$	&  $8.05\pm0.03$	& $14.47^{+0.02}_{-0.01}$	 &   $69.09^{+0.14}_{-0.12}$	 &   $0.544 ^{+0.326}_{-0.326}$  &   \text{\cite{2018ApJ...857...56R}}   \\   
\label{table:sample_DBVS}
\end{longtable}

\end{landscape}

\begin{landscape}

\small\tabcolsep=1.8pt

\centering

%\movetabledown=3mm
\begin{longtable}{lcccccccccccccc}

%\caption{Input catalogue of DB white dwarfs studied in this work.} \\
\caption{Sample of the fourteen GW Vir stars considered in this work. Each column contains information similar to that presented in Table \ref{table:sample_DBVS}. The spectroscopic effective temperature and surface gravity are those tabulated by \cite{2019A&ARv..27....7C}, unless another reference is indicated. Since PG~2131+066  lacks parallax data from {\it Gaia}, we are unable to obtain the \cite{2021AJ....161..147B}'s distance, therefore, we include here the distance derived by \cite{2000ApJ...545..429R} using the spectroscopic parallax of the M star, the primary component of the binary system to which PG~2131+066 belongs.}\\

\toprule
                    Name &     RA   &   DEC   &    $V$  &    $G$  &  $G_{\rm BP}$  & $G_{\rm RP}$  & S. Type & $T_{\rm eff}$   & $\log{g}$    &   $\pi$  & $d_{\rm BJ}$ & $A_V$  & Reference   \\     
                         &   (deg)  &  (deg)  &   (mag) &   (mag) &   (mag)        &  (mag)        &         & (K)             &  (cgs)       &   (mas)  &  (pc)        &  (mag)    &             \\    
\hline                    
                    (a) &  (b) &  (c)  & (d) & (e)  & (f)  & (g) & (h) & (i)  & (j)  & (k) &  (l) & (m) & (n)      \\ 
                                          
\midrule
\endhead
\midrule
\multicolumn{9}{r}{{Continued on next page}} \\
\midrule
\endfoot

\bottomrule
\endlastfoot
PG~0122+200  	&  21.343833	&   20.299097	 & $16.845\pm0.019$ & 16.751 & 16.593 & 17.109 & PG~1159 & $80\,000\pm4000$     & $7.50\pm0.50$ & $1.64^{+0.08}_{-0.07}$ & $618^{+42}_{-32}$	   & $0.133^{+0.017}_{-0.008}$  &  \text{\cite{2014A&A...569A..99W}}  \\
PG~2131+066  	&  323.534167	&   6.849331	 & $16.577\pm0.100$ & 16.545 & 16.356	& 16.160 & PG~1159 & $95\,000\pm5000$     & $7.50\pm0.50$ & $\hdots$               & $681^{+170}_{-137}$   & $0.120^{+0.010}_{-0.022}$  &  \text{\cite{2014A&A...569A..99W}}  \\
PG~1707+427  	&  257.198625	&   42.683578	 & $16.757\pm0.027$ & 16.648 & 16.452	& 17.027 & PG~1159 & $85\,000\pm5000$     & $7.50\pm0.50$ & $1.40^{+0.05}_{-0.03}$ & $733^{+32}_{-26}$     & $0.145^{+0.006}_{-0.020}$  &  \text{\cite{2014A&A...569A..99W}} \\
SDSS J075415.12+085232.2	&  118.562991	&   8.875609	 & $19.186\pm0.046$ & 18.078 & 18.925 & 19.495 & PG~1159 & $120\,000\pm10\,000$ & $7.00\pm0.30$ & $0.57^{+0.23}_{-0.16}$ & $1800^{+1200}_{-500}$ & $0.095^{+0.014}_{-0.012}$  &  \text{\cite{2014A&A...564A..53W}} \\
SDSS J034917.41$-$005919.3 &  57.322537	& $-0.988675$	 & $17.845\pm0.026$ & 17.798 & 17.694 & 17.968 & PG~1159 & $90\,000\pm 900$     & $7.50\pm0.01$ & $1.15^{+0.12}_{-0.09}$ & $850^{+120}_{-80}$    & $0.392^{+0.020}_{-0.0.18}$  &  \text{\cite{2006A&A...454..617H}}\\
RX~J2117.1+3412	&  319.284507	&   34.207616	 & $13.120\pm0.009$ & 13.022 & 12.842 & 13.375 & PG~1159 & $170\,000\pm10\,000$ & $6.00\pm0.03$ & $1.99^{+0.04}_{-0.05}$ & $499^{+8}_{-9}$       & $0.164^{+0.027}_{-0.007}$  &  \text{\cite{2005A&A...433..641W}}\\
HS~2324+3944 	&  351.816397	&   40.023175	 & $14.821\pm0.012$ & 14.770 & 14.653	& 14.958 & PG41159 & $130\,000\pm10\,000$ & $6.20\pm0.20$ & $0.70^{+0.03}_{-0.02}$ & $1400^{+74}_{-54}$    & $0.474^{+0.022}_{-0.043}$  &  \text{\cite{1996A&A...309..820D}} \\
NGC~6905     	&  305.595800	&   20.104524	 & $14.634\pm0.026$ & 14.597 & 14.443 & 14.637 &  [WC3]  & $141\,000\pm10\,000$ & $6.00\pm0.20$ & $0.37^{+0.03}_{-0.03}$ & $2700^{+200}_{-200}$  & $0.266^{+0.013}_{-0.063}$  &  \text{\cite{1996AJ....111.2332C}}  \\% Montreal(Stilism)
NGC~2371     	&  111.394535	&   29.490675	 & $14.853\pm0.025$ & 14.778 & 14.620 & 15.053 &   [WO1]   & $135\,000\pm10\,000$ & $6.30\pm0.20$ & $0.58^{+0.05}_{-0.04}$ & $1692^{+143}_{-122}$  & $0.158^{+0.005}_{-0.018}$  &  \text{\cite{2004ApJ...609..378H}}  \\% Montreal(Stilism)
NGC~1501     	&  61.747467	&   60.920631	 & $14.329\pm0.023$ & 14.237 & 14.470	& 13.873 & [WO4]   & $134\,000\pm10\,000$ & $6.00\pm0.20$ & $0.58^{+0.02}_{-0.01}$ & $1645^{+54}_{-47}$    & $1.829^{+0.047}_{-0.115}$  &  \text{\cite{1997A&A...320...91K}} \\% Montreal(Stilism)
TIC~333432673	&  100.315183	&   $-13.689936$ & $15.290\pm0.021$ & 15.215 & 15.058	& 15.492 & PG~1159 & $120\,000\pm10\,000$ & $7.50\pm0.50$ & $2.55^{+0.04}_{-0.03}$ & $389^{+6}_{-5}$       & $0.458^{+0.011}_{-0.142}$  &  \text{\cite{2021A&A...655A..27U}} \\
TIC~095332541	&  90.687447	&   $-13.850989$ & $15.413\pm0.020$ & 15.319 & 15.139	& 15.656 & PG~1159 & $120\,000\pm10\,000$ & $7.50\pm0.50$ & $2.59^{+0.04}_{-0.03}$ & $384^{+6}_{-5}$       & $0.186^{+0.019}_{-0.026}$  &  \text{\cite{2021A&A...655A..27U}} \\
PG~1159$-$035  	&  180.441555	&   $-3.7612860$ & $15.808\pm0.015$ & 14.693 & 14.485	& 14.080 & PG~1159 & $140\,000\pm5000$    & $7.00\pm0.50$ & $1.69^{+0.06}_{-0.06}$ & $585^{+20}_{-21}$     & $0.063^{+0.010}_{-0.005}$  &  \text{\cite{2011A&A...531A.146W}} \\
NGC~246      	&  11.763925	&   $-11.871936$ & $11.905\pm0.012$ & 11.797 & 11.597	& 12.169 & PG~1159 & $150\,000\pm10\,000$ & $5.70\pm0.10$ & $1.88^{+0.02}_{-0.02}$ & $538^{+20}_{-20}$     & $0.148^{+0.120}_{-0.051}$  &  \text{\cite{2020IAUS..357..158L}}  \\                                                                         
\label{table:sample_GWVIR}
\end{longtable}

\end{landscape}

% Aca estan las referencias de los estudios astrosismologicos donde se analizaron
%
%(1) \cite{2012MNRAS.420.1462R}, (2) \cite{2013ApJ...779...5 & &8R}, (3) \cite{2017ApJ...851...60R}, (4) \cite{2019MNRAS.490.1803R}, (5) %\cite{2019A&A...632A.119C}, (6) \cite{2022MNRAS.511.1574R}, (7) %\cite{2023MNRAS.518.1448R}, (8) \cite{2023MNRAS.526.2846U},  

%(9) \cite{2012A&A...541A..42C}, (10) \cite{2014A&A...570A.116B}, (11) \cite{2019A&A...632A..42B}, (12) \cite{2022A&A...659A..30C}, (13) \cite{2022A&A...668A.161C}

%(1) \cite{2007A&A...475..619C}, (2) \cite{2009A&A...499..257C}, (3) \cite{2014MNRAS.442.2278K}, (4) \cite{2016A&A...589A..40C}, (5) \cite{2021A&A...645A.117C}, (6) \cite{2021A&A...655A..27U}, (7) \cite{2022ApJ...936..187O}, (8) Calcaferro et al. (2023)

\end{document}